\documentclass[aip,amsmath,amssymb, reprint]{revtex4-1}

\usepackage{graphicx}
\usepackage{dcolumn}
\usepackage{bm}
\usepackage{booktabs}
\usepackage{multirow}
\usepackage[utf8]{inputenc}
\usepackage[T1]{fontenc}
\usepackage[version=3]{mhchem}
\usepackage{hyperref} 

\newcommand{\Python}{\textsc{Python}}
\newcommand{\PySCF}{\textsc{PySCF}}
\newcommand{\PyFLOSIC}{\textsc{PyFLOSIC}}
\newcommand{\ERKALE}{\textsc{ERKALE}}
\newcommand{\PyCOM}{\textsc{PyCOM}}
\newcommand{\PyEFF}{\textsc{PyEFF}}
\newcommand{\PyLEWIS}{\textsc{PyLEWIS}}
\newcommand{\fodMC}{\textsc{fodMC}}
\newcommand{\NRLMOL}{\textsc{NRLMOL}}
\newcommand{\NWChem}{\textsc{NWChem}}
\newcommand{\PSIFour}{\textsc{PSI4}}

\newcommand*{\citeref}[1]{ref.~\citenum{#1}}
\newcommand*{\citerefs}[1]{refs.~\citenum{#1}}

\newcommand{\kcal}{\mbox{kcal/mol}}

\let\eqref\undefined
\newcommand*{\eqref}[1]{equation~(\ref{eq:#1})}
\newcommand*{\tabref}[1]{table~\ref{tab:#1}}
\newcommand*{\tabsref}[1]{tables~\ref{tab:#1}}
\newcommand*{\figref}[1]{figure~\ref{fig:#1}}
\newcommand*{\secref}[1]{Section~\ref{sec:#1}}
\newcommand*{\subsecref}[1]{Subsection~\ref{subsec:#1}}

\usepackage{braket}
\usepackage{xcolor}
\usepackage{listings}
\usepackage{bbold}
\usepackage{url}

\begin{document}

\title{\PyFLOSIC{}: \Python{}-based Fermi--L\"owdin orbital self-interaction correction}

\author{Sebastian Schwalbe}
\email{pyflosic@gmail.com}
\author{Lenz Fiedler}
\author{Jakob Kraus}
\author{Jens Kortus} 
\email{jens.kortus@physik.tu-freiberg.de}
\affiliation{
Institute of Theoretical Physics, TU Bergakademie Freiberg, \newline Leipziger Str. 23, D-09599 Freiberg, Germany}
\author{Kai Trepte} 
\affiliation{
Department of Physics, Central Michigan University, \newline Mount Pleasant, MI 48859, USA} 
\author{Susi Lehtola}
\email{susi.lehtola@alumni.helsinki.fi}
\affiliation{Department of Chemistry, University of Helsinki, \newline P.O. Box 55 (A. I. Virtasen aukio 1),  FI-00014 University of Helsinki, Finland
}

\begin{abstract}
We present \PyFLOSIC, an open-source, general-purpose \Python{} implementation of the Fermi--L\"owdin orbital self-interaction correction (FLO-SIC), which is based on the \Python{} simulation of chemistry framework~(\PySCF{}) electronic structure and quantum chemistry code. 
Thanks to \PySCF{}, \PyFLOSIC{} can be used with any kind of Gaussian-type basis set, various kinds of radial and angular quadrature grids, and all exchange-correlation functionals within the local density approximation~(LDA), generalized-gradient approximation~(GGA), and meta-GGA provided in the \textsc{Libxc} and \textsc{XCFun} libraries.
A central aspect of FLO-SIC are Fermi-orbital descriptors, which are used to estimate the self-interaction correction.
Importantly, they can be initialized automatically within \PyFLOSIC{}; they can also be optimized with an interface to the atomic simulation environment, a \Python{} library that provides a variety of powerful gradient-based algorithms for geometry optimization.
Although \PyFLOSIC{} has already facilitated applications of FLO-SIC to chemical studies, it offers an excellent starting point for further developments in FLO-SIC approaches, thanks to its use of a high-level programming language and pronounced modularity.
\end{abstract}

\maketitle

\section{Introduction \label{sec:intro}}

The continuously growing availability of open-source software has given rise to an ongoing paradigm shift in quantum chemical software development.
At variance to the traditional model of monolithic programs, where new algorithms need to be re-implemented separately from scratch in each code, present-day computational science is proceeding in leaps and bounds thanks to the thriving ecosystems of small projects dedicated to solving well-defined sub-problems\cite{Sun2015_1664, Kim2018_195901, Lehtola2018_1, Herbst2019_e1462, Koval2019_188, Iskakov2019_085112} that can be easily combined to form a code that is greater than the sum of its parts; see e.g. \citerefs{Sun2020_024109} and \citenum{Smith2020_184108} and references therein for further discussion.
The efficiency of the new development models comes from not having to "re-invent the wheel" within every program; instead, standardized, modular tools designed to solve specific problems can be reused across the board.
This kind of cleaner code design and efficient code reuse has enabled fast development, and streamlined the adoption of new methodologies, since features added to any one component become straightforwardly available in everything built on top of them.

This work describes \textsc{PyFLOSIC}, an open-source, \Python{}-based implementation of the Fermi--L\"owdin orbital self-interaction correction~(FLO-SIC).\cite{Pederson2014_121103,Pederson2015_064112,Pederson2015_153,Yang2017_052505}
The core routines of \PyFLOSIC{} were developed during Lenz Fiedler's master's thesis,\cite{Fiedler2018_1} and the code is freely available on \textsc{GitHub}~(\url{https://github.com/pyflosic/pyflosic}).
\PyFLOSIC{} builds on the \Python{} simulation of chemistry framework~(\PySCF{}),\cite{Sun2018_e1340,Sun2020_024109} which is an open-source electronic structure and quantum chemistry code written primarily in \Python{}. As a general-purpose quantum chemistry program \PySCF{} already contains a vast number of methods.\cite{Sun2018_e1340,Sun2020_024109} For example, self-consistent field~(SCF) approaches like Hartree--Fock and density functional theory\cite{Hohenberg1964_B864,Kohn1965_A1133}~(DFT) are available with various choices for spin treatment, density fitting techniques, as well as numerical quadrature methods. Second-order orbital optimization is also available in \PySCF{}.\cite{Sun2016_1}

Since \PySCF{} is highly modular, and its parts can be imported like any other \Python{} module, it is rather easy to add new functionality to it or to combine it with existing software,\cite{Sun2018_e1340} as has already been demonstrated by various authors.\cite{Kim2018_195901,Koval2019_188,Iskakov2019_085112} This is also the strategy that was adopted for \PyFLOSIC{}.
Before \PyFLOSIC{}, the FLO-SIC method has only been available in the reference implementation within the Naval Research Laboratory Molecular Orbital Library~(\textsc{NRLMOL}).\cite{Pederson1990_7453,Pederson1991_3891,Pederson1991_7312,Perdew1992_6671,Porezag1996_7830,Porezag1997_1,Porezag1999_2840,Kortus2000_5755,Pederson2000_197}
There is also a  FLO-SIC implementation in a developer branch of the \textsc{NWChem} program,\cite{Aquino2018_6456,Aquino2020_1200} but to the best of our knowledge this implementation is not currently publicly available. 

Similarly to \NRLMOL{} and \NWChem{}, \PySCF{}---and therefore \PyFLOSIC{}---uses a Gaussian-type orbital~(GTO) basis set.\cite{Sun2018_e1340,Sun2020_024109}
However, in contrast to the \textsc{NRLMOL} program, \NWChem{} and \PySCF{} (and \PyFLOSIC{}) are able to routinely handle basis functions with high angular momentum.
The most-commonly used all-electron as well as effective core potential GTO basis sets are already available within \PySCF{}, and additional ones can be downloaded from, e.g., the Basis Set Exchange\cite{Pritchard2019_4814} 
and parsed in from several formats. 

The importance of the tractability of calculations within large basis sets is underlined by recent fully numerical benchmark studies that have shown that the reproduction of atomization energies even within DFT may require several shells of polarization functions.\cite{Jensen2017_1449,Jensen2017_6104,Feller2018_2598,Lehtola2020_134108}
\PySCF{} contains fast routines for the evaluation of molecular integrals via the \textsc{Libcint} library,\cite{Sun2015_1664} which are also used in \PyFLOSIC{}. Moreover, \PyFLOSIC{} inherits OpenMP parallelization from \PySCF{}, enabling the efficient use of multi-core computation architectures.\cite{Sun2020_024109}
By enabling the use of large basis sets in FLO-SIC calculations thanks to the fast elementary routines in \PySCF{}, \PyFLOSIC{} enables reliable computational studies of even molecules for which very large and flexible basis sets are required, such as \ce{SO2} and \ce{SF6},\cite{Jensen2017_1449, Jensen2017_6104, Lehtola2020_134108} as will also be demonstrated later in this work.
Recently developed powerful approaches to handle significant linear dependencies in the underlying molecular basis set\cite{Lehtola2019_241102, Lehtola2020_032504} are also available in \PyFLOSIC{} through \PySCF{}, enabling accurate FLO-SIC calculations even in pathologically over-complete basis sets.\cite{Lehtola2019_241102, Lehtola2020_032504, Lehtola2020_134108}

Equally importantly, again at variance to the previous implementations of FLO-SIC that feature in-house implementations of only a handful of density functionals, \PyFLOSIC{} contains (again via \PySCF{}) interfaces to the \textsc{Libxc}\cite{Lehtola2018_1} and \textsc{XCFun}\cite{Ekstrom2010_1971}  libraries of exchange-correlation functionals, which grant access to a wide range of hundreds of local density approximations~(LDAs), generalized-gradient approximations~(GGAs), meta-GGAs, hybrid functionals, non-local correlation functionals, and range-separated hybrids.\cite{Sun2020_024109} At the moment, all LDAs, GGAs, and meta-GGAs provided through \PySCF{} can be used in \PyFLOSIC{}; for a thorough list of the functionals implemented in \textsc{Libxc} and \textsc{XCFun}, we refer to the libraries' respective documentations.\cite{funclist} 
Custom linear combinations of exchange-correlation functionals can also be used within \PySCF{}, further expanding the capabilities of \PyFLOSIC{}.

As an add-on, \PyFLOSIC{} inherits all of the important features of \PySCF{}. Because \PyFLOSIC{} is implemented as a collection of \Python{} modules like \PySCF{},\cite{Sun2018_e1340} those already familiar with \PySCF{} do not have to learn a new package-specific input format to run FLO-SIC calculations with \PyFLOSIC{}.
Users of \PyFLOSIC{} can also rely on the full power of \Python{}---one of today's most commonly used and taught programming languages that has a massive and vibrant community---to fulfill their every need. The availability of all \Python{} language tools within the input script allows for elaborate work schemes, e.g., the combination of calculation, evaluation, and plotting routines.\cite{Sun2018_e1340} Calculations can also be done interactively in the \Python{} interpreter shell.\cite{Sun2018_e1340} 

Having introduced and motivated \PyFLOSIC{}, we will next discuss the theory of FLO-SIC in detail in \secref{theory}.
We will start out by motivating the use of self-interaction corrections in general~(\subsecref{sic}), continue by presenting Perdew and Zunger's approach in specific~(\subsecref{pzsic}), introduce Fermi--L\"owdin orbitals to undertake the self-interaction correction following Perdew and Zunger's prescription~(\subsecref{flos}), and discuss the mandatory initialization and optimization of the Fermi-orbital descriptors that parametrize the Fermi--Löwdin orbitals~(\subsecref{fods}).
The schemes used in \PyFLOSIC{} for optimizing the FLO-SIC density are discussed in \subsecref{flosic-min}. 
The code is showcased in \secref{example}, starting with example inputs for running a calculation on tetracyanoethylene.
The structure of the tetracyanoethylene example follows that of the calculation: first, the FODs are initialized (\subsecref{autogen}), and then the electron density and the FODs are optimized (\subsecref{optimization}).
A discussion on repeated calculations follows in \subsecref{repetition}.
As a practical application of the code, in \subsecref{basis} we perform an in-depth basis set convergence study of the atomization energies of \ce{SO2} and \ce{SF6} that have been found to be challenging cases in the literature, as was already discussed above. The article concludes in a summary and outlook in \secref{summary}. Atomic units are used throughout the paper, if not stated otherwise.

\section{Fermi--L\"owdin orbital self-interaction correction\label{sec:theory}} 

We will adopt the notation introduced by \citet{Lehtola2014_5324} in the following.
However, in contrast to \citeref{Lehtola2014_5324} vectors and matrices are distinguished in the presently used notation: all vectors are expressed in bold, non-italic letters (\textbf{a}), and all matrices 
are given in bold, italic letters ($\boldsymbol{R}$).

\subsection{Self-interaction correction in density functional theory \label{subsec:sic}}

DFT has become one of the standard methods in computational materials science, condensed matter physics, as well as chemistry thanks to its combination of reasonable accuracy with computational efficiency.\cite{Sherrill2010_110902,Burke2012_150901}
However, currently available density functional approximations (DFAs) are well-known to fail in a number of situations.\cite{Cohen2008_792,Perdew2009_902,Burke2012_150901}
Although the description of systems exhibiting significant static correlation remains an open problem, posing limitations on studies of many transition metal complexes and molecules with stretched bonds, challenges exist also in the absence of static correlation.
For instance, localized states and negatively charged species such as \ce{F-} are incorrectly described by pure (i.e., local or semi-local) density functionals.

These shortcomings come from the fact that DFAs include spurious interactions of electrons with themselves, known as the self-interaction error, which causes the electron density to delocalize.
Delocalization error has been long identified as a key issue in the application of DFT onto the study of chemical systems; for instance, the barrier height of the simplest hydrogen abstraction reaction \ce{H + H2 <-> H2 + H} already poses a challenge, as many functionals overestimate the stability of the intermediate \ce{H3} state.\cite{Johnson1994_100}
The error can often be significantly reduced by including a fraction of exact exchange as in (possibly range-separated) hybrid functionals; however, this does not fully remove the problems with self-interaction.

An exchange-correlation functional that fulfills three constraints may be free of self-interaction for many-body systems.
It needs to (i) be free of self-interactions for all one-electron densities, (ii) provide the correct piecewise linearity (PWL),\cite{Kraisler2013_126403} and (iii) recover the correct asymptotic limit of the potential.
As no DFAs that fulfill all of these constraints completely are available, self-interaction corrections (SICs) aim to rectify some of the aforementioned issues for the available DFAs. Since the early formulation of SIC,\cite{Perdew1981_5048} SICs have been shown to be able to significantly reduce the errors of the underlying exchange-correlation functional for the first two of the aforementioned issues, namely the delocalization error (i) and the lack of piecewise linearity (ii).\cite{Pederson1988_1807,Vydrov2005_184107,Klupfel2011_050501,Gudmundsdottir2013_194102,Borghi2014_075135,Gudmundsdottir2014_234308,Lehtola2014_5324,Pederson2015_153,Perdew2015_1,Cheng2016_11013,Zhang2016_2068,Schwalbe2018_2463}
The lack of piecewise linearity is the reason the highest occupied molecular orbital (HOMO) energies from Kohn--Sham~(KS) DFAs yield poor estimates for the ionization potential as $\text{IP} = -\varepsilon_{\text{HOMO}}$; 
in contrast, the so-called Koopmans-compliant~(KC) functionals\cite{Borghi2014_075135,Borghi2015_155112} can deliver especially accurate results for IPs.

Among the multitude of current implementations of SIC, most employ real-valued orbitals (RSIC).\cite{Garza2000_7880,Garza2001_639,Patchkovskii2001_26,Patchkovskii2002_7806,Patchkovskii2002_1088,Vydrov2004_8187,Vydrov2006_094108,Pemmaraju2007_045101} 
However, it has been shown 
that SIC based on complex-valued orbitals (CSIC) has several advantages;\cite{Klupfel2011_050501,Klupfel2012_124102,Valdes2012_49,Gudmundsdottir2015_083006,Cheng2016_11013} in fact, \citet{Lehtola2016_3195} showed some time ago that the use of complex-valued orbitals is actually mandatory to properly minimize the SIC functional.
(Implementations of RSIC and CSIC that support arbitrary basis sets and exchange-correlation functionals similarly to \PyFLOSIC{} are freely available in the \ERKALE{} program.\cite{Lehtola2012_1572,Hel2016_1})

Another variant of SIC is the FLO-SIC approach mentioned in \secref{intro}, which has been the subject of various studies during recent years,\cite{Pederson2014_121103,Hahn2015_224104,Pederson2015_153,Pederson2015_064112,Pederson2016_164117,Hahn2017_5823,Yang2017_052505,Kao2017_164107,Sharkas2018_9307,Schwalbe2018_2463,Joshi2018_164101,Withanage2018_4122,Aquino2018_6456,Johnson2019_174106,Jackson2019_012002,Zope2019_214108,Withanage2019_012505,Santra2019_174106,Trepte2019_820,Schwalbe2019_2843,Yamamoto2019_154105,Aquino2020_1200,Vargas2020_3789} 
and is also the topic of the present work; it will be discussed in depth below in Subsections \ref{subsec:flos}, \ref{subsec:fods}, and \ref{subsec:flosic-min}.
It is first worth mentioning, however, that in spite of the many successes of SIC (some of which were referenced above), there are also several results that show SIC degrading the performance of higher-rung exchange-correlation functionals.\cite{Vydrov2004_8187,Borghi2014_075135,Lehtola2016_4296}
This is the so-called "paradox of self-interaction correction" discussed in the comprehensive summary of \citet{Perdew2015_1}, which remains a challenge for the future.
For instance, it has been pointed out that the Perdew--Zunger approach\cite{Perdew1981_5048} does not completely eliminate one-electron self-interaction;\cite{Lundin2001_247} however, the path for rectifying the remaining error is still unclear.
Despite these theoretical paradoxes, self-interaction corrections have been shown to be useful in many cases, which is why we will not consider this problem further in this work and will instead carry on with the Perdew--Zunger approach.

\subsection{Perdew--Zunger self-interaction correction \label{subsec:pzsic}}

The total energy in the KS-DFT formalism is given by
\begin{equation}
    E_{\text{KS}}[n^\alpha,n^\beta] = T_{\text{s}}[n^\alpha,n^\beta]+ V[n] + J[n] + K[n^\alpha,n^\beta]\;, \label{eq:ksE}
\end{equation}
where $E_{\text{KS}}[n^\alpha,n^\beta]$ is the total KS energy, $T_{\text{s}}[n^\alpha,n^\beta]$ is the kinetic energy of the non-interacting 
system, $V[n]$ is the external potential energy, $J[n]$ is the Coulomb energy, $K[n^\alpha,n^\beta]$ is the exchange-correlation energy, and $n^\sigma$ is the electron density for spin $\sigma$, with $\alpha$ and $\beta$ representing spin up and spin down, respectively.

The KS functional can be minimized with a SCF procedure,\cite{Lehtola2020_1218} in which the KS-Fock matrix
\begin{equation}
    \boldsymbol{F}_{\text{KS}}^{\sigma} = \boldsymbol{H}_{\text{core}} + \boldsymbol{J} + \boldsymbol{K}^{\sigma}\;, \label{eq:fock}
\end{equation}
is iteratively diagonalized, using appropriate measures to ensure the procedure becomes convergent.
The first term of \eqref{fock}, $\boldsymbol{H}_{\text{core}}$, is the core Hamiltonian that arises from the first two terms of \eqref{ksE} as
\begin{equation}
    \boldsymbol{H}_{\text{core}} = -\frac{1}{2}\nabla^2 - \sum_{A}^{\text{nuclei}} \frac{Z_{A}}{|\textbf{r}-\textbf{r}_{A}|}\;,
\end{equation}
where $Z_{A}$ is the nuclear charge of atom $A$ at position $\textbf{r}_{A}$. The last two terms in \eqref{fock}, $\boldsymbol{J}$ and $\boldsymbol{K}^\sigma$, are the Coulomb and exchange-correlation matrix, respectively.

As only the two first terms in \eqref{ksE}---or equivalently, the first term of \eqref{fock}---are necessary for exactness for one-electron systems,\cite{Perdew1981_5048} the exact KS exchange-correlation functional must perfectly cancel out the Coulomb term for any one-electron density $n_1^\sigma$ with spin $\sigma$
\begin{equation}
K[n_{1}^{\sigma},0] = -J[n_{1}^{\sigma}].
\end{equation}
DFAs violate this condition, leading to a spurious self-interaction~(SI) energy
\begin{equation}
E_{\text{SI}}[n_{1}^{\sigma}] = K[n_{1}^{\sigma},0] + J[n_{1}^{\sigma}].
\end{equation}
The Perdew--Zunger self-interaction correction\cite{Perdew1981_5048}~(PZ-SIC) enforces the correct behavior of the DFA by explicitly removing the SI energy orbital by orbital, leading to a corrected total energy functional
\begin{align}
    E_{\text{PZ}} &= E_{\text{KS}}[n^{\alpha},n^{\beta}] + E_\text{SIC} \nonumber \\
    &= E_{\text{KS}}[n^{\alpha},n^{\beta}] - \sum_\sigma \sum_{i}^{N^\sigma} E_{\text{SI}}[n_{i}^{\sigma}]\;, \label{eq:Epz}
\end{align}
where $N^\sigma$ is the number of electrons with spin $\sigma$.

Because of this correction, the PZ-SIC Hamiltonian depends not only on the total spin densities, but also on the individual orbital densities, at variance to standard KS-DFT.
Due to this, the unitary invariance of KS-DFT\cite{Lehtola2020_1218} is broken, and the functional is no longer invariant to orbital rotations within the occupied space.

The optimal PZ-SIC orbitals minimize $E_{\text{PZ}}$.
It has been shown that localized orbitals generally deliver lower self-interaction corrected total energies $E_{\text{PZ}}$ than delocalized ones.\cite{Perdew1981_5048,Harrison1983_2079,Pederson1984_1972}
Hence, the best way to minimize \eqref{Epz} is to start out with localized orbitals (see \citeref{Lehtola2013_5365} and references therein for discussion on this topic)
produced, e.g., by the Foster--Boys,\cite{Boys1960_296} Edmiston--Ruedenberg, \cite{Edmiston1963_457} or Pipek--Mezey \cite{Pipek1989_4916} method or generalizations thereof,\cite{Lehtola2014_642} and then iteratively rotate the orbitals until the derivative of the energy, the occupied-occupied orbital gradient \cite{Lehtola2014_5324}
\begin{equation}
    \kappa_{ij}^{\sigma} = (\textbf{c}_{j}^{\sigma})^{\text{T}}(\boldsymbol{f}_{i}^\sigma - \boldsymbol{f}_{j}^\sigma)\textbf{c}_{i}^\sigma
    \label{eq:kappa}
\end{equation}
vanishes; $\boldsymbol{\kappa}^\sigma = \boldsymbol{0}$ is known as the Pederson condition.
Here,  $\boldsymbol{f}_i^\sigma=\boldsymbol{J}(\boldsymbol{p}_{i}^{\sigma})+\boldsymbol{K}(\boldsymbol{p}_{i}^{\sigma})$ is the orbital Fock matrix (without the one-electron part) that arises from the density matrix $\boldsymbol{p}_{i}^{\sigma}$ corresponding to the $i$:th occupied orbital of spin $\sigma$
\begin{equation}
\label{eq:orbdens}
    \boldsymbol{p}_{i}^{\sigma} = \textbf{c}_{i}^{\sigma}(\textbf{c}_{i}^{\sigma})^{\text{T}}\;,
\end{equation}
where $\textbf{c}_{i}^{\sigma}$ are the coefficients for the $i$:th optimal orbital.
In addition to the occupied-occupied rotations, the gradient for which was given in \eqref{kappa}, the occupied-virtual rotations also have to be optimized, as discussed in \citeref{Lehtola2016_3195}.

Practical calculations based on orbital rotations, such as those with the \ERKALE{} program, pursue minimization only to a finite numerical threshold; that is, until the orbital gradient is small but still non-zero.
Still, the minimization of $E_{\text{PZ}}$ by orbital rotations is arduous due to slow convergence of the orbital optimization. 
Moreover, as was already mentioned in the Introduction, the correct minimization of \eqref{Epz} has been shown to require complex-valued orbitals and to exhibit a plethora of local minima.\cite{Lehtola2016_3195}

The asymptotic scaling of the orbital rotation approach is determined by the calculation of the orbital gradient in \eqref{kappa}. Assuming $K$ basis functions, $\mathbf{c}^\sigma$ and $\{\mathbf{f}_i^\sigma\}_{i=1}^{N^\sigma}$ are all $K \times K$ matrices, with $K\geq N^\sigma$. Disregarding the cost to compute the $N^\sigma$ orbital Fock matrices (which should overall scale linearly with system size in an optimal implementation due to the localized nature of the optimal orbitals), the evaluation of  $\boldsymbol{\kappa}$ carries an iterative  $\mathcal{O}(N^\sigma K^3)$ cost, while orbital optimization using the approach of \citeref{Lehtola2016_3195} carries an iterative $\mathcal{O}(K^3)$ cost like conventional Kohn--Sham density functional theory. Still, the bottleneck in practical calculations is typically in the evaluation of the orbital Fock matrices, and alternative approaches that avoid the irrelevant virtual-virtual block can be pursued in the case $K \gg N^\sigma$.
However, due to the presence of many local minima and saddle point solutions, the approach of \citeref{Lehtola2016_3195} recommends the use of stability analysis; since there are $\mathcal{O}((N^\sigma)^2)$ rotation angles, stability analysis employing iterative diagonalization carries a $\mathcal{O}((N^\sigma)^4)$ cost.

\subsection{Fermi--L\"owdin orbitals \label{subsec:flos}}

At variance to the direct minimization of $E_{\text{PZ}}$ with orbital rotations, Pederson and coworkers proposed using Fermi--L\"owdin orbitals (FLOs) together with PZ-SIC, giving rise to the FLO-SIC method.\cite{Pederson2014_121103,Pederson2015_064112,Pederson2015_153,Yang2017_052505}
FLOs were originally introduced by Luken and coworkers in a series of papers that dealt with the exchange or Fermi hole in the context of many-electron wave functions.\cite{Luken1982_265,Luken1982_265b,Luken1984_1283,Luken1984_279} In the FLO approach, one starts by building Fermi orbitals (FOs) $\phi_{\text{FO},i}^\sigma$  via
 \begin{equation}
\boldsymbol{c}_\text{FO}^\sigma = \boldsymbol{C}^\sigma \boldsymbol{R}^{\,\sigma}, 
\label{eq:foloc2}
\end{equation}
where $\boldsymbol{c}_\text{FO}^\sigma$ and $\boldsymbol{C}^\sigma$ contain the FO and KS orbital coefficients, respectively. The transformation matrix $\boldsymbol{R}^{\,\sigma}$ is defined as
\begin{equation}
	R^{\,\sigma}_{ji} = \frac{\braket{\psi^\sigma_j|\textbf{a}^\sigma_i}}{\sqrt{n^\sigma ({\bf a}_i^\sigma)}} \; . \label{eq:rotation}
\end{equation}
where $|\textbf{a}_i^\sigma\rangle$ denotes a position eigenstate localized at the so-called Fermi-orbital descriptor (FOD) $\textbf{a}_i^\sigma$, and $\psi_j^\sigma$ are KS orbitals.
The FOs are normalized, but they do not form an orthonormal set in general. As a consequence, Luken proposed using L\"owdin's method of symmetric orthonormalization~\cite{Lowdin1950_365} to end up with an orthonormalized set of Fermi--L\"owdin orbitals (FLOs)~$\phi_k^\sigma$ as
\begin{equation}
	\boldsymbol{c}^\sigma = \boldsymbol{c}_\text{FO}^\sigma [\boldsymbol{T}^\sigma  (\boldsymbol{Q}^\sigma)^{-1/2} (\boldsymbol{T}^{\sigma})^{\text{T}} ],
\label{eq:unitarytrafo}
\end{equation}
where $\boldsymbol{c}^\sigma$ holds the FLO coefficients, while $\boldsymbol{T}^{\sigma}$ and $\boldsymbol{Q}^\sigma$ contain the eigenvectors and eigenvalues of the FO overlap matrix, respectively.

The usefulness of the FOs and FLOs arises from the property of the FO. It has the value $\braket{\textbf{a}^\sigma_i|\phi_{\text{FO},i}^\sigma}=\sqrt{n^\sigma ({\bf a}_i^\sigma)}$ at its descriptor, meaning that \emph{all} the electron density~(for a given spin channel) at this point in space comes from a \emph{single} orbital. Thus, the FOs are localized around their corresponding FODs, in contrast to the typically delocalized KS orbitals. As the orthonormalization mixes the FOs, the FLOs are slightly less localized than the original non-orthogonal FOs.

Another key feature of the FLO approach is that since the FLOs are uniquely determined by equations~(\ref{eq:rotation}) and (\ref{eq:unitarytrafo}) for a given set of FODs $\boldsymbol{a}_i^\sigma$ and a set of occupied orbitals $\boldsymbol{C}^\sigma$, the FLO-SIC approach turns out to restore the unitary invariance of KS-DFT:
rotations of the occupied orbitals in $\boldsymbol{C}^\sigma$ are countermanded by an inverse rotation occurring in $\boldsymbol{R}^\sigma$ in \eqref{rotation}.
This feature means that the optimization of the FODs and that of the density matrix can be decoupled.

\subsection{Initialization and optimization of Fermi-orbital descriptors \label{subsec:fods}}
The FODs $\textbf{a}^\sigma_i$  formally introduced in \eqref{rotation} are a key feature of the FLO-SIC approach.
In FLO-SIC, the optimization of the \mbox{$N^\sigma(N^\sigma -1)/2$} orbital rotation angles within the occupied space (twice that if imaginary rotations are also included as in the CSIC approach) is replaced by the optimization of $3N^\sigma$ FOD coordinates.\cite{Pederson2014_121103, Pederson2015_064112, Pederson2015_153, Yang2017_052505, Aquino2018_6456}
Although full minimization of the PZ-SIC energy functional used in FLO-SIC requires optimization of the FODs, the interesting part about the model is that since plausible FODs can be generated in an automatic fashion,\cite{Schwalbe2019_2843} qualitative calculations may be performed without having to optimize the FODs, although the predictive power of such calculations is limited in the same way as the use of, e.g., Foster--Boys orbitals for evaluating the self-interaction correction criticized in \citeref{Lehtola2014_5324}.

Surprisingly, the optimization of this small number of FOD coordinates turns out to be at least as challenging as the optimization of the far more numerous possible orbital rotations.
As previously shown in \citerefs{Pederson2015_064112} and \citenum{Pederson2015_153}, the construction of the FOD derivatives from the orbital Fock matrices leads to $\mathcal{O}((N^\sigma)^4)$ scaling and $\mathcal{O}((N^\sigma)^3)$ data storage cost; this computational scaling and data storage behavior is also realized within \PyFLOSIC{}.
The gradient of the energy with respect to the FODs can be expressed as \cite{Pederson2015_064112,Pederson2015_153} \begin{widetext}
\begin{equation}
	\frac{\partial E_\text{PZ}}{\partial\,\textbf{a}^\sigma_i} = 	\frac{\partial E_\text{SIC}}{\partial\,\textbf{a}^\sigma_i}  = \sum\limits_{k,l}^{N^\sigma} \varepsilon^{k,\sigma}_{kl} \left(\Braket{\frac{\partial\phi^\sigma_k}{\partial\,\textbf{a}^\sigma_i} | \phi^\sigma_l } + \Braket{ \phi^\sigma_l | \frac{\partial\phi^\sigma_k}{\partial\,\textbf{a}^\sigma_i}}\right) \;,
\label{eq:sicforces} \end{equation}
\end{widetext}
where the elements of the Lagrange multiplier matrix $\varepsilon^{k,\sigma}_{kl}$ are defined by
\begin{equation}
\varepsilon^{k,\sigma}_{ij}  = -(\textbf{c}_{i}^{\sigma})^{\text{T}}\boldsymbol{f}_{k}^\sigma\textbf{c}_{j}^\sigma\,.
\label{eq:sicepsilon}
\end{equation}

As always, the FOD optimization starts from some initial values for the FODs.
As was discussed in \subsecref{pzsic}, minimization of the PZ functional by orbital rotation techniques is typically started from localized orbitals.
Now, since FLOs are highly localized around their FODs, this initialization can also be carried over to FLO-SIC by retrieving initial FODs from the centroids of the localized orbitals from the Foster--Boys, Edmiston--Ruedenberg, or (generalized) Pipek--Mezey.\cite{Schwalbe2019_2843}
Such a procedure is implemented in the \textsc{Python} center of mass~(\textsc{PyCOM}) module within \PyFLOSIC{}, which employs the second-order orbital localizer\cite{Sun2016_1} in \PySCF{}.

Several other ways to initialize the FODs have also been recently discussed in \citeref{Schwalbe2019_2843}. They include an electronic force field which yields quasi-classical positions for the electrons to which the FODs can be assigned (\PyEFF{}); using Lewis-like bonding information to place FODs accordingly (\PyLEWIS{}); or a Thomson problem-like procedure of obtaining the FODs from Monte-Carlo minimization of a distribution of point charges under the restriction of a certain bond order (\fodMC{}).
The initial FODs from any of these alternative generators can be easily read in by \PyFLOSIC{}.
Please refer to \citeref{Schwalbe2019_2843} for further details on automatic FOD initialization.

Regardless of the procedure used to initialize the FODs, the initial FOD geometries should always be visualized; an example will be discussed below in \secref{example}.
By visualizing the resulting \emph{electronic geometry}, one should check whether it is in reasonable agreement with Lewis~\cite{Lewis1916_762} or Linnett double-quartet~(LDQ) theory,\cite{Linnett1960_859,Linnett1961_2643,Linnett1964_1,Luder1964_55} as good-natured FODs should generally correspond to these theories.\cite{Kraus2017_1,Schwalbe2019_2843}

The gradients of these initial FODs are usually non-negligible, which is why FOD optimization is an important part of FLO-SIC calculations.
FOD optimization is carried out automatically in \PyFLOSIC{} through an interface to the atomic simulation environment (\textsc{ASE}\cite{Larsen2017_273002}), which has ample algorithms for geometry optimization with forces, including conjugate gradients (\textsc{CG}), the (limited memory) Broyden--Fletcher--Goldfarb--Shanno scheme ((\textsc{L}-)\textsc{BFGS}),\cite{Broyden1970_76,Fletcher1970_317,Goldfarb1970_23,Shanno1970_647,Nocedal1980_773,Liu1989_503,Byrd1995_1190,Zhu1997_550} and the fast inertial relaxation engine~(\textsc{FIRE}).\cite{Bitzek2006_170201}
In order to use \textsc{ASE}, the FOD gradient $\partial E_\text{SIC}/\partial\,\textbf{a}^\sigma_i$ is simply expressed in terms of a fictitious force $\textbf{f}^{\,\sigma}_i = - \partial E_\text{PZ}/\partial\,\textbf{a}^\sigma_i$ acting on the FODs, and the optimization is continued until some numerical threshold is reached for these FOD forces.

\subsection{Optimization of the density with FLO-SIC\label{subsec:flosic-min}}

As was discussed in \subsecref{flos}, the optimization of the energy functional used in FLO-SIC can be split into two sub-problems: optimization of the density matrix for a fixed set of FODs, and optimization of the FODs for a fixed density matrix, which was already discussed in \subsecref{fods}.
Although the PZ-SIC Hamiltonian $\boldsymbol{F}^\sigma_\text{PZ}$ based on \eqref{Epz} implies that each orbital experiences a different potential, all the orbital-dependent Hamiltonians can be cast in a common form by employing a unified Hamiltonian approach.\cite{Heaton1982_827, Harrison1983_2079, Lehtola2014_5324}
Assuming such a unified Hamiltonian, the SCF equations that determine the optimal occupied orbitals for PZ-SIC can be written as
\begin{equation}
\boldsymbol{F}^\sigma_\text{PZ}\boldsymbol{C}^\sigma = 
    (\boldsymbol{F}_{\text{KS}}^{\sigma} + \boldsymbol{F}_{\text{SIC}}^{\sigma})\boldsymbol{C}^{\sigma} = \boldsymbol{S}\boldsymbol{C}^{\sigma}\boldsymbol{E}^{\sigma}\;, 
    \label{eq:scf_SIC}
\end{equation}
where $\boldsymbol{S}$ is the basis set's overlap matrix, and $\boldsymbol{E}^\sigma$ is a diagonal matrix that holds the energy eigenvalues.

There are at least two different ways to construct a unified SIC Hamiltonian $\boldsymbol{F}^\sigma_\text{SIC}$,\cite{Heaton1982_827,Harrison1983_2079,Lehtola2014_5324} and both Hamiltonians presented here have been implemented in \PyFLOSIC{}. The first SIC Hamiltonian  
\begin{equation}
    \boldsymbol{F}_{\text{SIC,OO}}^{\sigma} = -\frac{1}{2}\sum_{i}^{N^{\sigma}}(\boldsymbol{f}_{i}^{\sigma}\boldsymbol{p}_{i}^{\sigma}\boldsymbol{S} + \boldsymbol{S}\boldsymbol{p}_{i}^{\sigma}\boldsymbol{f}_{i}^{\sigma})
\end{equation}
only contains operators that project to and from the space of occupied orbitals, as indicated by the identifier OO. This is equivalent to ignoring the frequently small off-diagonal Lagrange multipliers that ensure orbital orthonormality.\cite{Lehtola2014_5324}
The second SIC Hamiltonian is \begin{widetext}
\begin{equation}
   \boldsymbol{F}_{\text{SIC,OOOV}}^{\sigma} = -\boldsymbol{S} \sum_{i}^{N^{\sigma}}(\boldsymbol{p}_{i}^{\sigma}\boldsymbol{f}_{i}^{\sigma}\boldsymbol{p}_{i}^{\sigma}+\boldsymbol{v}^{\sigma}\boldsymbol{f}_{i}^{\sigma}\boldsymbol{p}_{i}^{\sigma}+\boldsymbol{p}_{i}^{\sigma}\boldsymbol{f}_{i}^{\sigma}\boldsymbol{v}^{\sigma})\boldsymbol{S},
\end{equation}
\end{widetext}
where the virtual space projector is defined as
\begin{equation}
\boldsymbol{v}^{\sigma}=\sum_{i}^{\text{virtual}}\textbf{C}_i^\sigma (\textbf{C}_{i}^{\sigma})^\text{T}\;.
\end{equation}
Therefore, $\boldsymbol{F}_{\text{SIC,OOOV}}^{\sigma}$ also allows for projections between the occupied and virtual orbital spaces, leading to the identifier OOOV; this Hamiltonian does not assume a diagonal Lagrange multiplier matrix.
The second SIC Hamiltonian can also be written as a block matrix in the occupied and virtual spaces as
\begin{align}
\nonumber    \boldsymbol{F}^\sigma_{\text{SIC,OOOV}} & = 
    \begin{pmatrix}
    \boldsymbol{F}^\sigma_{\text{SIC,OO}} & \boldsymbol{F}^\sigma_{\text{SIC,OV}} \\
    \boldsymbol{F}^\sigma_{\text{SIC,VO}} & \boldsymbol{F}^\sigma_{\text{SIC,VV}} 
    \end{pmatrix} \\
    & = \begin{pmatrix}
    \boldsymbol{F}^\sigma_{\text{SIC,OO}} & \boldsymbol{F}^\sigma_{\text{SIC,OV}} \\
    \boldsymbol{F}^\sigma_{\text{SIC,VO}} & \boldsymbol{0} 
    \end{pmatrix} \;, 
\end{align}
as the operator includes no terms that couple virtual orbitals together.

\section{Example calculations with \PyFLOSIC{} \label{sec:example}}

The features of \PyFLOSIC{} are first showcased using the tetracyanoethylene (\ce{C2(CN)4}) molecule. In addition to specifying the nuclear geometry for the molecule, as is necessary for any electronic structure calculation, the initial FODs have to be specified in order to perform a FLO-SIC calculation, as was discussed in \subsecref{fods}; that is, an \emph{electronic} geometry needs to be specified as well.

\subsection{Automatic generation of electronic geometry \label{subsec:autogen}}

Besides the versatile core functionality inherited from \PySCF{}, \PyFLOSIC{} has unique features specific to FLO-SIC calculations.
The first of these is the automatic generation of the electronic geometry, i.e., the FODs. The necessary code to obtain initial FODs with the \PyCOM{} approach is presented in \figref{initialfods}, and the resulting electronic geometry is visualized in \figref{guess}.
Note that only \emph{two lines of code} are needed to generate a reasonable initial guess for the FODs, not counting the import of the necessary \Python{} modules.

Alternatively to the use of \PyCOM{}, the FODs can also be initialized in \PyFLOSIC{} by reading in the output of any of the other methods described in \citeref{Schwalbe2019_2843}.
The necessary code for the read-in is shown in \figref{optimizefods} that will be discussed in more detail below.

Optimized FODs have been shown to carry bonding information in the context of Lewis/LDQ theory, which is discussed in detail in \citeref{Schwalbe2019_2843}. For example, a simple FOD bond order (BO$_{\text{FOD}}$) can be evaluated via
\begin{equation}
    \text{BO}_{\text{FOD}} = m_{\text{electrons}} / n_{\text{centers}} \label{eq:BO_FOD}
\end{equation}
just by counting the number of FODs $m$ between the bonded atoms, and then dividing by the number of atoms  $n$ that partake in this bond.
The initial set of FODs shown in \figref{guess} appears to be reasonable, as it is in excellent agreement with Lewis/LDQ theory; it can thus be expected that the optimization of these FODs will yield useful results as well.
It is, however, important to note  that since the exchange-correlation functional affects the electron density, the reasonableness of the \PyCOM{} guess may depend on the system and the functional, which is why we recommend to always visualize the initial FODs before running calculations.
As the FODs change during the optimization, the final geometries should be inspected as well.

\begin{figure}
  \begin{lstlisting}[language=Python]
from ase.io import read
from pycom import pycom_guess

# Load nuclear information
ase_nuclei = read('Conformer3D_CID_12635.sdf')
# Start FOD generation
pycom_guess(ase_nuclei=ase_nuclei,
            charge=0,
            spin=0,
            xc='pbesol',
            basis='pc-0',
            method='fb')
  \end{lstlisting}
  \caption{Generation of initial FODs using \PyCOM{}.
  After importing the needed modules, the nuclear information is read from an \textsc{.sdf} file containing data for tetracyanoethylene (\ce{C2(CN)4}), which has been downloaded from the \textsc{PubChem}  database.\cite{Kim2019_D1102}
  Next, \PyCOM{} generates an FOD guess for the molecule by running a conventional KS-DFT calculation with the PBEsol functional\cite{Perdew2008_136406} and the pc-0 basis set,\cite{Jensen2001_9113,Jensen2002_7372} localizing the resulting occupied orbitals with the Foster--Boys ('fb') method, and calculating their centroids. 
  The resulting FODs are stored in an \textsc{.xyz} file together with the nuclear information, and are shown for this example in \figref{guess}.}
  \label{fig:initialfods}
\end{figure}

\begin{figure}
\centering
  \includegraphics[scale=0.15]{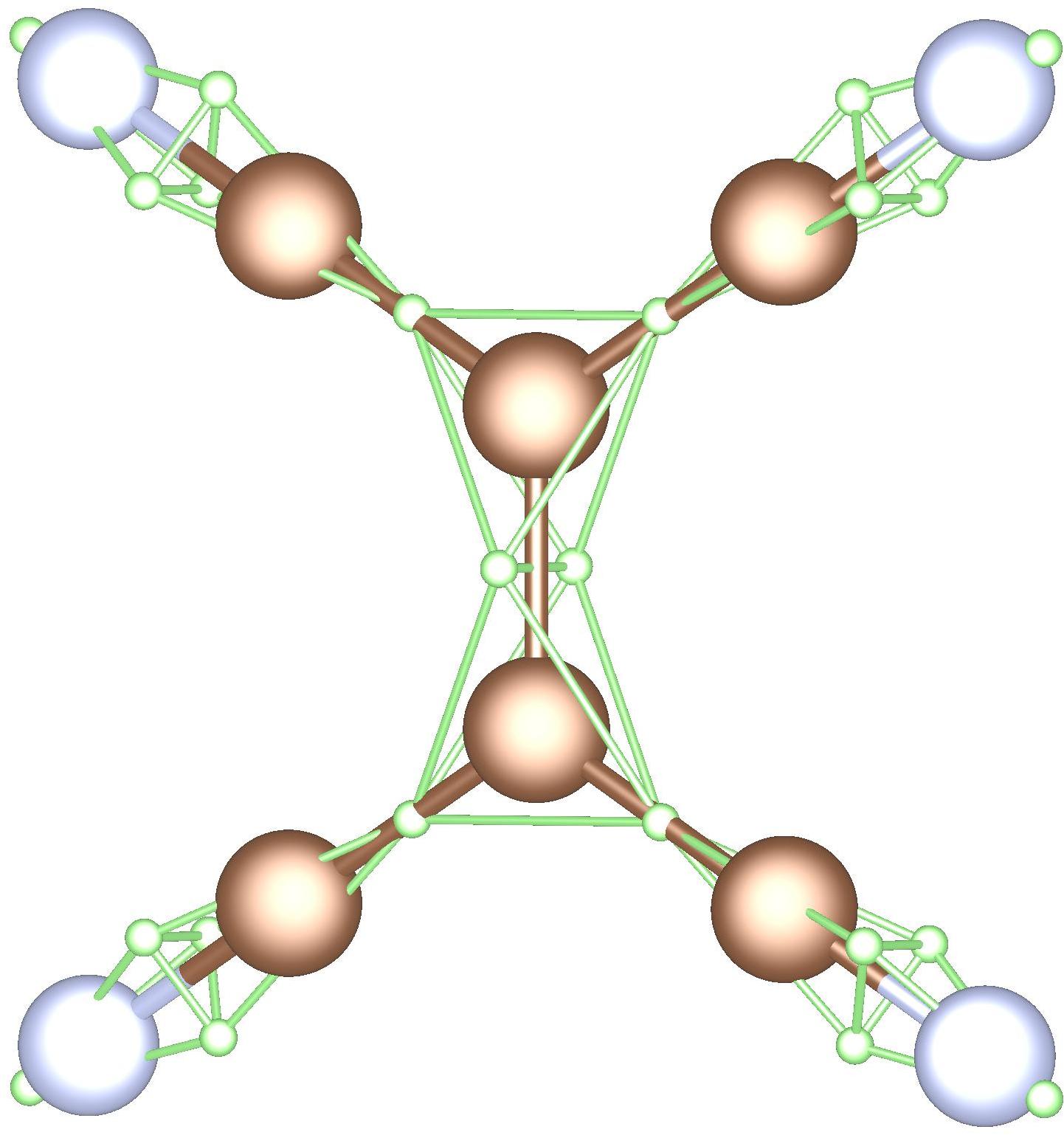}
  \caption{The nuclear geometry for tetracyanoethylene (\ce{C2(CN)4}) and the FODs arising from the \PyCOM{} script shown in \figref{initialfods}.
  Color code: C: brown, N: blue, FODs: green.
  Note: The FODs for spin up and spin down electrons are identical in this case.
  Solid lines between nuclei indicate bonds and tiny solid lines between the FODs indicate the valence electronic geometry, in this case spanned by tetrahedra.
  Face-sharing tetrahedra between C and N atoms indicate a triple bond, i.e., $\text{BO}_{\text{FOD}}=3$. 
  Edge-sharing tetrahedra between C and C atoms indicate a double bond, i.e., $\text{BO}_{\text{FOD}}=2$.
  Corner-sharing tetrahedra between C atoms indicate a single bond, i.e., $\text{BO}_{\text{FOD}}=1$.
  Therefore, these initial FODs agree well with Lewis/LDQ theory.
  }
  \label{fig:guess}
\end{figure}

\subsection{FOD and density optimization \label{subsec:optimization}}

\begin{figure}
  \centering
 \includegraphics[width=0.51\textwidth]{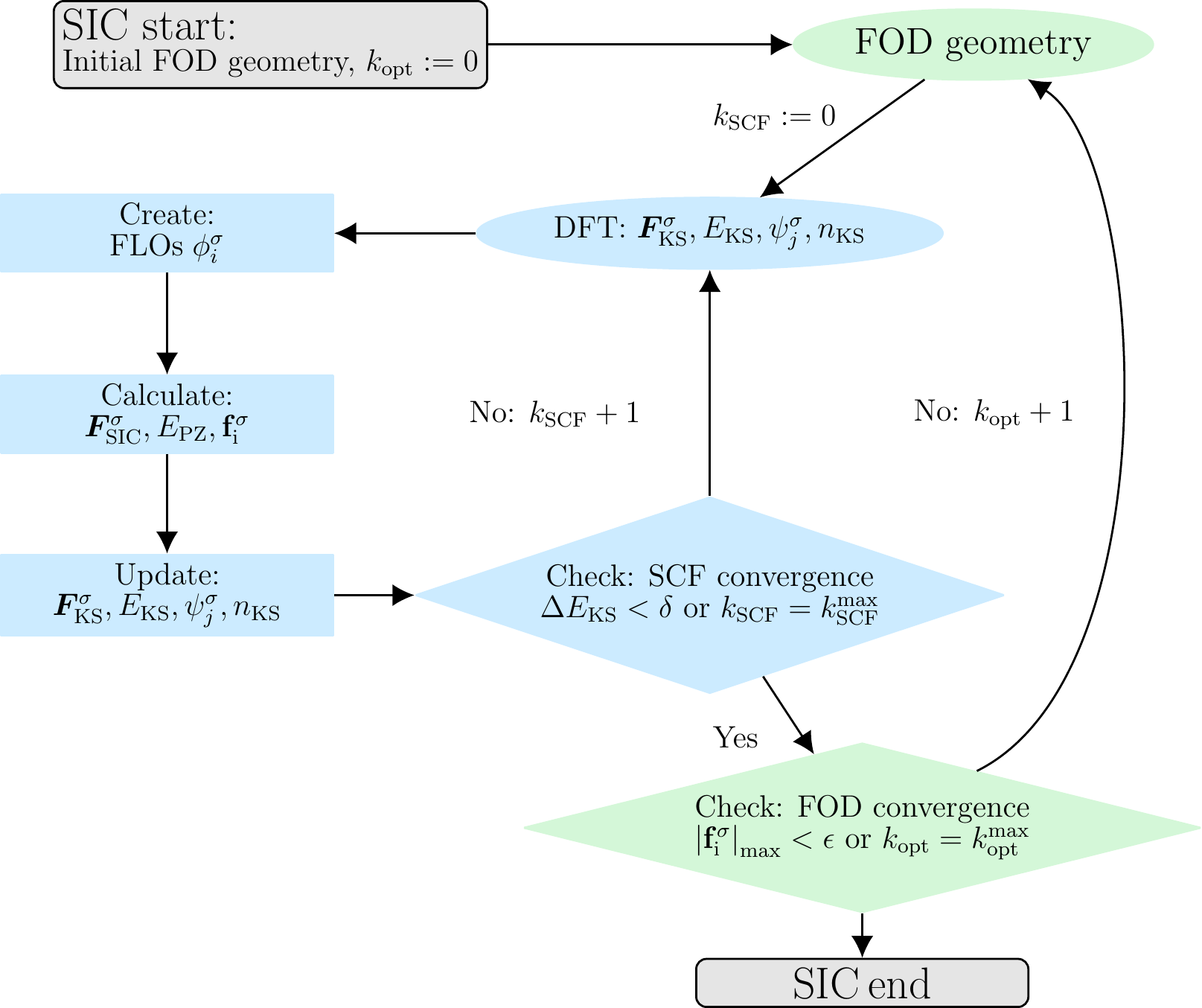} 
 \caption{FLO-SIC workflow for optimizing both the density and the FODs. $n_\text{KS}$ refers to the KS density. Initial FODs can be generated with the built-in \PyCOM{} routine. Here, $k_\text{SCF}$ and $k_\text{opt}$ refer to the number of SCF iterations and FOD optimization steps, respectively, whereas $\delta$ and $\epsilon$ are user-defined numerical thresholds for the energy and the FOD forces. 
 The inner loop optimizes the density as well as the SIC total energy $E_\text{PZ}$, while the outer loop optimizes the FODs by using the FOD forces $\textbf{f}^{\,\sigma}_i$ as input for any of the gradient-based algorithms found in \textsc{ASE}.}
  \label{fig:pyflosic_scf}
\end{figure}

Having established the necessary nuclear and electronic geometries, the FLO-SIC calculation can begin.
As was discussed in \secref{theory}, FLO-SIC has two classes of degrees of freedom: the occupied orbitals, i.e., the electron density, and the FODs.
Both should be optimized in order to minimize the PZ-SIC energy functional; the workflow for such FLO-SIC calculations is illustrated in \figref{pyflosic_scf}.
Note that the self-interaction correction is re-evaluated at every SCF iteration with the up-to-date density matrix, ensuring the self-consistency of the orbitals and the correction.

However, in some cases one may want to keep the density or the FODs fixed; this may be useful, e.g., for exploratory FLO-SIC calculations.
The default mode in \PyFLOSIC{} is to optimize both the FODs and the density, but fixed-density and fixed-FOD calculations are also supported.

As was discussed in \secref{theory}, \PyFLOSIC{} optimizes the FODs via \textsc{ASE}, whereas the electron density is optimized through a unified Hamiltonian approach.
Example code for the optimization of both the density and the FODs with \PyFLOSIC{} is shown in \figref{optimizefods}.
For this example, the FODs generated with \PyCOM{} (see \figref{guess}) are used as starting points for the FOD optimization, and the default unified SIC Hamiltonian $ \boldsymbol{F}^\sigma_{\text{SIC,OOOV}}$ is used for the density optimization within the SCF approach.
Note that again, \emph{only two function calls} are needed to run the calculation. 
FOD optimization can be analyzed just like any other geometry optimization in \textsc{ASE}, e.g., by visualizing the trajectory of the system generated by \textsc{ASE}.

\subsection{Repeated calculations \label{subsec:repetition}}

Having the optimized FODs for a given level of theory (including, e.g., the exchange-correlation functional, quadrature grid, and the orbital basis set), calculations for other levels of theory can easily be performed by starting the calculations from the preoptimized FODs.
Because the cost of the FLO-SIC calculation is heavily dependent on the size of the orbital basis set, it is recommended to start out with preoptimized FODs from calculations using small basis sets before going to expensive calculations in large but more accurate basis sets.

Note, however, that if the level of theory is changed significantly, e.g., by going from an LDA to a GGA or a meta-GGA, the optimal FOD positions may also experience large changes, decreasing the value of the preoptimization. 
The changes in the optimal FODs are related not only to the effect of the total density changing with the exchange-correlation functional, but also to the FLO single electron densities having much sharper features than the total density,\cite{Shahi2019_174102} which accentuates the sensitivity to the DFA.
FLO-SIC may thus be more sensitive to the exchange-correlation functional than KS-DFT.
This is also reflected in the functional requirements for PZ-SIC, which differ from KS-DFT. For example, the Perdew--Burke--Ernzerhof functional 
with adjustments for accuracy for solids\cite{Perdew2008_136406} (PBEsol) has been found to afford better accuracy for molecular PZ-SIC calculations than 
the original functional,\cite{Perdew1996_3865} at variance to calculations at the KS-DFT level.\cite{Jonsson2015_1858,Lehtola2016_4296}

\begin{figure}
  \begin{lstlisting}[language=Python]
from ase.io import read
from ase_pyflosic_optimizer import flosic_optimize

# Load combined nuclear/FOD information
ase_atoms = read('FB_GUESS_COM.xyz')
# Start optimization of density and FODs 
flosic = flosic_optimize(mode='flosic-scf', 
                         atoms=ase_atoms,
                         charge=0,
                         spin=0,
                         xc='pbesol',
                         basis='ccpvqz',
                         opt='fire',
                         maxstep=0.1,
                         fmax=0.001)
  \end{lstlisting} 
   \caption{\textsc{Python} script file for a FLO-SIC optimization of both the density and the FODs (indicated by mode='flosic-scf'), using the \textsc{ASE} interface. After importing the needed modules, the nuclear and FOD information are read from a file, in this case the \textsc{.xyz} file that was created by the script in \figref{initialfods}. Afterwards, the optimization is started using the nuclear and FOD information, the charge and spin of the system, the chosen exchange-correlation functional (PBEsol) and basis set (cc-pVQZ\cite{Dunning1989_1007}), the optimizer (\textsc{FIRE}) with its maximum step size in \AA{}, and the numerical threshold for the maximum absolute FOD force in eV/\AA{} as input.}
 \label{fig:optimizefods}
\end{figure}

\subsection{Basis set convergence of the atomization energy of \ce{SO2} and \ce{SF6} \label{subsec:basis}}

Having exemplified the use of the novel code, we proceed with a quantitative application.
As was already mentioned in the Introduction, reproducing the atomization energies of \ce{SO2} and \ce{SF6} is known to require large and flexible basis sets at the Kohn--Sham level of theory,\cite{Jensen2017_1449,Jensen2017_6104,Lehtola2020_134108} making these molecules ideal candidates for a basis set convergence study of FLO-SIC.
For simplicity, we chose to use fixed high-level \emph{ab initio} geometries from the W4-17 database\cite{Karton2017_2063} for the molecules, as fixed nuclear geometries suffice for the present purposes of establishing convergence to the complete basis set limit.
We also chose the polarization-consistent pc-$n$ family of basis sets for this study, since these basis sets are designed for achieving optimal convergence to the basis set limit in DFT and Hartree--Fock calculations.\cite{Jensen2001_9113, Jensen2002_7372} We furthermore chose to study the PBEsol functional, as it has been found to yield good accuracy in PZ-SIC calculations as was already mentioned in the Introduction;\cite{Jonsson2015_1858,Lehtola2016_4296} however, basis set convergence patterns are well-known to be similar for both Hartree--Fock and all density functionals in the lack of post-Hartree--Fock correlation contributions.

However, as it is well-known that SIC methods require much larger quadrature grids than KS-DFT to reach similar levels of convergence,\cite{Vydrov2004_8187, Lehtola2014_5324}
a (200,590) quadrature grid was used for all calculations, because preliminary tests revealed that such a grid was necessary to converge molecular FLO-SIC total energies of \ce{SF6} and \ce{SO2} to roughly $\mu E_\text{h}$ accuracy. 
Unpruned grids are used in this work because at variance to KS-DFT, grid pruning leads to significant errors in FLO-SIC calculations: the orbital densities are not spherically symmetric near the nuclei in contrast to the total electron density used in KS-DFT.

For comparison, we also include results for the default basis sets used in \NRLMOL{}, i.e., the DFO and DFO+ basis sets for density functional calculations which have been used in a number of FLO-SIC studies in the literature.\cite{Pederson2014_121103, Hahn2015_224104, Pederson2015_153, Pederson2015_064112, Pederson2016_164117, Hahn2017_5823, Yang2017_052505, Kao2017_164107, Sharkas2018_9307, Schwalbe2018_2463, Joshi2018_164101, Withanage2018_4122, Johnson2019_174106, Jackson2019_012002, Zope2019_214108, Withanage2019_012505, Santra2019_174106, Trepte2019_820, Yamamoto2019_154105, Vargas2020_3789}
The DFO basis set is recommended for general use,\cite{Porezag1997_1} while the DFO+ basis set is obtained from the DFO basis by adding further polarization functions aimed to improve the accuracy of polarizability calculations;\cite{Porezag1997_1} naturally, the additional functions in DFO+ may also improve the convergence of other properties.
For consistency with previous literature, the DFO and DFO+ basis sets were extracted from NRLMOL for this work; the converted sets are available on \textsc{GitHub} as part of \PyFLOSIC{}.

Four sets of computational models will be considered, the first one being Kohn--Sham DFT, and the next three being variants of FLO-SIC.
In the guess-pc-0 scheme, the FODs are fixed to the \PyCOM{} starting guess values, computed in the pc-0 basis set.
Next, in the opt-pc-0 scheme, the FODs are frozen to the variationally optimized pc-0 values.
Last, in the opt-basis scheme, the FODs are fully optimized.
All FOD optimizations are carried out until a force convergence criterion of $\max_{i\sigma} |\textbf{f}^{\,\sigma}_i | < 10^{-3}~E_\text{h}/a_{0}$ is satisfied.

The atomization energies resulting from the previously described procedures are shown in \tabref{ae}.
The data show remarkable variations of hundreds of \kcal{} when the size of the basis set is increased.
Because larger basis sets are successively better at describing polarization effects in the molecules---thus decreasing the total energy of the molecule---the fully variational atomization energies (i.e. KS-DFT and the opt-basis model) increase with increasing basis set size.
An acceptable level of convergence (basis set truncation error smaller than 1 \kcal{}) is only reached with the quadruple-$\zeta$ pc-3 basis set, highlighting the need to support large basis sets in FLO-SIC calculations which is routinely available in \PyFLOSIC{}.
In contrast, the DFO and DFO+ basis sets produce results between the double-$\zeta$ pc-1 and the triple-$\zeta$ pc-2 basis sets, suggesting the DFO and DFO+ basis sets are merely of polarized double-zeta quality.
The DFO and DFO+ basis sets have a basis set truncation error of roughly 14 \kcal{} and 30 \kcal{} in KS-DFT calculations on \ce{SO2} and \ce{SF6}, respectively, while in FLO-SIC calculations the truncation error for \ce{SO2} increases to 17 \kcal{}, the one for \ce{SF6} remaining 30 \kcal{}.
These results indicate that the DFO and DFO+ basis sets are woefully inaccurate for quantitative applications in chemistry, underlining the need to support high-angular momentum basis sets in KS-DFT as well as FLO-SIC calculations.

\begin{table*}
    \centering
    \caption{Basis set convergence for the atomization energy of SO$_{2}$ and SF$_{6}$ in \kcal{}, calculated with an unpruned (200,590) quadrature grid and the PBEsol exchange-correlation functional. W4-17\cite{Karton2017_2063} nuclear geometries were used.}
    \label{tab:ae}
\begin{tabular}{lcccccccc}
\hline
&  \multicolumn{2}{c}{KS-DFT} & \multicolumn{2}{c}{FLO-SIC (guess-pc-0)}   & \multicolumn{2}{c}{FLO-SIC (opt-pc-0)}   & \multicolumn{2}{l}{FLO-SIC (opt-basis)}  \\
basis & SO$_2$ & SF$_6$ & SO$_2$ & SF$_6$ & SO$_2$ & SF$_6$ & SO$_2$ & SF$_6$ \\
\hline
\hline
  pc-0 &  155.876 &  396.551 &  298.089 &  574.227 &   56.087 &  241.725 &   56.087 &  241.725 \\
  pc-1 &  268.312 &  512.775 &  412.687 &  729.099 &  184.844 &  388.415 &  185.488 &  388.740 \\
   DFO &  289.147 &  538.089 &  434.999 &  763.010 &  207.368 &  422.575 &  207.406 &  422.623 \\
  DFO+ &  289.261 &  538.638 &  435.792 &  763.500 &  207.205 &  423.141 &  207.626 &  423.183 \\
  pc-2 &  294.159 &  556.651 &  440.800 &  786.006 &  215.682 &  442.813 &  214.234 &  443.019 \\
  pc-3 &  302.640 &  568.079 &  448.222 &  792.663 &  220.803 &  452.028 &  224.012 &  452.425 \\
  pc-4 &  303.179 &  568.534 &  448.727 &  792.518 &  224.647 &  452.562 &  224.629 &  452.898 \\
\hline
\end{tabular}
\end{table*}

An examination into the three flavors of FLO-SIC studied in \tabref{ae} shows that optimization of the FODs is clearly necessary, the guess-pc-0 atomization energies being very far from the fully optimized FLO-SIC values, guess-pc-0 overestimating the atomization energy of \ce{SF6} by over 300 kcal/mol.
In contrast, the atomization energies obtained with FODs frozen to the pc-0 optimal values are surprisingly close to the fully variational results, showing differences below 1 \kcal{} for all values except for the pc-3 value for \ce{SO2} that differs from the fully variational one by 3.2~\kcal{}.
These results suggest using FODs frozen to preoptimized values for a small basis set may be an acceptable alternative to fully variational FLO-SIC calculations, if the goal is to estimate the importance of self-interaction corrections to e.g. atomization energies.
The preoptimized FODs also allow for straightforward basis set convergence studies, as omitting the optimization of the FODs implies a significant speed-up of FLO-SIC calculations in large basis sets.

\begin{table*}
    \centering
    \caption{Atomization energies of SO$_{2}$ and SF$_{6}$ in \kcal{} with uncontracted basis sets, calculated with an unpruned (200,590) quadrature grid and the PBEsol exchange-correlation functional. W4-17 nuclear geometries were used.}
    \label{tab:ae2}
\begin{tabular}{lcccc}
\hline
& \multicolumn{2}{c}{KS-DFT} & \multicolumn{2}{l}{FLO-SIC (opt-basis)}  \\
basis & SO$_2$ & SF$_6$ & SO$_2$ & SF$_6$ \\
\hline
\hline
unc-pc-0 &  158.414 &  400.937 &   59.557 &  245.947 \\
unc-pc-1 &  269.065 &  515.849 &  186.428 &  395.124 \\
unc-pc-2 &  294.361 &  557.464 &  215.115 &  443.889 \\
unc-pc-3 &  302.614 &  568.003 &  224.295 &  452.023 \\
unc-pc-4 &  303.126 &  568.411 &  224.235 &  452.456 \\
\hline
\end{tabular}
\end{table*}

As a further topic, the contraction error in the pc-$n$ basis sets for DFT calculations and fully optimized FLO-SIC calculations were also studied, since the self-interaction correction may affect the core orbitals significantly.
The atomization energies in the uncontracted pc-$n$ (unc-pc-$n$) basis sets are shown in \tabref{ae2}.
As shown by the comparison of the results in \tabsref{ae} and \ref{tab:ae2}, the contraction errors are under 1 \kcal{} for all basis sets larger than pc-1, while the contraction error for the largest pc-3 and pc-4 basis sets is in the order of 0.4 \kcal{} for FLO-SIC calculations, which is (unsurprisingly) several times larger than the contraction error in the KS-DFT calculations.
Uncontracting the basis set allows for an improved description of the core orbitals, and the unc-pc-4 results in \tabref{ae2} are our best estimates for the FLO-SIC atomization energies of \ce{SO2} and \ce{SF6} with the PBEsol functional.
The W4-17 atomization energies for \ce{SO2} and \ce{SF6} are 260.580 \kcal{} and 485.425 \kcal{}. As can be seen from \tabref{ae2}, FLO-SIC reduces the error in the PBEsol functional from 42.5 \kcal{} and 83.0 \kcal{} for \ce{SO2} and \ce{SF6} at the KS-DFT level of theory to -36.3 \kcal{} and -33.0 \kcal{} at the FLO-SIC level of theory, respectively, analogously to the findings for the PZ-SIC level of theory in \citeref{Lehtola2016_4296}. 

\section{Summary and Outlook \label{sec:summary}}

\begin{figure*}
\centering
  \includegraphics[scale=0.4]{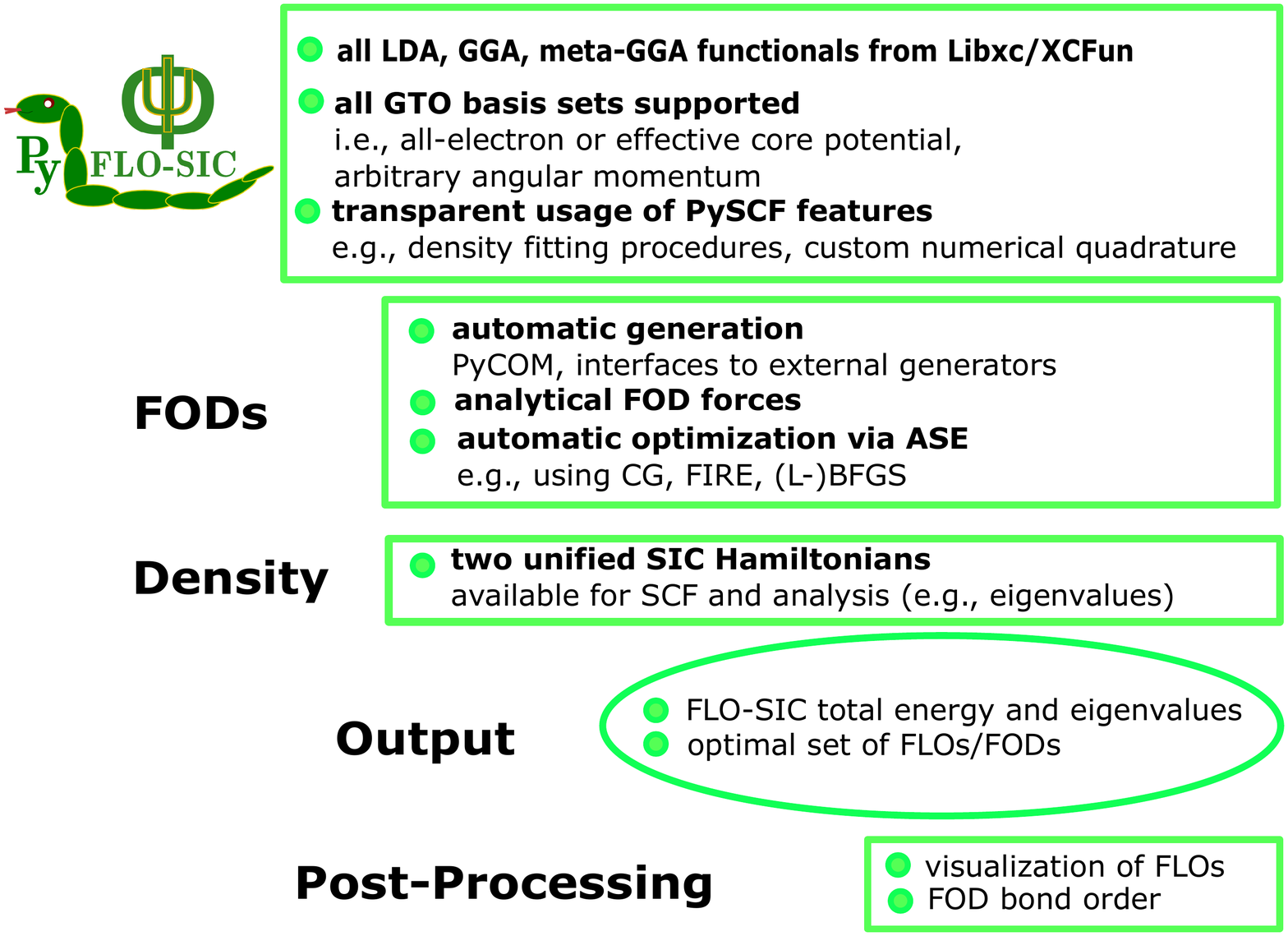}
  \caption{\PyFLOSIC{}~code features.}
  \label{fig:pyflosic_features}
\end{figure*}

We have presented the implementation of FLO-SIC in \PyFLOSIC{} as an extension to the \PySCF{} code, as well as given instructive examples of code usage.
We have also demonstrated the need to be able to use large and flexible basis sets in FLO-SIC calculations with a basis set convergence study of the atomization energies of \ce{SO2} and \ce{SF6}, for which we showed that the DFO basis sets that have been commonly used in FLO-SIC calculations in the literature exhibit truncation errors of tens of \kcal{}.

The most important features of \PyFLOSIC{} are summarized in \figref{pyflosic_features}. 
Our implementation offers the possibility to use FLO-SIC with any GTO basis set, various types of 
quadrature grids, and hundreds of LDA, GGA, and meta-GGA exchange-correlation functionals available in the \textsc{Libxc} and \textsc{XCFun} libraries.
FLO-SIC depends on FODs, which need to be initialized and optimized. 
Importantly, the FODs can be initialized automatically in \PyFLOSIC{} with \PyCOM{}---the internal routine based on localized orbitals' centroids---or read in from other FOD generators, such as the \PyEFF{}, \PyLEWIS{}, and \fodMC{} routines presented in \citeref{Schwalbe2019_2843}.
FOD optimization is carried out in \PyFLOSIC{} through an interface to \textsc{ASE}.
The electron density is optimized in \PyFLOSIC{} with an SCF procedure employing a unified Hamiltonian.

The version of \PyFLOSIC{} described in this work has already been used in some applications.\cite{Shahi2019_174102,Schwalbe2019_2843} However, \PySCF{} offers features that have not yet been exploited in FLO-SIC calculations.
For instance, \PySCF{} implements solvation models, such as the domain-decomposed conductor-like screening model\cite{Cances2013_054111,Lipparini2013_3637,Lipparini2014_184108}~(COSMO) as well as the domain-decomposed polarizable continuum model\cite{Stamm2016_054101,Lipparini2016_160901}~(PCM), which could be combined with FLO-SIC in \PyFLOSIC{} for studying systems in solution.

For the next iteration of \PyFLOSIC{}, we also aim to extend the program to include fully variational FLO-SIC, as well as conventional RSIC and CSIC approaches based on orbital rotation techniques which are presently available only in \ERKALE{}. Since SIC is sensitive to the details of the evaluation of the density functional (e.g., the numerical quadrature), consistent implementations of the various schemes will allow for unbiased comparison of various SIC methods. In addition, FLO-SIC stability analysis along the lines of \citeref{Lehtola2016_3195} will be investigated.
Due to the simple interfacing required for SIC calculations and the modularity of \PyFLOSIC{}, additional interfaces to other electronic structure codes such as \PSIFour{}\cite{Smith2020_184108} could be provided in the future.

An important feature of \PySCF{} are calculations on crystalline systems through the use of periodic boundary conditions.\cite{Sun2018_e1340,Sun2020_024109}
Although calculations on periodic solids with exact exchange are tractable within a GTO basis set as in \PySCF{},\cite{Pisani1988_1}  allowing the use of various hybrid functionals that are less prone to self-interaction error, exact exchange is undesirable in many cases in the study of the solid state due to problems with, e.g., metallic systems.
SIC does not pose such problems, and the application of SIC to solid-state systems has been an active field of research for several decades.\cite{Perdew1981_5048,Heaton1983_5992,Heaton1984_1853,Svane1988_171,Svane1988_9919,Erwin1988_4285,Szotek1990_1031,Szotek1990_275,Svane1990_851,Svane1990_1148,Svane1992_1900,Rieger1995_16567,Vogel1995_R14316,Arai1995_1477,Vogel1996_5495,Svane1996_4275,Svane2000_799,Filippetti2003_125109,Bylaska2004_1,Luders2005_205109,Bylaska2006_86,Pemmaraju2007_045101,Hourahine2007_5671,Stengel2008_155106,Dane2009_045604,Gudmundsdottir2015_083006,Nguyen2018_021051}
\PyFLOSIC{} could be extended to solids in the future as well.
The necessity of a localized orbital picture within a periodic system is a complication for pushing FLO-SIC to the solid state, requiring changes to the algorithms as well as to the initialization of FODs.
Although orbital localization methods are less developed for the solid state than for molecular studies, solid-state FODs could be initialized in analogy to the present work either with maximally localized Foster--Boys\cite{Marzari1997_12847} or generalized Pipek--Mezey Wannier functions.\cite{Jonsson2017_460} The other methods of \citeref{Schwalbe2019_2843} might also be extended for the solid state; for instance, developmental versions of \PyEFF{} and \fodMC{} are already able to provide FODs that appear reasonable for simple solids (e.g., solid deuterium~\cite{Su2007_185003} or lithium) and metal-organic frameworks (MOFs).\cite{Trepte2015_17122,Schwalbe2016_8075,Trepte2017_10020,Trepte2018_25039,Trepte2019_1}

\section*{Data availability}
The \PyFLOSIC{} code, presented and used within this study, is openly available on \textsc{GitHub}~(\url{https://github.com/pyflosic/pyflosic}), and can be referenced via \url{https://doi.org/10.5281/zenodo.3948143}.

\begin{acknowledgments}
J. Kortus and S. Schwalbe have been funded by the
Deutsche Forschungsgemeinschaft (DFG, German Research Foundation)
- Project-ID 421663657 - KO 1924/9-1. 
J. Kortus and J. Kraus have been funded by the DFG  - Project-ID 169148856 - SFB 920, subproject A04. K. Trepte has been supported by the U.S. Department of Energy, Office of Science, Office of Basic Energy Sciences, as part of the Computational Chemical Sciences Program under Award Number \#DE-SC0018331.
S. Lehtola has been supported by the Academy of Finland (Suomen Akatemia) through project number 311149.
We also thank the ZIH in Dresden for computational time and support, Prof. Mark R. Pederson for creating the FLO-SIC method that is the core of everything discussed within the manuscript, and Dr. Torsten Hahn for various discussions and his contributions in an earlier stage of the project.
\end{acknowledgments}

\bibliography{PyFLOSIC_manuscript_JCP_revised.bib}

\begin{thebibliography}{165}%
\makeatletter
\providecommand \@ifxundefined [1]{%
 \@ifx{#1\undefined}
}%
\providecommand \@ifnum [1]{%
 \ifnum #1\expandafter \@firstoftwo
 \else \expandafter \@secondoftwo
 \fi
}%
\providecommand \@ifx [1]{%
 \ifx #1\expandafter \@firstoftwo
 \else \expandafter \@secondoftwo
 \fi
}%
\providecommand \natexlab [1]{#1}%
\providecommand \enquote  [1]{``#1''}%
\providecommand \bibnamefont  [1]{#1}%
\providecommand \bibfnamefont [1]{#1}%
\providecommand \citenamefont [1]{#1}%
\providecommand \href@noop [0]{\@secondoftwo}%
\providecommand \href [0]{\begingroup \@sanitize@url \@href}%
\providecommand \@href[1]{\@@startlink{#1}\@@href}%
\providecommand \@@href[1]{\endgroup#1\@@endlink}%
\providecommand \@sanitize@url [0]{\catcode `\\12\catcode `\$12\catcode
  `\&12\catcode `\#12\catcode `\^12\catcode `\_12\catcode `\%12\relax}%
\providecommand \@@startlink[1]{}%
\providecommand \@@endlink[0]{}%
\providecommand \url  [0]{\begingroup\@sanitize@url \@url }%
\providecommand \@url [1]{\endgroup\@href {#1}{\urlprefix }}%
\providecommand \urlprefix  [0]{URL }%
\providecommand \Eprint [0]{\href }%
\providecommand \doibase [0]{http://dx.doi.org/}%
\providecommand \selectlanguage [0]{\@gobble}%
\providecommand \bibinfo  [0]{\@secondoftwo}%
\providecommand \bibfield  [0]{\@secondoftwo}%
\providecommand \translation [1]{[#1]}%
\providecommand \BibitemOpen [0]{}%
\providecommand \bibitemStop [0]{}%
\providecommand \bibitemNoStop [0]{.\EOS\space}%
\providecommand \EOS [0]{\spacefactor3000\relax}%
\providecommand \BibitemShut  [1]{\csname bibitem#1\endcsname}%
\let\auto@bib@innerbib\@empty
\bibitem [{\citenamefont {Sun}(2015)}]{Sun2015_1664}%
  \BibitemOpen
  \bibfield  {author} {\bibinfo {author} {\bibfnamefont {Q.}~\bibnamefont
  {Sun}},\ }\bibfield  {title} {\enquote {\bibinfo {title} {{Libcint: An
  efficient general integral library for Gaussian basis functions}},}\ }\href
  {\doibase 10.1002/jcc.23981} {\bibfield  {journal} {\bibinfo  {journal} {J.
  Comput. Chem.}\ }\textbf {\bibinfo {volume} {36}},\ \bibinfo {pages} {1664}
  (\bibinfo {year} {2015})},\ \Eprint {http://arxiv.org/abs/1412.0649}
  {arXiv:1412.0649} \BibitemShut {NoStop}%
\bibitem [{\citenamefont {Kim}\ \emph {et~al.}(2018)\citenamefont {Kim},
  \citenamefont {Baczewski}, \citenamefont {Beaudet}, \citenamefont {Benali},
  \citenamefont {Bennett}, \citenamefont {Berrill}, \citenamefont {Blunt},
  \citenamefont {Borda}, \citenamefont {Casula}, \citenamefont {Ceperley} \emph
  {et~al.}}]{Kim2018_195901}%
  \BibitemOpen
  \bibfield  {author} {\bibinfo {author} {\bibfnamefont {J.}~\bibnamefont
  {Kim}}, \bibinfo {author} {\bibfnamefont {A.~D.}\ \bibnamefont {Baczewski}},
  \bibinfo {author} {\bibfnamefont {T.~D.}\ \bibnamefont {Beaudet}}, \bibinfo
  {author} {\bibfnamefont {A.}~\bibnamefont {Benali}}, \bibinfo {author}
  {\bibfnamefont {M.~C.}\ \bibnamefont {Bennett}}, \bibinfo {author}
  {\bibfnamefont {M.~A.}\ \bibnamefont {Berrill}}, \bibinfo {author}
  {\bibfnamefont {N.~S.}\ \bibnamefont {Blunt}}, \bibinfo {author}
  {\bibfnamefont {E.~J.~L.}\ \bibnamefont {Borda}}, \bibinfo {author}
  {\bibfnamefont {M.}~\bibnamefont {Casula}}, \bibinfo {author} {\bibfnamefont
  {D.~M.}\ \bibnamefont {Ceperley}},  \emph {et~al.},\ }\bibfield  {title}
  {\enquote {\bibinfo {title} {{QMCPACK: an open source ab initio quantum Monte
  Carlo package for the electronic structure of atoms, molecules and
  solids}},}\ }\href {\doibase 10.1088/1361-648X/aab9c3} {\bibfield  {journal}
  {\bibinfo  {journal} {{J. Phys. Condens. Matter}}\ }\textbf {\bibinfo
  {volume} {30}},\ \bibinfo {pages} {195901} (\bibinfo {year} {2018})},\
  \Eprint {http://arxiv.org/abs/1802.06922} {arXiv:1802.06922} \BibitemShut
  {NoStop}%
\bibitem [{\citenamefont {Lehtola}\ \emph {et~al.}(2018)\citenamefont
  {Lehtola}, \citenamefont {Steigemann}, \citenamefont {Oliveira},\ and\
  \citenamefont {Marques}}]{Lehtola2018_1}%
  \BibitemOpen
  \bibfield  {author} {\bibinfo {author} {\bibfnamefont {S.}~\bibnamefont
  {Lehtola}}, \bibinfo {author} {\bibfnamefont {C.}~\bibnamefont {Steigemann}},
  \bibinfo {author} {\bibfnamefont {M.~J.~T.}\ \bibnamefont {Oliveira}}, \ and\
  \bibinfo {author} {\bibfnamefont {M.~A.~L.}\ \bibnamefont {Marques}},\
  }\bibfield  {title} {\enquote {\bibinfo {title} {{Recent developments in
  LIBXC -- A comprehensive library of functionals for density functional
  theory}},}\ }\href {\doibase 10.1016/j.softx.2017.11.002} {\bibfield
  {journal} {\bibinfo  {journal} {SoftwareX}\ }\textbf {\bibinfo {volume}
  {7}},\ \bibinfo {pages} {1} (\bibinfo {year} {2018})}\BibitemShut {NoStop}%
\bibitem [{\citenamefont {Herbst}\ \emph {et~al.}(2019)\citenamefont {Herbst},
  \citenamefont {Scheurer}, \citenamefont {Fransson}, \citenamefont {Rehn},\
  and\ \citenamefont {Dreuw}}]{Herbst2019_e1462}%
  \BibitemOpen
  \bibfield  {author} {\bibinfo {author} {\bibfnamefont {M.~F.}\ \bibnamefont
  {Herbst}}, \bibinfo {author} {\bibfnamefont {M.}~\bibnamefont {Scheurer}},
  \bibinfo {author} {\bibfnamefont {T.}~\bibnamefont {Fransson}}, \bibinfo
  {author} {\bibfnamefont {D.~R.}\ \bibnamefont {Rehn}}, \ and\ \bibinfo
  {author} {\bibfnamefont {A.}~\bibnamefont {Dreuw}},\ }\bibfield  {title}
  {\enquote {\bibinfo {title} {adcc: A versatile toolkit for rapid development
  of algebraic-diagrammatic construction methods},}\ }\href {\doibase
  10.1002/wcms.1462} {\bibfield  {journal} {\bibinfo  {journal} {{Wiley
  Interdiscip. Rev.: Comput. Mol. Sci.}}\ ,\ \bibinfo {pages} {e1462}}
  (\bibinfo {year} {2019})},\ \Eprint {http://arxiv.org/abs/1910.07757}
  {arXiv:1910.07757} \BibitemShut {NoStop}%
\bibitem [{\citenamefont {Koval}, \citenamefont {Barbry},\ and\ \citenamefont
  {S{\'a}nchez-Portal}(2019)}]{Koval2019_188}%
  \BibitemOpen
  \bibfield  {author} {\bibinfo {author} {\bibfnamefont {P.}~\bibnamefont
  {Koval}}, \bibinfo {author} {\bibfnamefont {M.}~\bibnamefont {Barbry}}, \
  and\ \bibinfo {author} {\bibfnamefont {D.}~\bibnamefont
  {S{\'a}nchez-Portal}},\ }\bibfield  {title} {\enquote {\bibinfo {title}
  {{PySCF-NAO: An efficient and flexible implementation of linear response
  time-dependent density functional theory with numerical atomic orbitals}},}\
  }\href {\doibase 10.1016/j.cpc.2018.08.004} {\bibfield  {journal} {\bibinfo
  {journal} {Comput. Phys. Commun.}\ }\textbf {\bibinfo {volume} {236}},\
  \bibinfo {pages} {188} (\bibinfo {year} {2019})}\BibitemShut {NoStop}%
\bibitem [{\citenamefont {Iskakov}\ \emph {et~al.}(2019)\citenamefont
  {Iskakov}, \citenamefont {Rusakov}, \citenamefont {Zgid},\ and\ \citenamefont
  {Gull}}]{Iskakov2019_085112}%
  \BibitemOpen
  \bibfield  {author} {\bibinfo {author} {\bibfnamefont {S.}~\bibnamefont
  {Iskakov}}, \bibinfo {author} {\bibfnamefont {A.~A.}\ \bibnamefont
  {Rusakov}}, \bibinfo {author} {\bibfnamefont {D.}~\bibnamefont {Zgid}}, \
  and\ \bibinfo {author} {\bibfnamefont {E.}~\bibnamefont {Gull}},\ }\bibfield
  {title} {\enquote {\bibinfo {title} {Effect of propagator renormalization on
  the band gap of insulating solids},}\ }\href {\doibase
  10.1103/PhysRevB.100.085112} {\bibfield  {journal} {\bibinfo  {journal}
  {{Phys. Rev. B}}\ }\textbf {\bibinfo {volume} {100}},\ \bibinfo {pages}
  {085112} (\bibinfo {year} {2019})},\ \Eprint
  {http://arxiv.org/abs/1812.07027} {arXiv:1812.07027} \BibitemShut {NoStop}%
\bibitem [{\citenamefont {Sun}\ \emph {et~al.}(2020)\citenamefont {Sun},
  \citenamefont {Zhang}, \citenamefont {Banerjee}, \citenamefont {Bao},
  \citenamefont {Barbry}, \citenamefont {Blunt}, \citenamefont {Bogdanov},
  \citenamefont {Booth}, \citenamefont {Chen}, \citenamefont {Cui},
  \citenamefont {Eriksen}, \citenamefont {Gao}, \citenamefont {Guo},
  \citenamefont {Hermann}, \citenamefont {Hermes}, \citenamefont {Koh},
  \citenamefont {Koval}, \citenamefont {Lehtola}, \citenamefont {Li},
  \citenamefont {Liu}, \citenamefont {Mardirossian}, \citenamefont {McClain},
  \citenamefont {Motta}, \citenamefont {Mussard}, \citenamefont {Pham},
  \citenamefont {Pulkin}, \citenamefont {Purwanto}, \citenamefont {Robinson},
  \citenamefont {Ronca}, \citenamefont {Sayfutyarova}, \citenamefont
  {Scheurer}, \citenamefont {Schurkus}, \citenamefont {Smith}, \citenamefont
  {Sun}, \citenamefont {Sun}, \citenamefont {Upadhyay}, \citenamefont {Wagner},
  \citenamefont {Wang}, \citenamefont {White}, \citenamefont {Whitfield},
  \citenamefont {Williamson}, \citenamefont {Wouters}, \citenamefont {Yang},
  \citenamefont {Yu}, \citenamefont {Zhu}, \citenamefont {Berkelbach},
  \citenamefont {Sharma}, \citenamefont {Sokolov},\ and\ \citenamefont
  {Chan}}]{Sun2020_024109}%
  \BibitemOpen
  \bibfield  {author} {\bibinfo {author} {\bibfnamefont {Q.}~\bibnamefont
  {Sun}}, \bibinfo {author} {\bibfnamefont {X.}~\bibnamefont {Zhang}}, \bibinfo
  {author} {\bibfnamefont {S.}~\bibnamefont {Banerjee}}, \bibinfo {author}
  {\bibfnamefont {P.}~\bibnamefont {Bao}}, \bibinfo {author} {\bibfnamefont
  {M.}~\bibnamefont {Barbry}}, \bibinfo {author} {\bibfnamefont {N.~S.}\
  \bibnamefont {Blunt}}, \bibinfo {author} {\bibfnamefont {N.~A.}\ \bibnamefont
  {Bogdanov}}, \bibinfo {author} {\bibfnamefont {G.~H.}\ \bibnamefont {Booth}},
  \bibinfo {author} {\bibfnamefont {J.}~\bibnamefont {Chen}}, \bibinfo {author}
  {\bibfnamefont {Z.-H.}\ \bibnamefont {Cui}}, \bibinfo {author} {\bibfnamefont
  {J.~J.}\ \bibnamefont {Eriksen}}, \bibinfo {author} {\bibfnamefont
  {Y.}~\bibnamefont {Gao}}, \bibinfo {author} {\bibfnamefont {S.}~\bibnamefont
  {Guo}}, \bibinfo {author} {\bibfnamefont {J.}~\bibnamefont {Hermann}},
  \bibinfo {author} {\bibfnamefont {M.~R.}\ \bibnamefont {Hermes}}, \bibinfo
  {author} {\bibfnamefont {K.}~\bibnamefont {Koh}}, \bibinfo {author}
  {\bibfnamefont {P.}~\bibnamefont {Koval}}, \bibinfo {author} {\bibfnamefont
  {S.}~\bibnamefont {Lehtola}}, \bibinfo {author} {\bibfnamefont
  {Z.}~\bibnamefont {Li}}, \bibinfo {author} {\bibfnamefont {J.}~\bibnamefont
  {Liu}}, \bibinfo {author} {\bibfnamefont {N.}~\bibnamefont {Mardirossian}},
  \bibinfo {author} {\bibfnamefont {J.~D.}\ \bibnamefont {McClain}}, \bibinfo
  {author} {\bibfnamefont {M.}~\bibnamefont {Motta}}, \bibinfo {author}
  {\bibfnamefont {B.}~\bibnamefont {Mussard}}, \bibinfo {author} {\bibfnamefont
  {H.~Q.}\ \bibnamefont {Pham}}, \bibinfo {author} {\bibfnamefont
  {A.}~\bibnamefont {Pulkin}}, \bibinfo {author} {\bibfnamefont
  {W.}~\bibnamefont {Purwanto}}, \bibinfo {author} {\bibfnamefont {P.~J.}\
  \bibnamefont {Robinson}}, \bibinfo {author} {\bibfnamefont {E.}~\bibnamefont
  {Ronca}}, \bibinfo {author} {\bibfnamefont {E.~R.}\ \bibnamefont
  {Sayfutyarova}}, \bibinfo {author} {\bibfnamefont {M.}~\bibnamefont
  {Scheurer}}, \bibinfo {author} {\bibfnamefont {H.~F.}\ \bibnamefont
  {Schurkus}}, \bibinfo {author} {\bibfnamefont {J.~E.~T.}\ \bibnamefont
  {Smith}}, \bibinfo {author} {\bibfnamefont {C.}~\bibnamefont {Sun}}, \bibinfo
  {author} {\bibfnamefont {S.-N.}\ \bibnamefont {Sun}}, \bibinfo {author}
  {\bibfnamefont {S.}~\bibnamefont {Upadhyay}}, \bibinfo {author}
  {\bibfnamefont {L.~K.}\ \bibnamefont {Wagner}}, \bibinfo {author}
  {\bibfnamefont {X.}~\bibnamefont {Wang}}, \bibinfo {author} {\bibfnamefont
  {A.}~\bibnamefont {White}}, \bibinfo {author} {\bibfnamefont {J.~D.}\
  \bibnamefont {Whitfield}}, \bibinfo {author} {\bibfnamefont {M.~J.}\
  \bibnamefont {Williamson}}, \bibinfo {author} {\bibfnamefont
  {S.}~\bibnamefont {Wouters}}, \bibinfo {author} {\bibfnamefont
  {J.}~\bibnamefont {Yang}}, \bibinfo {author} {\bibfnamefont {J.~M.}\
  \bibnamefont {Yu}}, \bibinfo {author} {\bibfnamefont {T.}~\bibnamefont
  {Zhu}}, \bibinfo {author} {\bibfnamefont {T.~C.}\ \bibnamefont {Berkelbach}},
  \bibinfo {author} {\bibfnamefont {S.}~\bibnamefont {Sharma}}, \bibinfo
  {author} {\bibfnamefont {A.~Y.}\ \bibnamefont {Sokolov}}, \ and\ \bibinfo
  {author} {\bibfnamefont {G.~K.-L.}\ \bibnamefont {Chan}},\ }\bibfield
  {title} {\enquote {\bibinfo {title} {Recent developments in the pyscf program
  package},}\ }\href {\doibase 10.1063/5.0006074} {\bibfield  {journal}
  {\bibinfo  {journal} {J. Chem. Phys.}\ }\textbf {\bibinfo {volume} {153}},\
  \bibinfo {pages} {024109} (\bibinfo {year} {2020})},\ \Eprint
  {http://arxiv.org/abs/2002.12531} {arXiv:2002.12531} \BibitemShut {NoStop}%
\bibitem [{\citenamefont {Smith}\ \emph {et~al.}(2020)\citenamefont {Smith},
  \citenamefont {Burns}, \citenamefont {Simmonett}, \citenamefont {Parrish},
  \citenamefont {Schieber}, \citenamefont {Galvelis}, \citenamefont {Kraus},
  \citenamefont {Kruse}, \citenamefont {Di~Remigio}, \citenamefont {Alenaizan},
  \citenamefont {James}, \citenamefont {Lehtola}, \citenamefont {Misiewicz},
  \citenamefont {Scheurer}, \citenamefont {Shaw}, \citenamefont {Schriber},
  \citenamefont {Xie}, \citenamefont {Glick}, \citenamefont {Sirianni},
  \citenamefont {O’Brien}, \citenamefont {Waldrop}, \citenamefont {Kumar},
  \citenamefont {Hohenstein}, \citenamefont {Pritchard}, \citenamefont
  {Brooks}, \citenamefont {Schaefer}, \citenamefont {Sokolov}, \citenamefont
  {Patkowski}, \citenamefont {DePrince}, \citenamefont {Bozkaya}, \citenamefont
  {King}, \citenamefont {Evangelista}, \citenamefont {Turney}, \citenamefont
  {Crawford},\ and\ \citenamefont {Sherrill}}]{Smith2020_184108}%
  \BibitemOpen
  \bibfield  {author} {\bibinfo {author} {\bibfnamefont {D.~G.~A.}\
  \bibnamefont {Smith}}, \bibinfo {author} {\bibfnamefont {L.~A.}\ \bibnamefont
  {Burns}}, \bibinfo {author} {\bibfnamefont {A.~C.}\ \bibnamefont
  {Simmonett}}, \bibinfo {author} {\bibfnamefont {R.~M.}\ \bibnamefont
  {Parrish}}, \bibinfo {author} {\bibfnamefont {M.~C.}\ \bibnamefont
  {Schieber}}, \bibinfo {author} {\bibfnamefont {R.}~\bibnamefont {Galvelis}},
  \bibinfo {author} {\bibfnamefont {P.}~\bibnamefont {Kraus}}, \bibinfo
  {author} {\bibfnamefont {H.}~\bibnamefont {Kruse}}, \bibinfo {author}
  {\bibfnamefont {R.}~\bibnamefont {Di~Remigio}}, \bibinfo {author}
  {\bibfnamefont {A.}~\bibnamefont {Alenaizan}}, \bibinfo {author}
  {\bibfnamefont {A.~M.}\ \bibnamefont {James}}, \bibinfo {author}
  {\bibfnamefont {S.}~\bibnamefont {Lehtola}}, \bibinfo {author} {\bibfnamefont
  {J.~P.}\ \bibnamefont {Misiewicz}}, \bibinfo {author} {\bibfnamefont
  {M.}~\bibnamefont {Scheurer}}, \bibinfo {author} {\bibfnamefont {R.~A.}\
  \bibnamefont {Shaw}}, \bibinfo {author} {\bibfnamefont {J.~B.}\ \bibnamefont
  {Schriber}}, \bibinfo {author} {\bibfnamefont {Y.}~\bibnamefont {Xie}},
  \bibinfo {author} {\bibfnamefont {Z.~L.}\ \bibnamefont {Glick}}, \bibinfo
  {author} {\bibfnamefont {D.~A.}\ \bibnamefont {Sirianni}}, \bibinfo {author}
  {\bibfnamefont {J.~S.}\ \bibnamefont {O’Brien}}, \bibinfo {author}
  {\bibfnamefont {J.~M.}\ \bibnamefont {Waldrop}}, \bibinfo {author}
  {\bibfnamefont {A.}~\bibnamefont {Kumar}}, \bibinfo {author} {\bibfnamefont
  {E.~G.}\ \bibnamefont {Hohenstein}}, \bibinfo {author} {\bibfnamefont
  {B.~P.}\ \bibnamefont {Pritchard}}, \bibinfo {author} {\bibfnamefont {B.~R.}\
  \bibnamefont {Brooks}}, \bibinfo {author} {\bibfnamefont {H.~F.}\
  \bibnamefont {Schaefer}}, \bibinfo {author} {\bibfnamefont {A.~Y.}\
  \bibnamefont {Sokolov}}, \bibinfo {author} {\bibfnamefont {K.}~\bibnamefont
  {Patkowski}}, \bibinfo {author} {\bibfnamefont {A.~E.}\ \bibnamefont
  {DePrince}}, \bibinfo {author} {\bibfnamefont {U.}~\bibnamefont {Bozkaya}},
  \bibinfo {author} {\bibfnamefont {R.~A.}\ \bibnamefont {King}}, \bibinfo
  {author} {\bibfnamefont {F.~A.}\ \bibnamefont {Evangelista}}, \bibinfo
  {author} {\bibfnamefont {J.~M.}\ \bibnamefont {Turney}}, \bibinfo {author}
  {\bibfnamefont {T.~D.}\ \bibnamefont {Crawford}}, \ and\ \bibinfo {author}
  {\bibfnamefont {C.~D.}\ \bibnamefont {Sherrill}},\ }\bibfield  {title}
  {\enquote {\bibinfo {title} {{PSI4 1.4: Open-source software for
  high-throughput quantum chemistry}},}\ }\href {\doibase 10.1063/5.0006002}
  {\bibfield  {journal} {\bibinfo  {journal} {J. Chem. Phys.}\ }\textbf
  {\bibinfo {volume} {152}},\ \bibinfo {pages} {184108} (\bibinfo {year}
  {2020})}\BibitemShut {NoStop}%
\bibitem [{\citenamefont {Pederson}, \citenamefont {Ruzsinszky},\ and\
  \citenamefont {Perdew}(2014)}]{Pederson2014_121103}%
  \BibitemOpen
  \bibfield  {author} {\bibinfo {author} {\bibfnamefont {M.~R.}\ \bibnamefont
  {Pederson}}, \bibinfo {author} {\bibfnamefont {A.}~\bibnamefont
  {Ruzsinszky}}, \ and\ \bibinfo {author} {\bibfnamefont {J.~P.}\ \bibnamefont
  {Perdew}},\ }\bibfield  {title} {\enquote {\bibinfo {title} {{Communication:
  Self-interaction correction with unitary invariance in density functional
  theory}},}\ }\href {\doibase 10.1063/1.4869581} {\bibfield  {journal}
  {\bibinfo  {journal} {J. Chem. Phys.}\ }\textbf {\bibinfo {volume} {140}},\
  \bibinfo {pages} {121103} (\bibinfo {year} {2014})}\BibitemShut {NoStop}%
\bibitem [{\citenamefont {Pederson}(2015)}]{Pederson2015_064112}%
  \BibitemOpen
  \bibfield  {author} {\bibinfo {author} {\bibfnamefont {M.~R.}\ \bibnamefont
  {Pederson}},\ }\bibfield  {title} {\enquote {\bibinfo {title} {{Fermi orbital
  derivatives in self-interaction corrected density functional theory:
  Applications to closed shell atoms}},}\ }\href {\doibase 10.1063/1.4907592}
  {\bibfield  {journal} {\bibinfo  {journal} {J. Chem. Phys.}\ }\textbf
  {\bibinfo {volume} {142}},\ \bibinfo {pages} {064112} (\bibinfo {year}
  {2015})},\ \Eprint {http://arxiv.org/abs/1412.3101} {arXiv:1412.3101}
  \BibitemShut {NoStop}%
\bibitem [{\citenamefont {Pederson}\ and\ \citenamefont
  {Baruah}(2015)}]{Pederson2015_153}%
  \BibitemOpen
  \bibfield  {author} {\bibinfo {author} {\bibfnamefont {M.~R.}\ \bibnamefont
  {Pederson}}\ and\ \bibinfo {author} {\bibfnamefont {T.}~\bibnamefont
  {Baruah}},\ }\bibfield  {title} {\enquote {\bibinfo {title}
  {{Self-interaction corrections within the Fermi-orbital-based formalism}},}\
  }\href {\doibase 10.1016/bs.aamop.2015.06.005} {\bibfield  {journal}
  {\bibinfo  {journal} {{Adv. At., Mol., Opt. Phys.}}\ }\textbf {\bibinfo
  {volume} {64}},\ \bibinfo {pages} {153} (\bibinfo {year} {2015})}\BibitemShut
  {NoStop}%
\bibitem [{\citenamefont {Yang}, \citenamefont {Pederson},\ and\ \citenamefont
  {Perdew}(2017)}]{Yang2017_052505}%
  \BibitemOpen
  \bibfield  {author} {\bibinfo {author} {\bibfnamefont {Z.-h.}\ \bibnamefont
  {Yang}}, \bibinfo {author} {\bibfnamefont {M.~R.}\ \bibnamefont {Pederson}},
  \ and\ \bibinfo {author} {\bibfnamefont {J.~P.}\ \bibnamefont {Perdew}},\
  }\bibfield  {title} {\enquote {\bibinfo {title} {{Full self-consistency in
  the Fermi-orbital self-interaction correction}},}\ }\href {\doibase
  10.1103/PhysRevA.95.052505} {\bibfield  {journal} {\bibinfo  {journal} {Phys.
  Rev. A}\ }\textbf {\bibinfo {volume} {95}},\ \bibinfo {pages} {052505}
  (\bibinfo {year} {2017})}\BibitemShut {NoStop}%
\bibitem [{\citenamefont {Fiedler}(2018)}]{Fiedler2018_1}%
  \BibitemOpen
  \bibfield  {author} {\bibinfo {author} {\bibfnamefont {L.}~\bibnamefont
  {Fiedler}},\ }\href {\doibase 10.13140/RG.2.2.22369.86884} {\enquote
  {\bibinfo {title} {{\emph{Implementation and reassessment of the
  Fermi--L\"owdin orbital self-interaction correction for LDA, GGA and mGGA
  functionals}, Master's thesis, TU Bergakademie Freiberg}},}\ } (\bibinfo
  {year} {2018})\BibitemShut {NoStop}%
\bibitem [{\citenamefont {Sun}\ \emph {et~al.}(2018)\citenamefont {Sun},
  \citenamefont {Berkelbach}, \citenamefont {Blunt}, \citenamefont {Booth},
  \citenamefont {Guo}, \citenamefont {Li}, \citenamefont {Liu}, \citenamefont
  {McClain}, \citenamefont {Sayfutyarova}, \citenamefont {Sharma} \emph
  {et~al.}}]{Sun2018_e1340}%
  \BibitemOpen
  \bibfield  {author} {\bibinfo {author} {\bibfnamefont {Q.}~\bibnamefont
  {Sun}}, \bibinfo {author} {\bibfnamefont {T.~C.}\ \bibnamefont {Berkelbach}},
  \bibinfo {author} {\bibfnamefont {N.~S.}\ \bibnamefont {Blunt}}, \bibinfo
  {author} {\bibfnamefont {G.~H.}\ \bibnamefont {Booth}}, \bibinfo {author}
  {\bibfnamefont {S.}~\bibnamefont {Guo}}, \bibinfo {author} {\bibfnamefont
  {Z.}~\bibnamefont {Li}}, \bibinfo {author} {\bibfnamefont {J.}~\bibnamefont
  {Liu}}, \bibinfo {author} {\bibfnamefont {J.~D.}\ \bibnamefont {McClain}},
  \bibinfo {author} {\bibfnamefont {E.~R.}\ \bibnamefont {Sayfutyarova}},
  \bibinfo {author} {\bibfnamefont {S.}~\bibnamefont {Sharma}},  \emph
  {et~al.},\ }\bibfield  {title} {\enquote {\bibinfo {title} {{PySCF: The
  Python-based simulations of chemistry framework}},}\ }\href {\doibase
  10.1002/wcms.1340} {\bibfield  {journal} {\bibinfo  {journal} {{Wiley
  Interdiscip. Rev. Comput. Mol. Sci.}}\ }\textbf {\bibinfo {volume} {8}},\
  \bibinfo {pages} {e1340} (\bibinfo {year} {2018})}\BibitemShut {NoStop}%
\bibitem [{\citenamefont {Hohenberg}\ and\ \citenamefont
  {Kohn}(1964)}]{Hohenberg1964_B864}%
  \BibitemOpen
  \bibfield  {author} {\bibinfo {author} {\bibfnamefont {P.}~\bibnamefont
  {Hohenberg}}\ and\ \bibinfo {author} {\bibfnamefont {W.}~\bibnamefont
  {Kohn}},\ }\bibfield  {title} {\enquote {\bibinfo {title} {Inhomogeneous
  electron gas},}\ }\href {\doibase 10.1103/PhysRev.136.B864} {\bibfield
  {journal} {\bibinfo  {journal} {Phys. Rev.}\ }\textbf {\bibinfo {volume}
  {136}},\ \bibinfo {pages} {B864} (\bibinfo {year} {1964})}\BibitemShut
  {NoStop}%
\bibitem [{\citenamefont {Kohn}\ and\ \citenamefont
  {Sham}(1965)}]{Kohn1965_A1133}%
  \BibitemOpen
  \bibfield  {author} {\bibinfo {author} {\bibfnamefont {W.}~\bibnamefont
  {Kohn}}\ and\ \bibinfo {author} {\bibfnamefont {L.~J.}\ \bibnamefont
  {Sham}},\ }\bibfield  {title} {\enquote {\bibinfo {title} {Self-consistent
  equations including exchange and correlation effects},}\ }\href {\doibase
  10.1103/PhysRev.140.A1133} {\bibfield  {journal} {\bibinfo  {journal} {Phys.
  Rev.}\ }\textbf {\bibinfo {volume} {140}},\ \bibinfo {pages} {A1133}
  (\bibinfo {year} {1965})}\BibitemShut {NoStop}%
\bibitem [{\citenamefont {Sun}(2016)}]{Sun2016_1}%
  \BibitemOpen
  \bibfield  {author} {\bibinfo {author} {\bibfnamefont {Q.}~\bibnamefont
  {Sun}},\ }\bibfield  {title} {\enquote {\bibinfo {title} {{Co-iterative
  augmented Hessian method for orbital optimization}},}\ }\href
  {https://arxiv.org/abs/1610.08423} {\bibfield  {journal} {\bibinfo  {journal}
  {{arXiv:1610.08423}}\ } (\bibinfo {year} {2016})}\BibitemShut {NoStop}%
\bibitem [{\citenamefont {Pederson}\ and\ \citenamefont
  {Jackson}(1990)}]{Pederson1990_7453}%
  \BibitemOpen
  \bibfield  {author} {\bibinfo {author} {\bibfnamefont {M.~R.}\ \bibnamefont
  {Pederson}}\ and\ \bibinfo {author} {\bibfnamefont {K.~A.}\ \bibnamefont
  {Jackson}},\ }\bibfield  {title} {\enquote {\bibinfo {title} {Variational
  mesh for quantum-mechanical simulations},}\ }\href {\doibase
  10.1103/PhysRevB.41.7453} {\bibfield  {journal} {\bibinfo  {journal} {Phys.
  Rev. B}\ }\textbf {\bibinfo {volume} {41}},\ \bibinfo {pages} {7453}
  (\bibinfo {year} {1990})}\BibitemShut {NoStop}%
\bibitem [{\citenamefont {Pederson}, \citenamefont {Jackson},\ and\
  \citenamefont {Pickett}(1991)}]{Pederson1991_3891}%
  \BibitemOpen
  \bibfield  {author} {\bibinfo {author} {\bibfnamefont {M.~R.}\ \bibnamefont
  {Pederson}}, \bibinfo {author} {\bibfnamefont {K.~A.}\ \bibnamefont
  {Jackson}}, \ and\ \bibinfo {author} {\bibfnamefont {W.~E.}\ \bibnamefont
  {Pickett}},\ }\bibfield  {title} {\enquote {\bibinfo {title}
  {Local-density-approximation-based simulations of hydrocarbon interactions
  with applications to diamond chemical vapor deposition},}\ }\href {\doibase
  10.1103/PhysRevB.44.3891} {\bibfield  {journal} {\bibinfo  {journal} {Phys.
  Rev. B}\ }\textbf {\bibinfo {volume} {44}},\ \bibinfo {pages} {3891}
  (\bibinfo {year} {1991})}\BibitemShut {NoStop}%
\bibitem [{\citenamefont {Pederson}\ and\ \citenamefont
  {Jackson}(1991)}]{Pederson1991_7312}%
  \BibitemOpen
  \bibfield  {author} {\bibinfo {author} {\bibfnamefont {M.~R.}\ \bibnamefont
  {Pederson}}\ and\ \bibinfo {author} {\bibfnamefont {K.~A.}\ \bibnamefont
  {Jackson}},\ }\bibfield  {title} {\enquote {\bibinfo {title} {Pseudoenergies
  for simulations on metallic systems},}\ }\href {\doibase
  10.1103/PhysRevB.43.7312} {\bibfield  {journal} {\bibinfo  {journal} {Phys.
  Rev. B}\ }\textbf {\bibinfo {volume} {43}},\ \bibinfo {pages} {7312}
  (\bibinfo {year} {1991})}\BibitemShut {NoStop}%
\bibitem [{\citenamefont {Perdew}\ \emph {et~al.}(1992)\citenamefont {Perdew},
  \citenamefont {Chevary}, \citenamefont {Vosko}, \citenamefont {Jackson},
  \citenamefont {Pederson}, \citenamefont {Singh},\ and\ \citenamefont
  {Fiolhais}}]{Perdew1992_6671}%
  \BibitemOpen
  \bibfield  {author} {\bibinfo {author} {\bibfnamefont {J.~P.}\ \bibnamefont
  {Perdew}}, \bibinfo {author} {\bibfnamefont {J.~A.}\ \bibnamefont {Chevary}},
  \bibinfo {author} {\bibfnamefont {S.~H.}\ \bibnamefont {Vosko}}, \bibinfo
  {author} {\bibfnamefont {K.~A.}\ \bibnamefont {Jackson}}, \bibinfo {author}
  {\bibfnamefont {M.~R.}\ \bibnamefont {Pederson}}, \bibinfo {author}
  {\bibfnamefont {D.~J.}\ \bibnamefont {Singh}}, \ and\ \bibinfo {author}
  {\bibfnamefont {C.}~\bibnamefont {Fiolhais}},\ }\bibfield  {title} {\enquote
  {\bibinfo {title} {{Atoms, molecules, solids, and surfaces: Applications of
  the generalized gradient approximation for exchange and correlation}},}\
  }\href {\doibase 10.1103/PhysRevB.46.6671} {\bibfield  {journal} {\bibinfo
  {journal} {Phys. Rev. B}\ }\textbf {\bibinfo {volume} {46}},\ \bibinfo
  {pages} {6671} (\bibinfo {year} {1992})}\BibitemShut {NoStop}%
\bibitem [{\citenamefont {Porezag}\ and\ \citenamefont
  {Pederson}(1996)}]{Porezag1996_7830}%
  \BibitemOpen
  \bibfield  {author} {\bibinfo {author} {\bibfnamefont {D.}~\bibnamefont
  {Porezag}}\ and\ \bibinfo {author} {\bibfnamefont {M.~R.}\ \bibnamefont
  {Pederson}},\ }\bibfield  {title} {\enquote {\bibinfo {title} {{Infrared
  intensities and Raman-scattering activities within density-functional
  theory}},}\ }\href {\doibase 10.1103/PhysRevB.54.7830} {\bibfield  {journal}
  {\bibinfo  {journal} {Phys. Rev. B}\ }\textbf {\bibinfo {volume} {54}},\
  \bibinfo {pages} {7830} (\bibinfo {year} {1996})}\BibitemShut {NoStop}%
\bibitem [{\citenamefont {Porezag}(1997)}]{Porezag1997_1}%
  \BibitemOpen
  \bibfield  {author} {\bibinfo {author} {\bibfnamefont {D.~V.}\ \bibnamefont
  {Porezag}},\ }\href {https://nbn-resolving.org/urn:nbn:de:bsz:ch1-199700253}
  {\enquote {\bibinfo {title} {{\emph{Development of ab-initio and approximate
  density functional methods and their application to complex fullerene
  systems}, PhD thesis, TU Chemnitz}},}\ } (\bibinfo {year} {1997})\BibitemShut
  {NoStop}%
\bibitem [{\citenamefont {Porezag}\ and\ \citenamefont
  {Pederson}(1999)}]{Porezag1999_2840}%
  \BibitemOpen
  \bibfield  {author} {\bibinfo {author} {\bibfnamefont {D.}~\bibnamefont
  {Porezag}}\ and\ \bibinfo {author} {\bibfnamefont {M.~R.}\ \bibnamefont
  {Pederson}},\ }\bibfield  {title} {\enquote {\bibinfo {title} {{Optimization
  of Gaussian basis sets for density-functional calculations}},}\ }\href
  {\doibase 10.1103/PhysRevA.60.2840} {\bibfield  {journal} {\bibinfo
  {journal} {Phys. Rev. A}\ }\textbf {\bibinfo {volume} {60}},\ \bibinfo
  {pages} {2840} (\bibinfo {year} {1999})}\BibitemShut {NoStop}%
\bibitem [{\citenamefont {Kortus}\ and\ \citenamefont
  {Pederson}(2000)}]{Kortus2000_5755}%
  \BibitemOpen
  \bibfield  {author} {\bibinfo {author} {\bibfnamefont {J.}~\bibnamefont
  {Kortus}}\ and\ \bibinfo {author} {\bibfnamefont {M.~R.}\ \bibnamefont
  {Pederson}},\ }\bibfield  {title} {\enquote {\bibinfo {title} {Magnetic and
  vibrational properties of the uniaxial {Fe}$_{\textrm{13}}${O}$_{\textrm{8}}$
  cluster},}\ }\href {\doibase 10.1103/PhysRevB.62.5755} {\bibfield  {journal}
  {\bibinfo  {journal} {Phys. Rev. B}\ }\textbf {\bibinfo {volume} {62}},\
  \bibinfo {pages} {5755} (\bibinfo {year} {2000})}\BibitemShut {NoStop}%
\bibitem [{\citenamefont {Pederson}\ \emph {et~al.}(2000)\citenamefont
  {Pederson}, \citenamefont {Porezag}, \citenamefont {Kortus},\ and\
  \citenamefont {Patton}}]{Pederson2000_197}%
  \BibitemOpen
  \bibfield  {author} {\bibinfo {author} {\bibfnamefont {M.~R.}\ \bibnamefont
  {Pederson}}, \bibinfo {author} {\bibfnamefont {D.~V.}\ \bibnamefont
  {Porezag}}, \bibinfo {author} {\bibfnamefont {J.}~\bibnamefont {Kortus}}, \
  and\ \bibinfo {author} {\bibfnamefont {D.~C.}\ \bibnamefont {Patton}},\
  }\bibfield  {title} {\enquote {\bibinfo {title} {Strategies for massively
  parallel local-orbital-based electronic structure methods},}\ }\href
  {\doibase 10.1002/(SICI)1521-3951(200001)217:1<197::AID-PSSB197>3.0.CO;2-B}
  {\bibfield  {journal} {\bibinfo  {journal} {{Phys. Status Solidi B}}\
  }\textbf {\bibinfo {volume} {217}},\ \bibinfo {pages} {197} (\bibinfo {year}
  {2000})}\BibitemShut {NoStop}%
\bibitem [{\citenamefont {Aquino}\ and\ \citenamefont
  {Wong}(2018)}]{Aquino2018_6456}%
  \BibitemOpen
  \bibfield  {author} {\bibinfo {author} {\bibfnamefont {F.~W.}\ \bibnamefont
  {Aquino}}\ and\ \bibinfo {author} {\bibfnamefont {B.~M.}\ \bibnamefont
  {Wong}},\ }\bibfield  {title} {\enquote {\bibinfo {title} {{Additional
  insights between Fermi--L{\"o}wdin orbital SIC and the localization equation
  constraints in SIC-DFT}},}\ }\href {\doibase 10.1021/acs.jpclett.8b02786}
  {\bibfield  {journal} {\bibinfo  {journal} {J. Phys. Chem. Lett.}\ }\textbf
  {\bibinfo {volume} {9}},\ \bibinfo {pages} {6456} (\bibinfo {year}
  {2018})}\BibitemShut {NoStop}%
\bibitem [{\citenamefont {Aquino}, \citenamefont {Shinde},\ and\ \citenamefont
  {Wong}(2020)}]{Aquino2020_1200}%
  \BibitemOpen
  \bibfield  {author} {\bibinfo {author} {\bibfnamefont {F.~W.}\ \bibnamefont
  {Aquino}}, \bibinfo {author} {\bibfnamefont {R.}~\bibnamefont {Shinde}}, \
  and\ \bibinfo {author} {\bibfnamefont {B.~M.}\ \bibnamefont {Wong}},\
  }\bibfield  {title} {\enquote {\bibinfo {title} {{Fractional occupation
  numbers and self-interaction correction-scaling methods with the
  Fermi--L{\"o}wdin orbital self-interaction correction approach}},}\ }\href
  {\doibase 10.1002/jcc.26168} {\bibfield  {journal} {\bibinfo  {journal} {J.
  Comput. Chem.}\ }\textbf {\bibinfo {volume} {41}},\ \bibinfo {pages} {1200}
  (\bibinfo {year} {2020})}\BibitemShut {NoStop}%
\bibitem [{\citenamefont {Pritchard}\ \emph {et~al.}(2019)\citenamefont
  {Pritchard}, \citenamefont {Altarawy}, \citenamefont {Didier}, \citenamefont
  {Gibson},\ and\ \citenamefont {Windus}}]{Pritchard2019_4814}%
  \BibitemOpen
  \bibfield  {author} {\bibinfo {author} {\bibfnamefont {B.~P.}\ \bibnamefont
  {Pritchard}}, \bibinfo {author} {\bibfnamefont {D.}~\bibnamefont {Altarawy}},
  \bibinfo {author} {\bibfnamefont {B.~T.}\ \bibnamefont {Didier}}, \bibinfo
  {author} {\bibfnamefont {T.~D.}\ \bibnamefont {Gibson}}, \ and\ \bibinfo
  {author} {\bibfnamefont {T.~L.}\ \bibnamefont {Windus}},\ }\bibfield  {title}
  {\enquote {\bibinfo {title} {{A new basis set exchange: An open, up-to-date
  resource for the molecular sciences community}},}\ }\href {\doibase
  10.1021/acs.jcim.9b00725} {\bibfield  {journal} {\bibinfo  {journal} {{J.
  Chem. Inf. Model.}}\ }\textbf {\bibinfo {volume} {59}},\ \bibinfo {pages}
  {4814} (\bibinfo {year} {2019})}\BibitemShut {NoStop}%
\bibitem [{\citenamefont {Jensen}\ \emph {et~al.}(2017)\citenamefont {Jensen},
  \citenamefont {Saha}, \citenamefont {Flores-Livas}, \citenamefont {Huhn},
  \citenamefont {Blum}, \citenamefont {Goedecker},\ and\ \citenamefont
  {Frediani}}]{Jensen2017_1449}%
  \BibitemOpen
  \bibfield  {author} {\bibinfo {author} {\bibfnamefont {S.~R.}\ \bibnamefont
  {Jensen}}, \bibinfo {author} {\bibfnamefont {S.}~\bibnamefont {Saha}},
  \bibinfo {author} {\bibfnamefont {J.~A.}\ \bibnamefont {Flores-Livas}},
  \bibinfo {author} {\bibfnamefont {W.}~\bibnamefont {Huhn}}, \bibinfo {author}
  {\bibfnamefont {V.}~\bibnamefont {Blum}}, \bibinfo {author} {\bibfnamefont
  {S.}~\bibnamefont {Goedecker}}, \ and\ \bibinfo {author} {\bibfnamefont
  {L.}~\bibnamefont {Frediani}},\ }\bibfield  {title} {\enquote {\bibinfo
  {title} {{The elephant in the room of density functional theory
  calculations}},}\ }\href {\doibase 10.1021/acs.jpclett.7b00255} {\bibfield
  {journal} {\bibinfo  {journal} {J. Phys. Chem. Lett.}\ }\textbf {\bibinfo
  {volume} {8}},\ \bibinfo {pages} {1449} (\bibinfo {year} {2017})},\ \Eprint
  {http://arxiv.org/abs/1702.00957} {arXiv:1702.00957} \BibitemShut {NoStop}%
\bibitem [{\citenamefont {Jensen}(2017)}]{Jensen2017_6104}%
  \BibitemOpen
  \bibfield  {author} {\bibinfo {author} {\bibfnamefont {F.}~\bibnamefont
  {Jensen}},\ }\bibfield  {title} {\enquote {\bibinfo {title} {{How large is
  the elephant in the density functional theory room?}}}\ }\href {\doibase
  10.1021/acs.jpca.7b04760} {\bibfield  {journal} {\bibinfo  {journal} {J.
  Phys. Chem. A}\ }\textbf {\bibinfo {volume} {121}},\ \bibinfo {pages} {6104}
  (\bibinfo {year} {2017})},\ \Eprint {http://arxiv.org/abs/1704.08832}
  {arXiv:1704.08832} \BibitemShut {NoStop}%
\bibitem [{\citenamefont {Feller}\ and\ \citenamefont
  {Dixon}(2018)}]{Feller2018_2598}%
  \BibitemOpen
  \bibfield  {author} {\bibinfo {author} {\bibfnamefont {D.}~\bibnamefont
  {Feller}}\ and\ \bibinfo {author} {\bibfnamefont {D.~A.}\ \bibnamefont
  {Dixon}},\ }\bibfield  {title} {\enquote {\bibinfo {title} {{Density
  functional theory and the basis set truncation problem with correlation
  consistent basis sets: Elephant in the room or mouse in the closet?}}}\
  }\href {\doibase 10.1021/acs.jpca.8b00392} {\bibfield  {journal} {\bibinfo
  {journal} {J. Phys. Chem. A}\ }\textbf {\bibinfo {volume} {122}},\ \bibinfo
  {pages} {2598} (\bibinfo {year} {2018})}\BibitemShut {NoStop}%
\bibitem [{\citenamefont {Lehtola}(2020{\natexlab{a}})}]{Lehtola2020_134108}%
  \BibitemOpen
  \bibfield  {author} {\bibinfo {author} {\bibfnamefont {S.}~\bibnamefont
  {Lehtola}},\ }\bibfield  {title} {\enquote {\bibinfo {title} {{Polarized
  Gaussian basis sets from one-electron ions}},}\ }\href {\doibase
  10.1063/1.5144964} {\bibfield  {journal} {\bibinfo  {journal} {J. Chem.
  Phys.}\ }\textbf {\bibinfo {volume} {152}},\ \bibinfo {pages} {134108}
  (\bibinfo {year} {2020}{\natexlab{a}})},\ \Eprint
  {http://arxiv.org/abs/2001.04224} {arXiv:2001.04224} \BibitemShut {NoStop}%
\bibitem [{\citenamefont {Lehtola}(2019)}]{Lehtola2019_241102}%
  \BibitemOpen
  \bibfield  {author} {\bibinfo {author} {\bibfnamefont {S.}~\bibnamefont
  {Lehtola}},\ }\bibfield  {title} {\enquote {\bibinfo {title} {{Curing basis
  set overcompleteness with pivoted Cholesky decompositions}},}\ }\href
  {\doibase 10.1063/1.5139948} {\bibfield  {journal} {\bibinfo  {journal} {J.
  Chem. Phys.}\ }\textbf {\bibinfo {volume} {151}},\ \bibinfo {pages} {241102}
  (\bibinfo {year} {2019})},\ \Eprint {http://arxiv.org/abs/1911.10372}
  {arXiv:1911.10372} \BibitemShut {NoStop}%
\bibitem [{\citenamefont {Lehtola}(2020{\natexlab{b}})}]{Lehtola2020_032504}%
  \BibitemOpen
  \bibfield  {author} {\bibinfo {author} {\bibfnamefont {S.}~\bibnamefont
  {Lehtola}},\ }\bibfield  {title} {\enquote {\bibinfo {title} {{Accurate
  reproduction of strongly repulsive interatomic potentials}},}\ }\href
  {\doibase 10.1103/PhysRevA.101.032504} {\bibfield  {journal} {\bibinfo
  {journal} {Phys. Rev. A}\ }\textbf {\bibinfo {volume} {101}},\ \bibinfo
  {pages} {032504} (\bibinfo {year} {2020}{\natexlab{b}})},\ \Eprint
  {http://arxiv.org/abs/1912.12624} {arXiv:1912.12624} \BibitemShut {NoStop}%
\bibitem [{\citenamefont {Ekstr\"om}\ \emph {et~al.}(2010)\citenamefont
  {Ekstr\"om}, \citenamefont {Visscher}, \citenamefont {Bast}, \citenamefont
  {Thorvaldsen},\ and\ \citenamefont {Ruud}}]{Ekstrom2010_1971}%
  \BibitemOpen
  \bibfield  {author} {\bibinfo {author} {\bibfnamefont {U.}~\bibnamefont
  {Ekstr\"om}}, \bibinfo {author} {\bibfnamefont {L.}~\bibnamefont {Visscher}},
  \bibinfo {author} {\bibfnamefont {R.}~\bibnamefont {Bast}}, \bibinfo {author}
  {\bibfnamefont {A.~J.}\ \bibnamefont {Thorvaldsen}}, \ and\ \bibinfo {author}
  {\bibfnamefont {K.}~\bibnamefont {Ruud}},\ }\bibfield  {title} {\enquote
  {\bibinfo {title} {Arbitrary-order density functional response theory from
  automatic differentiation},}\ }\href {\doibase 10.1021/ct100117s} {\bibfield
  {journal} {\bibinfo  {journal} {{J. Chem. Theory Comput.}}\ }\textbf
  {\bibinfo {volume} {6}},\ \bibinfo {pages} {1971} (\bibinfo {year}
  {2010})}\BibitemShut {NoStop}%
\bibitem [{fun()}]{funclist}%
  \BibitemOpen
  \href@noop {} {\enquote {\bibinfo {title} {See
  \url{https://www.tddft.org/programs/libxc/functionals/} for list of
  functionals in \textsc{Libxc} (accessed 17 {J}uly 2020) and
  \url{https://xcfun.readthedocs.io/en/latest/functionals.html} for those in
  \textsc{XCFun} (accessed 17 {J}uly 2020).}}\ }\BibitemShut {NoStop}%
\bibitem [{\citenamefont {Lehtola}\ and\ \citenamefont
  {J\'onsson}(2014)}]{Lehtola2014_5324}%
  \BibitemOpen
  \bibfield  {author} {\bibinfo {author} {\bibfnamefont {S.}~\bibnamefont
  {Lehtola}}\ and\ \bibinfo {author} {\bibfnamefont {H.}~\bibnamefont
  {J\'onsson}},\ }\bibfield  {title} {\enquote {\bibinfo {title} {{Variational,
  self-consistent implementation of the Perdew--Zunger self-interaction
  correction with complex optimal orbitals}},}\ }\href {\doibase
  10.1021/ct500637x} {\bibfield  {journal} {\bibinfo  {journal} {J. Chem.
  Theory Comput.}\ }\textbf {\bibinfo {volume} {10}},\ \bibinfo {pages} {5324}
  (\bibinfo {year} {2014})}\BibitemShut {NoStop}%
\bibitem [{\citenamefont {Sherrill}(2010)}]{Sherrill2010_110902}%
  \BibitemOpen
  \bibfield  {author} {\bibinfo {author} {\bibfnamefont {C.~D.}\ \bibnamefont
  {Sherrill}},\ }\bibfield  {title} {\enquote {\bibinfo {title} {Frontiers in
  electronic structure theory},}\ }\href {\doibase 10.1063/1.3369628}
  {\bibfield  {journal} {\bibinfo  {journal} {J. Chem. Phys.}\ }\textbf
  {\bibinfo {volume} {132}},\ \bibinfo {pages} {110902} (\bibinfo {year}
  {2010})}\BibitemShut {NoStop}%
\bibitem [{\citenamefont {Burke}(2012)}]{Burke2012_150901}%
  \BibitemOpen
  \bibfield  {author} {\bibinfo {author} {\bibfnamefont {K.}~\bibnamefont
  {Burke}},\ }\bibfield  {title} {\enquote {\bibinfo {title} {Perspective on
  density functional theory},}\ }\href {\doibase 10.1063/1.4704546} {\bibfield
  {journal} {\bibinfo  {journal} {J. Chem. Phys.}\ }\textbf {\bibinfo {volume}
  {136}},\ \bibinfo {pages} {150901} (\bibinfo {year} {2012})},\ \Eprint
  {http://arxiv.org/abs/1201.3679} {arXiv:1201.3679} \BibitemShut {NoStop}%
\bibitem [{\citenamefont {Cohen}, \citenamefont {Mori-S{\'a}nchez},\ and\
  \citenamefont {Yang}(2008)}]{Cohen2008_792}%
  \BibitemOpen
  \bibfield  {author} {\bibinfo {author} {\bibfnamefont {A.~J.}\ \bibnamefont
  {Cohen}}, \bibinfo {author} {\bibfnamefont {P.}~\bibnamefont
  {Mori-S{\'a}nchez}}, \ and\ \bibinfo {author} {\bibfnamefont
  {W.}~\bibnamefont {Yang}},\ }\bibfield  {title} {\enquote {\bibinfo {title}
  {{Insights into current limitations of density functional theory}},}\ }\href
  {\doibase 10.1126/science.1158722} {\bibfield  {journal} {\bibinfo  {journal}
  {Science}\ }\textbf {\bibinfo {volume} {321}},\ \bibinfo {pages} {792}
  (\bibinfo {year} {2008})}\BibitemShut {NoStop}%
\bibitem [{\citenamefont {Perdew}\ \emph {et~al.}(2009)\citenamefont {Perdew},
  \citenamefont {Ruzsinszky}, \citenamefont {Constantin}, \citenamefont {Sun},\
  and\ \citenamefont {Csonka}}]{Perdew2009_902}%
  \BibitemOpen
  \bibfield  {author} {\bibinfo {author} {\bibfnamefont {J.~P.}\ \bibnamefont
  {Perdew}}, \bibinfo {author} {\bibfnamefont {A.}~\bibnamefont {Ruzsinszky}},
  \bibinfo {author} {\bibfnamefont {L.~A.}\ \bibnamefont {Constantin}},
  \bibinfo {author} {\bibfnamefont {J.}~\bibnamefont {Sun}}, \ and\ \bibinfo
  {author} {\bibfnamefont {G.~I.}\ \bibnamefont {Csonka}},\ }\bibfield  {title}
  {\enquote {\bibinfo {title} {{Some fundamental issues in ground-state density
  functional theory: A guide for the perplexed}},}\ }\href {\doibase
  10.1021/ct800531s} {\bibfield  {journal} {\bibinfo  {journal} {J. Chem.
  Theory Comput.}\ }\textbf {\bibinfo {volume} {5}},\ \bibinfo {pages} {902}
  (\bibinfo {year} {2009})}\BibitemShut {NoStop}%
\bibitem [{\citenamefont {Johnson}\ \emph {et~al.}(1994)\citenamefont
  {Johnson}, \citenamefont {Gonzales}, \citenamefont {Gill},\ and\
  \citenamefont {Pople}}]{Johnson1994_100}%
  \BibitemOpen
  \bibfield  {author} {\bibinfo {author} {\bibfnamefont {B.~G.}\ \bibnamefont
  {Johnson}}, \bibinfo {author} {\bibfnamefont {C.~A.}\ \bibnamefont
  {Gonzales}}, \bibinfo {author} {\bibfnamefont {P.~M.~W.}\ \bibnamefont
  {Gill}}, \ and\ \bibinfo {author} {\bibfnamefont {J.~A.}\ \bibnamefont
  {Pople}},\ }\bibfield  {title} {\enquote {\bibinfo {title} {{A density
  functional study of the simplest hydrogen abstraction reaction. Effect of
  self-interaction correction}},}\ }\href {\doibase
  10.1016/0009-2614(94)87024-1} {\bibfield  {journal} {\bibinfo  {journal}
  {Chem. Phys. Lett.}\ }\textbf {\bibinfo {volume} {221}},\ \bibinfo {pages}
  {100} (\bibinfo {year} {1994})}\BibitemShut {NoStop}%
\bibitem [{\citenamefont {Kraisler}\ and\ \citenamefont
  {Kronik}(2013)}]{Kraisler2013_126403}%
  \BibitemOpen
  \bibfield  {author} {\bibinfo {author} {\bibfnamefont {E.}~\bibnamefont
  {Kraisler}}\ and\ \bibinfo {author} {\bibfnamefont {L.}~\bibnamefont
  {Kronik}},\ }\bibfield  {title} {\enquote {\bibinfo {title} {{Piecewise
  linearity of approximate density functionals revisited: Implications for
  frontier orbital energies}},}\ }\href {\doibase
  10.1103/PhysRevLett.110.126403} {\bibfield  {journal} {\bibinfo  {journal}
  {Phys. Rev. Lett.}\ }\textbf {\bibinfo {volume} {110}},\ \bibinfo {pages}
  {126403} (\bibinfo {year} {2013})},\ \Eprint {http://arxiv.org/abs/1211.5950}
  {arXiv:1211.5950} \BibitemShut {NoStop}%
\bibitem [{\citenamefont {Perdew}\ and\ \citenamefont
  {Zunger}(1981)}]{Perdew1981_5048}%
  \BibitemOpen
  \bibfield  {author} {\bibinfo {author} {\bibfnamefont {J.~P.}\ \bibnamefont
  {Perdew}}\ and\ \bibinfo {author} {\bibfnamefont {A.}~\bibnamefont
  {Zunger}},\ }\bibfield  {title} {\enquote {\bibinfo {title} {Self-interaction
  correction to density-functional approximations for many-electron systems},}\
  }\href {\doibase 10.1103/PhysRevB.23.5048} {\bibfield  {journal} {\bibinfo
  {journal} {Phys. Rev. B}\ }\textbf {\bibinfo {volume} {23}},\ \bibinfo
  {pages} {5048} (\bibinfo {year} {1981})}\BibitemShut {NoStop}%
\bibitem [{\citenamefont {Pederson}\ and\ \citenamefont
  {Lin}(1988)}]{Pederson1988_1807}%
  \BibitemOpen
  \bibfield  {author} {\bibinfo {author} {\bibfnamefont {M.~R.}\ \bibnamefont
  {Pederson}}\ and\ \bibinfo {author} {\bibfnamefont {C.~C.}\ \bibnamefont
  {Lin}},\ }\bibfield  {title} {\enquote {\bibinfo {title} {{Localized and
  canonical atomic orbitals in self‐interaction corrected local density
  functional approximation}},}\ }\href {\doibase 10.1063/1.454104} {\bibfield
  {journal} {\bibinfo  {journal} {J. Chem. Phys.}\ }\textbf {\bibinfo {volume}
  {88}},\ \bibinfo {pages} {1807} (\bibinfo {year} {1988})}\BibitemShut
  {NoStop}%
\bibitem [{\citenamefont {Vydrov}\ and\ \citenamefont
  {Scuseria}(2005)}]{Vydrov2005_184107}%
  \BibitemOpen
  \bibfield  {author} {\bibinfo {author} {\bibfnamefont {O.~A.}\ \bibnamefont
  {Vydrov}}\ and\ \bibinfo {author} {\bibfnamefont {G.~E.}\ \bibnamefont
  {Scuseria}},\ }\bibfield  {title} {\enquote {\bibinfo {title} {{Ionization
  potentials and electron affinities in the Perdew–-Zunger self-interaction
  corrected density-functional theory}},}\ }\href {\doibase 10.1063/1.1897378}
  {\bibfield  {journal} {\bibinfo  {journal} {J. Chem. Phys.}\ }\textbf
  {\bibinfo {volume} {122}},\ \bibinfo {pages} {184107} (\bibinfo {year}
  {2005})}\BibitemShut {NoStop}%
\bibitem [{\citenamefont {Kl\"upfel}, \citenamefont {Kl\"upfel},\ and\
  \citenamefont {J\'onsson}(2011)}]{Klupfel2011_050501}%
  \BibitemOpen
  \bibfield  {author} {\bibinfo {author} {\bibfnamefont {S.}~\bibnamefont
  {Kl\"upfel}}, \bibinfo {author} {\bibfnamefont {P.}~\bibnamefont
  {Kl\"upfel}}, \ and\ \bibinfo {author} {\bibfnamefont {H.}~\bibnamefont
  {J\'onsson}},\ }\bibfield  {title} {\enquote {\bibinfo {title} {Importance of
  complex orbitals in calculating the self-interaction-corrected ground state
  of atoms},}\ }\href {\doibase 10.1103/PhysRevA.84.050501} {\bibfield
  {journal} {\bibinfo  {journal} {Phys. Rev. A}\ }\textbf {\bibinfo {volume}
  {84}},\ \bibinfo {pages} {050501} (\bibinfo {year} {2011})},\ \Eprint
  {http://arxiv.org/abs/1308.6063} {arXiv:1308.6063} \BibitemShut {NoStop}%
\bibitem [{\citenamefont {Gudmundsd{\'{o}}ttir}\ \emph
  {et~al.}(2013)\citenamefont {Gudmundsd{\'{o}}ttir}, \citenamefont {Zhang},
  \citenamefont {Weber},\ and\ \citenamefont
  {J{\'{o}}nsson}}]{Gudmundsdottir2013_194102}%
  \BibitemOpen
  \bibfield  {author} {\bibinfo {author} {\bibfnamefont {H.}~\bibnamefont
  {Gudmundsd{\'{o}}ttir}}, \bibinfo {author} {\bibfnamefont {Y.}~\bibnamefont
  {Zhang}}, \bibinfo {author} {\bibfnamefont {P.~M.}\ \bibnamefont {Weber}}, \
  and\ \bibinfo {author} {\bibfnamefont {H.}~\bibnamefont {J{\'{o}}nsson}},\
  }\bibfield  {title} {\enquote {\bibinfo {title} {{Self-interaction corrected
  density functional calculations of molecular Rydberg states}},}\ }\href
  {\doibase 10.1063/1.4829539} {\bibfield  {journal} {\bibinfo  {journal} {J.
  Chem. Phys.}\ }\textbf {\bibinfo {volume} {139}},\ \bibinfo {pages} {194102}
  (\bibinfo {year} {2013})}\BibitemShut {NoStop}%
\bibitem [{\citenamefont {Borghi}\ \emph {et~al.}(2014)\citenamefont {Borghi},
  \citenamefont {Ferretti}, \citenamefont {Nguyen}, \citenamefont {Dabo},\ and\
  \citenamefont {Marzari}}]{Borghi2014_075135}%
  \BibitemOpen
  \bibfield  {author} {\bibinfo {author} {\bibfnamefont {G.}~\bibnamefont
  {Borghi}}, \bibinfo {author} {\bibfnamefont {A.}~\bibnamefont {Ferretti}},
  \bibinfo {author} {\bibfnamefont {N.~L.}\ \bibnamefont {Nguyen}}, \bibinfo
  {author} {\bibfnamefont {I.}~\bibnamefont {Dabo}}, \ and\ \bibinfo {author}
  {\bibfnamefont {N.}~\bibnamefont {Marzari}},\ }\bibfield  {title} {\enquote
  {\bibinfo {title} {Koopmans-compliant functionals and their performance
  against reference molecular data},}\ }\href {\doibase
  10.1103/PhysRevB.90.075135} {\bibfield  {journal} {\bibinfo  {journal} {Phys.
  Rev. B}\ }\textbf {\bibinfo {volume} {90}},\ \bibinfo {pages} {075135}
  (\bibinfo {year} {2014})},\ \Eprint {http://arxiv.org/abs/1405.4635}
  {arXiv:1405.4635} \BibitemShut {NoStop}%
\bibitem [{\citenamefont {Gudmundsd{\'{o}}ttir}\ \emph
  {et~al.}(2014)\citenamefont {Gudmundsd{\'{o}}ttir}, \citenamefont {Zhang},
  \citenamefont {Weber},\ and\ \citenamefont
  {J{\'{o}}nsson}}]{Gudmundsdottir2014_234308}%
  \BibitemOpen
  \bibfield  {author} {\bibinfo {author} {\bibfnamefont {H.}~\bibnamefont
  {Gudmundsd{\'{o}}ttir}}, \bibinfo {author} {\bibfnamefont {Y.}~\bibnamefont
  {Zhang}}, \bibinfo {author} {\bibfnamefont {P.~M.}\ \bibnamefont {Weber}}, \
  and\ \bibinfo {author} {\bibfnamefont {H.}~\bibnamefont {J{\'{o}}nsson}},\
  }\bibfield  {title} {\enquote {\bibinfo {title} {{Self-interaction corrected
  density functional calculations of Rydberg states of molecular clusters:
  N,N-dimethylisopropylamine}},}\ }\href {\doibase 10.1063/1.4902383}
  {\bibfield  {journal} {\bibinfo  {journal} {J. Chem. Phys.}\ }\textbf
  {\bibinfo {volume} {141}},\ \bibinfo {pages} {234308} (\bibinfo {year}
  {2014})}\BibitemShut {NoStop}%
\bibitem [{\citenamefont {Perdew}\ \emph {et~al.}(2015)\citenamefont {Perdew},
  \citenamefont {Ruzsinszky}, \citenamefont {Sun},\ and\ \citenamefont
  {Pederson}}]{Perdew2015_1}%
  \BibitemOpen
  \bibfield  {author} {\bibinfo {author} {\bibfnamefont {J.~P.}\ \bibnamefont
  {Perdew}}, \bibinfo {author} {\bibfnamefont {A.}~\bibnamefont {Ruzsinszky}},
  \bibinfo {author} {\bibfnamefont {J.}~\bibnamefont {Sun}}, \ and\ \bibinfo
  {author} {\bibfnamefont {M.~R.}\ \bibnamefont {Pederson}},\ }\bibfield
  {title} {\enquote {\bibinfo {title} {{Paradox of self-interaction correction:
  How can anything so right be so wrong?}}}\ }\href {\doibase
  10.1016/bs.aamop.2015.06.004} {\bibfield  {journal} {\bibinfo  {journal}
  {{Adv. At., Mol., Opt. Phys.}}\ }\textbf {\bibinfo {volume} {64}},\ \bibinfo
  {pages} {1} (\bibinfo {year} {2015})}\BibitemShut {NoStop}%
\bibitem [{\citenamefont {Cheng}\ \emph {et~al.}(2016)\citenamefont {Cheng},
  \citenamefont {Zhang}, \citenamefont {J{\'o}nsson}, \citenamefont
  {J{\'o}nsson},\ and\ \citenamefont {Weber}}]{Cheng2016_11013}%
  \BibitemOpen
  \bibfield  {author} {\bibinfo {author} {\bibfnamefont {X.}~\bibnamefont
  {Cheng}}, \bibinfo {author} {\bibfnamefont {Y.}~\bibnamefont {Zhang}},
  \bibinfo {author} {\bibfnamefont {E.}~\bibnamefont {J{\'o}nsson}}, \bibinfo
  {author} {\bibfnamefont {H.}~\bibnamefont {J{\'o}nsson}}, \ and\ \bibinfo
  {author} {\bibfnamefont {P.~M.}\ \bibnamefont {Weber}},\ }\bibfield  {title}
  {\enquote {\bibinfo {title} {Charge localization in a diamine cation provides
  a test of energy functionals and self-interaction correction},}\ }\href
  {\doibase 10.1038/ncomms11013} {\bibfield  {journal} {\bibinfo  {journal}
  {Nat. Commun.}\ }\textbf {\bibinfo {volume} {7}},\ \bibinfo {pages} {11013}
  (\bibinfo {year} {2016})}\BibitemShut {NoStop}%
\bibitem [{\citenamefont {Zhang}, \citenamefont {Weber},\ and\ \citenamefont
  {J{\'{o}}nsson}(2016)}]{Zhang2016_2068}%
  \BibitemOpen
  \bibfield  {author} {\bibinfo {author} {\bibfnamefont {Y.}~\bibnamefont
  {Zhang}}, \bibinfo {author} {\bibfnamefont {P.~M.}\ \bibnamefont {Weber}}, \
  and\ \bibinfo {author} {\bibfnamefont {H.}~\bibnamefont {J{\'{o}}nsson}},\
  }\bibfield  {title} {\enquote {\bibinfo {title} {{Self-interaction corrected
  functional calculations of a dipole-bound molecular anion}},}\ }\href
  {\doibase 10.1021/acs.jpclett.6b00742} {\bibfield  {journal} {\bibinfo
  {journal} {J. Phys. Chem. Lett.}\ }\textbf {\bibinfo {volume} {7}},\ \bibinfo
  {pages} {2068} (\bibinfo {year} {2016})}\BibitemShut {NoStop}%
\bibitem [{\citenamefont {Schwalbe}\ \emph {et~al.}(2018)\citenamefont
  {Schwalbe}, \citenamefont {Hahn}, \citenamefont {Liebing}, \citenamefont
  {Trepte},\ and\ \citenamefont {Kortus}}]{Schwalbe2018_2463}%
  \BibitemOpen
  \bibfield  {author} {\bibinfo {author} {\bibfnamefont {S.}~\bibnamefont
  {Schwalbe}}, \bibinfo {author} {\bibfnamefont {T.}~\bibnamefont {Hahn}},
  \bibinfo {author} {\bibfnamefont {S.}~\bibnamefont {Liebing}}, \bibinfo
  {author} {\bibfnamefont {K.}~\bibnamefont {Trepte}}, \ and\ \bibinfo {author}
  {\bibfnamefont {J.}~\bibnamefont {Kortus}},\ }\bibfield  {title} {\enquote
  {\bibinfo {title} {{Fermi--L\"owdin orbital self-interaction corrected
  density functional theory: Ionization potentials and enthalpies of
  formation}},}\ }\href {\doibase 10.1002/jcc.25586} {\bibfield  {journal}
  {\bibinfo  {journal} {J. Comput. Chem.}\ }\textbf {\bibinfo {volume} {39}},\
  \bibinfo {pages} {2463} (\bibinfo {year} {2018})}\BibitemShut {NoStop}%
\bibitem [{\citenamefont {Borghi}\ \emph {et~al.}(2015)\citenamefont {Borghi},
  \citenamefont {Park}, \citenamefont {Nguyen}, \citenamefont {Ferretti},\ and\
  \citenamefont {Marzari}}]{Borghi2015_155112}%
  \BibitemOpen
  \bibfield  {author} {\bibinfo {author} {\bibfnamefont {G.}~\bibnamefont
  {Borghi}}, \bibinfo {author} {\bibfnamefont {C.-H.}\ \bibnamefont {Park}},
  \bibinfo {author} {\bibfnamefont {N.~L.}\ \bibnamefont {Nguyen}}, \bibinfo
  {author} {\bibfnamefont {A.}~\bibnamefont {Ferretti}}, \ and\ \bibinfo
  {author} {\bibfnamefont {N.}~\bibnamefont {Marzari}},\ }\bibfield  {title}
  {\enquote {\bibinfo {title} {Variational minimization of
  orbital-density-dependent functionals},}\ }\href {\doibase
  10.1103/PhysRevB.91.155112} {\bibfield  {journal} {\bibinfo  {journal} {Phys.
  Rev. B}\ }\textbf {\bibinfo {volume} {91}},\ \bibinfo {pages} {155112}
  (\bibinfo {year} {2015})}\BibitemShut {NoStop}%
\bibitem [{\citenamefont {Garza}, \citenamefont {Nichols},\ and\ \citenamefont
  {Dixon}(2000)}]{Garza2000_7880}%
  \BibitemOpen
  \bibfield  {author} {\bibinfo {author} {\bibfnamefont {J.}~\bibnamefont
  {Garza}}, \bibinfo {author} {\bibfnamefont {J.~A.}\ \bibnamefont {Nichols}},
  \ and\ \bibinfo {author} {\bibfnamefont {D.~A.}\ \bibnamefont {Dixon}},\
  }\bibfield  {title} {\enquote {\bibinfo {title} {{The optimized effective
  potential and the self-interaction correction in density functional theory:
  Application to molecules}},}\ }\href {\doibase 10.1063/1.481421} {\bibfield
  {journal} {\bibinfo  {journal} {J. Chem. Phys.}\ }\textbf {\bibinfo {volume}
  {112}},\ \bibinfo {pages} {7880} (\bibinfo {year} {2000})}\BibitemShut
  {NoStop}%
\bibitem [{\citenamefont {Garza}\ \emph {et~al.}(2001)\citenamefont {Garza},
  \citenamefont {Vargas}, \citenamefont {Nichols},\ and\ \citenamefont
  {Dixon}}]{Garza2001_639}%
  \BibitemOpen
  \bibfield  {author} {\bibinfo {author} {\bibfnamefont {J.}~\bibnamefont
  {Garza}}, \bibinfo {author} {\bibfnamefont {R.}~\bibnamefont {Vargas}},
  \bibinfo {author} {\bibfnamefont {J.~A.}\ \bibnamefont {Nichols}}, \ and\
  \bibinfo {author} {\bibfnamefont {D.~A.}\ \bibnamefont {Dixon}},\ }\bibfield
  {title} {\enquote {\bibinfo {title} {{Orbital energy analysis with respect to
  LDA and self-interaction corrected exchange-only potentials}},}\ }\href
  {\doibase 10.1063/1.1327269} {\bibfield  {journal} {\bibinfo  {journal} {J.
  Chem. Phys.}\ }\textbf {\bibinfo {volume} {114}},\ \bibinfo {pages} {639}
  (\bibinfo {year} {2001})}\BibitemShut {NoStop}%
\bibitem [{\citenamefont {Patchkovskii}, \citenamefont {Autschbach},\ and\
  \citenamefont {Ziegler}(2001)}]{Patchkovskii2001_26}%
  \BibitemOpen
  \bibfield  {author} {\bibinfo {author} {\bibfnamefont {S.}~\bibnamefont
  {Patchkovskii}}, \bibinfo {author} {\bibfnamefont {J.}~\bibnamefont
  {Autschbach}}, \ and\ \bibinfo {author} {\bibfnamefont {T.}~\bibnamefont
  {Ziegler}},\ }\bibfield  {title} {\enquote {\bibinfo {title} {Curing
  difficult cases in magnetic properties prediction with self-interaction
  corrected density functional theory},}\ }\href {\doibase 10.1063/1.1370527}
  {\bibfield  {journal} {\bibinfo  {journal} {J. Chem. Phys.}\ }\textbf
  {\bibinfo {volume} {115}},\ \bibinfo {pages} {26} (\bibinfo {year}
  {2001})}\BibitemShut {NoStop}%
\bibitem [{\citenamefont {Patchkovskii}\ and\ \citenamefont
  {Ziegler}(2002{\natexlab{a}})}]{Patchkovskii2002_7806}%
  \BibitemOpen
  \bibfield  {author} {\bibinfo {author} {\bibfnamefont {S.}~\bibnamefont
  {Patchkovskii}}\ and\ \bibinfo {author} {\bibfnamefont {T.}~\bibnamefont
  {Ziegler}},\ }\bibfield  {title} {\enquote {\bibinfo {title} {Improving
  “difficult” reaction barriers with self-interaction corrected density
  functional theory},}\ }\href {\doibase 10.1063/1.1468640} {\bibfield
  {journal} {\bibinfo  {journal} {J. Chem. Phys.}\ }\textbf {\bibinfo {volume}
  {116}},\ \bibinfo {pages} {7806} (\bibinfo {year}
  {2002}{\natexlab{a}})}\BibitemShut {NoStop}%
\bibitem [{\citenamefont {Patchkovskii}\ and\ \citenamefont
  {Ziegler}(2002{\natexlab{b}})}]{Patchkovskii2002_1088}%
  \BibitemOpen
  \bibfield  {author} {\bibinfo {author} {\bibfnamefont {S.}~\bibnamefont
  {Patchkovskii}}\ and\ \bibinfo {author} {\bibfnamefont {T.}~\bibnamefont
  {Ziegler}},\ }\bibfield  {title} {\enquote {\bibinfo {title} {{Phosphorus NMR
  chemical shifts with self-interaction free, gradient-corrected DFT}},}\
  }\href {\doibase 10.1021/jp014184v} {\bibfield  {journal} {\bibinfo
  {journal} {J. Phys. Chem. A}\ }\textbf {\bibinfo {volume} {106}},\ \bibinfo
  {pages} {1088} (\bibinfo {year} {2002}{\natexlab{b}})}\BibitemShut {NoStop}%
\bibitem [{\citenamefont {Vydrov}\ and\ \citenamefont
  {Scuseria}(2004)}]{Vydrov2004_8187}%
  \BibitemOpen
  \bibfield  {author} {\bibinfo {author} {\bibfnamefont {O.~A.}\ \bibnamefont
  {Vydrov}}\ and\ \bibinfo {author} {\bibfnamefont {G.~E.}\ \bibnamefont
  {Scuseria}},\ }\bibfield  {title} {\enquote {\bibinfo {title} {{Effect of the
  Perdew-–Zunger self-interaction correction on the thermochemical
  performance of approximate density functionals}},}\ }\href {\doibase
  10.1063/1.1794633} {\bibfield  {journal} {\bibinfo  {journal} {J. Chem.
  Phys.}\ }\textbf {\bibinfo {volume} {121}},\ \bibinfo {pages} {8187}
  (\bibinfo {year} {2004})}\BibitemShut {NoStop}%
\bibitem [{\citenamefont {Vydrov}\ \emph {et~al.}(2006)\citenamefont {Vydrov},
  \citenamefont {Scuseria}, \citenamefont {Perdew}, \citenamefont
  {Ruzsinszky},\ and\ \citenamefont {Csonka}}]{Vydrov2006_094108}%
  \BibitemOpen
  \bibfield  {author} {\bibinfo {author} {\bibfnamefont {O.~A.}\ \bibnamefont
  {Vydrov}}, \bibinfo {author} {\bibfnamefont {G.~E.}\ \bibnamefont
  {Scuseria}}, \bibinfo {author} {\bibfnamefont {J.~P.}\ \bibnamefont
  {Perdew}}, \bibinfo {author} {\bibfnamefont {A.}~\bibnamefont {Ruzsinszky}},
  \ and\ \bibinfo {author} {\bibfnamefont {G.~I.}\ \bibnamefont {Csonka}},\
  }\bibfield  {title} {\enquote {\bibinfo {title} {{Scaling down the
  Perdew--Zunger self-interaction correction in many-electron regions}},}\
  }\href {\doibase 10.1063/1.2176608} {\bibfield  {journal} {\bibinfo
  {journal} {J. Chem. Phys.}\ }\textbf {\bibinfo {volume} {124}},\ \bibinfo
  {pages} {094108} (\bibinfo {year} {2006})}\BibitemShut {NoStop}%
\bibitem [{\citenamefont {Pemmaraju}\ \emph {et~al.}(2007)\citenamefont
  {Pemmaraju}, \citenamefont {Archer}, \citenamefont {S\'anchez-Portal},\ and\
  \citenamefont {Sanvito}}]{Pemmaraju2007_045101}%
  \BibitemOpen
  \bibfield  {author} {\bibinfo {author} {\bibfnamefont {C.~D.}\ \bibnamefont
  {Pemmaraju}}, \bibinfo {author} {\bibfnamefont {T.}~\bibnamefont {Archer}},
  \bibinfo {author} {\bibfnamefont {D.}~\bibnamefont {S\'anchez-Portal}}, \
  and\ \bibinfo {author} {\bibfnamefont {S.}~\bibnamefont {Sanvito}},\
  }\bibfield  {title} {\enquote {\bibinfo {title} {Atomic-orbital-based
  approximate self-interaction correction scheme for molecules and solids},}\
  }\href {\doibase 10.1103/PhysRevB.75.045101} {\bibfield  {journal} {\bibinfo
  {journal} {Phys. Rev. B}\ }\textbf {\bibinfo {volume} {75}},\ \bibinfo
  {pages} {045101} (\bibinfo {year} {2007})}\BibitemShut {NoStop}%
\bibitem [{\citenamefont {Kl{\"u}pfel}, \citenamefont {Kl{\"u}pfel},\ and\
  \citenamefont {J{\'o}nsson}(2012)}]{Klupfel2012_124102}%
  \BibitemOpen
  \bibfield  {author} {\bibinfo {author} {\bibfnamefont {S.}~\bibnamefont
  {Kl{\"u}pfel}}, \bibinfo {author} {\bibfnamefont {P.}~\bibnamefont
  {Kl{\"u}pfel}}, \ and\ \bibinfo {author} {\bibfnamefont {H.}~\bibnamefont
  {J{\'o}nsson}},\ }\bibfield  {title} {\enquote {\bibinfo {title} {{The effect
  of the Perdew--Zunger self-interaction correction to density functionals on
  the energetics of small molecules}},}\ }\href {\doibase 10.1063/1.4752229}
  {\bibfield  {journal} {\bibinfo  {journal} {J. Chem. Phys.}\ }\textbf
  {\bibinfo {volume} {137}},\ \bibinfo {pages} {124102} (\bibinfo {year}
  {2012})}\BibitemShut {NoStop}%
\bibitem [{\citenamefont {Valdes}\ \emph {et~al.}(2012)\citenamefont {Valdes},
  \citenamefont {Brillet}, \citenamefont {Gratzel}, \citenamefont
  {Gudmundsd\'ottir}, \citenamefont {Hansen}, \citenamefont {J{\'o}nsson},
  \citenamefont {Kl{\"u}pfel}, \citenamefont {Kroes}, \citenamefont
  {Le~Formal}, \citenamefont {Man}, \citenamefont {Martins}, \citenamefont
  {Norskov}, \citenamefont {Rossmeisl}, \citenamefont {Sivula}, \citenamefont
  {Vojvodic},\ and\ \citenamefont {Zach}}]{Valdes2012_49}%
  \BibitemOpen
  \bibfield  {author} {\bibinfo {author} {\bibfnamefont {A.}~\bibnamefont
  {Valdes}}, \bibinfo {author} {\bibfnamefont {J.}~\bibnamefont {Brillet}},
  \bibinfo {author} {\bibfnamefont {M.}~\bibnamefont {Gratzel}}, \bibinfo
  {author} {\bibfnamefont {H.}~\bibnamefont {Gudmundsd\'ottir}}, \bibinfo
  {author} {\bibfnamefont {H.~A.}\ \bibnamefont {Hansen}}, \bibinfo {author}
  {\bibfnamefont {H.}~\bibnamefont {J{\'o}nsson}}, \bibinfo {author}
  {\bibfnamefont {P.}~\bibnamefont {Kl{\"u}pfel}}, \bibinfo {author}
  {\bibfnamefont {G.-J.}\ \bibnamefont {Kroes}}, \bibinfo {author}
  {\bibfnamefont {F.}~\bibnamefont {Le~Formal}}, \bibinfo {author}
  {\bibfnamefont {I.~C.}\ \bibnamefont {Man}}, \bibinfo {author} {\bibfnamefont
  {R.~S.}\ \bibnamefont {Martins}}, \bibinfo {author} {\bibfnamefont {J.~K.}\
  \bibnamefont {Norskov}}, \bibinfo {author} {\bibfnamefont {J.}~\bibnamefont
  {Rossmeisl}}, \bibinfo {author} {\bibfnamefont {K.}~\bibnamefont {Sivula}},
  \bibinfo {author} {\bibfnamefont {A.}~\bibnamefont {Vojvodic}}, \ and\
  \bibinfo {author} {\bibfnamefont {M.}~\bibnamefont {Zach}},\ }\bibfield
  {title} {\enquote {\bibinfo {title} {{Solar hydrogen production with
  semiconductor metal oxides: New directions in experiment and theory}},}\
  }\href {\doibase 10.1039/c1cp23212f} {\bibfield  {journal} {\bibinfo
  {journal} {Phys. Chem. Chem. Phys.}\ }\textbf {\bibinfo {volume} {14}},\
  \bibinfo {pages} {49} (\bibinfo {year} {2012})}\BibitemShut {NoStop}%
\bibitem [{\citenamefont {Gudmundsd\'ottir}, \citenamefont {J\'onsson},\ and\
  \citenamefont {J\'onsson}(2015)}]{Gudmundsdottir2015_083006}%
  \BibitemOpen
  \bibfield  {author} {\bibinfo {author} {\bibfnamefont {H.}~\bibnamefont
  {Gudmundsd\'ottir}}, \bibinfo {author} {\bibfnamefont {E.~{\"O}.}\
  \bibnamefont {J\'onsson}}, \ and\ \bibinfo {author} {\bibfnamefont
  {H.}~\bibnamefont {J\'onsson}},\ }\bibfield  {title} {\enquote {\bibinfo
  {title} {{Calculations of Al dopant in {$\alpha$}-quartz using a variational
  implementation of the Perdew–Zunger self-interaction correction}},}\ }\href
  {\doibase 10.1088/1367-2630/17/8/083006} {\bibfield  {journal} {\bibinfo
  {journal} {New J. Phys.}\ }\textbf {\bibinfo {volume} {17}},\ \bibinfo
  {pages} {083006} (\bibinfo {year} {2015})}\BibitemShut {NoStop}%
\bibitem [{\citenamefont {Lehtola}, \citenamefont {Head-Gordon},\ and\
  \citenamefont {J\'onsson}(2016)}]{Lehtola2016_3195}%
  \BibitemOpen
  \bibfield  {author} {\bibinfo {author} {\bibfnamefont {S.}~\bibnamefont
  {Lehtola}}, \bibinfo {author} {\bibfnamefont {M.}~\bibnamefont
  {Head-Gordon}}, \ and\ \bibinfo {author} {\bibfnamefont {H.}~\bibnamefont
  {J\'onsson}},\ }\bibfield  {title} {\enquote {\bibinfo {title} {{Complex
  orbitals, multiple local minima, and symmetry breaking in Perdew–-Zunger
  self-interaction corrected density functional theory calculations}},}\ }\href
  {\doibase 10.1021/acs.jctc.6b00347} {\bibfield  {journal} {\bibinfo
  {journal} {J. Chem. Theory Comput.}\ }\textbf {\bibinfo {volume} {12}},\
  \bibinfo {pages} {3195} (\bibinfo {year} {2016})}\BibitemShut {NoStop}%
\bibitem [{\citenamefont {Lehtola}\ \emph {et~al.}(2012)\citenamefont
  {Lehtola}, \citenamefont {Hakala}, \citenamefont {Sakko},\ and\ \citenamefont
  {H\"am\"al\"ainen}}]{Lehtola2012_1572}%
  \BibitemOpen
  \bibfield  {author} {\bibinfo {author} {\bibfnamefont {J.}~\bibnamefont
  {Lehtola}}, \bibinfo {author} {\bibfnamefont {M.}~\bibnamefont {Hakala}},
  \bibinfo {author} {\bibfnamefont {A.}~\bibnamefont {Sakko}}, \ and\ \bibinfo
  {author} {\bibfnamefont {K.}~\bibnamefont {H\"am\"al\"ainen}},\ }\bibfield
  {title} {\enquote {\bibinfo {title} {{ERKALE - A flexible program package for
  X-ray properties of atoms and molecules}},}\ }\href {\doibase
  10.1002/jcc.22987} {\bibfield  {journal} {\bibinfo  {journal} {J. Comput.
  Chem.}\ }\textbf {\bibinfo {volume} {33}},\ \bibinfo {pages} {1572} (\bibinfo
  {year} {2012})}\BibitemShut {NoStop}%
\bibitem [{\citenamefont {{S. Lehtola. ERKALE -- HF/DFT from
  Hel}}(2016)}]{Hel2016_1}%
  \BibitemOpen
  \bibfield  {author} {\bibinfo {author} {\bibnamefont {{S. Lehtola. ERKALE --
  HF/DFT from Hel}}},\ }\href@noop {} {\enquote {\bibinfo {title}
  {\url{http://github.com/susilehtola/erkale}},}\ } (\bibinfo {year}
  {2016})\BibitemShut {NoStop}%
\bibitem [{\citenamefont {Hahn}\ \emph {et~al.}(2015)\citenamefont {Hahn},
  \citenamefont {Liebing}, \citenamefont {Kortus},\ and\ \citenamefont
  {Pederson}}]{Hahn2015_224104}%
  \BibitemOpen
  \bibfield  {author} {\bibinfo {author} {\bibfnamefont {T.}~\bibnamefont
  {Hahn}}, \bibinfo {author} {\bibfnamefont {S.}~\bibnamefont {Liebing}},
  \bibinfo {author} {\bibfnamefont {J.}~\bibnamefont {Kortus}}, \ and\ \bibinfo
  {author} {\bibfnamefont {M.~R.}\ \bibnamefont {Pederson}},\ }\bibfield
  {title} {\enquote {\bibinfo {title} {{Fermi orbital self-interaction
  corrected electronic structure of molecules beyond local density
  approximation}},}\ }\href {\doibase 10.1063/1.4936777} {\bibfield  {journal}
  {\bibinfo  {journal} {J. Chem. Phys.}\ }\textbf {\bibinfo {volume} {143}},\
  \bibinfo {pages} {224104} (\bibinfo {year} {2015})},\ \Eprint
  {http://arxiv.org/abs/1508.00745} {arXiv:1508.00745} \BibitemShut {NoStop}%
\bibitem [{\citenamefont {Pederson}\ \emph {et~al.}(2016)\citenamefont
  {Pederson}, \citenamefont {Baruah}, \citenamefont {Kao},\ and\ \citenamefont
  {Basurto}}]{Pederson2016_164117}%
  \BibitemOpen
  \bibfield  {author} {\bibinfo {author} {\bibfnamefont {M.~R.}\ \bibnamefont
  {Pederson}}, \bibinfo {author} {\bibfnamefont {T.}~\bibnamefont {Baruah}},
  \bibinfo {author} {\bibfnamefont {D.-y.}\ \bibnamefont {Kao}}, \ and\
  \bibinfo {author} {\bibfnamefont {L.}~\bibnamefont {Basurto}},\ }\bibfield
  {title} {\enquote {\bibinfo {title} {{Self-interaction corrections applied to
  Mg-porphyrin, C$_{\textrm{60}}$, and pentacene molecules}},}\ }\href
  {\doibase 10.1063/1.4947042} {\bibfield  {journal} {\bibinfo  {journal} {{J.
  Chem. Phys.}}\ }\textbf {\bibinfo {volume} {144}},\ \bibinfo {pages} {164117}
  (\bibinfo {year} {2016})}\BibitemShut {NoStop}%
\bibitem [{\citenamefont {Hahn}\ \emph {et~al.}(2017)\citenamefont {Hahn},
  \citenamefont {Schwalbe}, \citenamefont {Kortus},\ and\ \citenamefont
  {Pederson}}]{Hahn2017_5823}%
  \BibitemOpen
  \bibfield  {author} {\bibinfo {author} {\bibfnamefont {T.}~\bibnamefont
  {Hahn}}, \bibinfo {author} {\bibfnamefont {S.}~\bibnamefont {Schwalbe}},
  \bibinfo {author} {\bibfnamefont {J.}~\bibnamefont {Kortus}}, \ and\ \bibinfo
  {author} {\bibfnamefont {M.~R.}\ \bibnamefont {Pederson}},\ }\bibfield
  {title} {\enquote {\bibinfo {title} {{Symmetry breaking within
  Fermi--L\"owdin orbital self-interaction corrected density functional
  theory}},}\ }\href {\doibase 10.1021/acs.jctc.7b00604} {\bibfield  {journal}
  {\bibinfo  {journal} {J. Chem. Theory Comput.}\ }\textbf {\bibinfo {volume}
  {13}},\ \bibinfo {pages} {5823} (\bibinfo {year} {2017})}\BibitemShut
  {NoStop}%
\bibitem [{\citenamefont {Kao}\ \emph {et~al.}(2017)\citenamefont {Kao},
  \citenamefont {Withanage}, \citenamefont {Hahn}, \citenamefont {Batool},
  \citenamefont {Kortus},\ and\ \citenamefont {Jackson}}]{Kao2017_164107}%
  \BibitemOpen
  \bibfield  {author} {\bibinfo {author} {\bibfnamefont {D.-y.}\ \bibnamefont
  {Kao}}, \bibinfo {author} {\bibfnamefont {K.}~\bibnamefont {Withanage}},
  \bibinfo {author} {\bibfnamefont {T.}~\bibnamefont {Hahn}}, \bibinfo {author}
  {\bibfnamefont {J.}~\bibnamefont {Batool}}, \bibinfo {author} {\bibfnamefont
  {J.}~\bibnamefont {Kortus}}, \ and\ \bibinfo {author} {\bibfnamefont
  {K.}~\bibnamefont {Jackson}},\ }\bibfield  {title} {\enquote {\bibinfo
  {title} {{Self-consistent self-interaction corrected density functional
  theory calculations for atoms using Fermi--L\"owdin orbitals: Optimized
  Fermi-orbital descriptors for Li-Kr}},}\ }\href {\doibase 10.1063/1.4996498}
  {\bibfield  {journal} {\bibinfo  {journal} {J. Chem. Phys.}\ }\textbf
  {\bibinfo {volume} {147}},\ \bibinfo {pages} {164107} (\bibinfo {year}
  {2017})}\BibitemShut {NoStop}%
\bibitem [{\citenamefont {Sharkas}\ \emph {et~al.}(2018)\citenamefont
  {Sharkas}, \citenamefont {Li}, \citenamefont {Trepte}, \citenamefont
  {Withanage}, \citenamefont {Joshi}, \citenamefont {Zope}, \citenamefont
  {Baruah}, \citenamefont {Johnson}, \citenamefont {Jackson},\ and\
  \citenamefont {Peralta}}]{Sharkas2018_9307}%
  \BibitemOpen
  \bibfield  {author} {\bibinfo {author} {\bibfnamefont {K.}~\bibnamefont
  {Sharkas}}, \bibinfo {author} {\bibfnamefont {L.}~\bibnamefont {Li}},
  \bibinfo {author} {\bibfnamefont {K.}~\bibnamefont {Trepte}}, \bibinfo
  {author} {\bibfnamefont {K.~P.~K.}\ \bibnamefont {Withanage}}, \bibinfo
  {author} {\bibfnamefont {R.~P.}\ \bibnamefont {Joshi}}, \bibinfo {author}
  {\bibfnamefont {R.~R.}\ \bibnamefont {Zope}}, \bibinfo {author}
  {\bibfnamefont {T.}~\bibnamefont {Baruah}}, \bibinfo {author} {\bibfnamefont
  {J.~K.}\ \bibnamefont {Johnson}}, \bibinfo {author} {\bibfnamefont {K.~A.}\
  \bibnamefont {Jackson}}, \ and\ \bibinfo {author} {\bibfnamefont {J.~E.}\
  \bibnamefont {Peralta}},\ }\bibfield  {title} {\enquote {\bibinfo {title}
  {{Shrinking self-interaction errors with the Fermi--L\"owdin orbital
  self-interaction-corrected density functional approximation}},}\ }\href
  {\doibase 10.1021/acs.jpca.8b09940} {\bibfield  {journal} {\bibinfo
  {journal} {J. Phys. Chem. A}\ }\textbf {\bibinfo {volume} {122}},\ \bibinfo
  {pages} {9307} (\bibinfo {year} {2018})}\BibitemShut {NoStop}%
\bibitem [{\citenamefont {Joshi}\ \emph {et~al.}(2018)\citenamefont {Joshi},
  \citenamefont {Trepte}, \citenamefont {Withanage}, \citenamefont {Sharkas},
  \citenamefont {Yamamoto}, \citenamefont {Basurto}, \citenamefont {Zope},
  \citenamefont {Baruah}, \citenamefont {Jackson},\ and\ \citenamefont
  {Peralta}}]{Joshi2018_164101}%
  \BibitemOpen
  \bibfield  {author} {\bibinfo {author} {\bibfnamefont {R.~P.}\ \bibnamefont
  {Joshi}}, \bibinfo {author} {\bibfnamefont {K.}~\bibnamefont {Trepte}},
  \bibinfo {author} {\bibfnamefont {K.~P.~K.}\ \bibnamefont {Withanage}},
  \bibinfo {author} {\bibfnamefont {K.}~\bibnamefont {Sharkas}}, \bibinfo
  {author} {\bibfnamefont {Y.}~\bibnamefont {Yamamoto}}, \bibinfo {author}
  {\bibfnamefont {L.}~\bibnamefont {Basurto}}, \bibinfo {author} {\bibfnamefont
  {R.~R.}\ \bibnamefont {Zope}}, \bibinfo {author} {\bibfnamefont
  {T.}~\bibnamefont {Baruah}}, \bibinfo {author} {\bibfnamefont {K.~A.}\
  \bibnamefont {Jackson}}, \ and\ \bibinfo {author} {\bibfnamefont {J.~E.}\
  \bibnamefont {Peralta}},\ }\bibfield  {title} {\enquote {\bibinfo {title}
  {{Fermi--L\"owdin orbital self-interaction correction to magnetic exchange
  couplings}},}\ }\href {\doibase 10.1063/1.5050809} {\bibfield  {journal}
  {\bibinfo  {journal} {J. Chem. Phys.}\ }\textbf {\bibinfo {volume} {149}},\
  \bibinfo {pages} {164101} (\bibinfo {year} {2018})}\BibitemShut {NoStop}%
\bibitem [{\citenamefont {Withanage}\ \emph {et~al.}(2018)\citenamefont
  {Withanage}, \citenamefont {Trepte}, \citenamefont {Peralta}, \citenamefont
  {Baruah}, \citenamefont {Zope},\ and\ \citenamefont
  {Jackson}}]{Withanage2018_4122}%
  \BibitemOpen
  \bibfield  {author} {\bibinfo {author} {\bibfnamefont {K.~P.~K.}\
  \bibnamefont {Withanage}}, \bibinfo {author} {\bibfnamefont {K.}~\bibnamefont
  {Trepte}}, \bibinfo {author} {\bibfnamefont {J.~E.}\ \bibnamefont {Peralta}},
  \bibinfo {author} {\bibfnamefont {T.}~\bibnamefont {Baruah}}, \bibinfo
  {author} {\bibfnamefont {R.}~\bibnamefont {Zope}}, \ and\ \bibinfo {author}
  {\bibfnamefont {K.~A.}\ \bibnamefont {Jackson}},\ }\bibfield  {title}
  {\enquote {\bibinfo {title} {{On the question of the total energy in the
  Fermi--L\"owdin Orbital self-interaction correction method}},}\ }\href
  {\doibase 10.1021/acs.jctc.8b00344} {\bibfield  {journal} {\bibinfo
  {journal} {J. Chem. Theory Comput.}\ }\textbf {\bibinfo {volume} {14}},\
  \bibinfo {pages} {4122} (\bibinfo {year} {2018})}\BibitemShut {NoStop}%
\bibitem [{\citenamefont {Johnson}\ \emph {et~al.}(2019)\citenamefont
  {Johnson}, \citenamefont {Withanage}, \citenamefont {Sharkas}, \citenamefont
  {Yamamoto}, \citenamefont {Baruah}, \citenamefont {Zope}, \citenamefont
  {Peralta},\ and\ \citenamefont {Jackson}}]{Johnson2019_174106}%
  \BibitemOpen
  \bibfield  {author} {\bibinfo {author} {\bibfnamefont {A.~I.}\ \bibnamefont
  {Johnson}}, \bibinfo {author} {\bibfnamefont {K.~P.~K.}\ \bibnamefont
  {Withanage}}, \bibinfo {author} {\bibfnamefont {K.}~\bibnamefont {Sharkas}},
  \bibinfo {author} {\bibfnamefont {Y.}~\bibnamefont {Yamamoto}}, \bibinfo
  {author} {\bibfnamefont {T.}~\bibnamefont {Baruah}}, \bibinfo {author}
  {\bibfnamefont {R.~R.}\ \bibnamefont {Zope}}, \bibinfo {author}
  {\bibfnamefont {J.~E.}\ \bibnamefont {Peralta}}, \ and\ \bibinfo {author}
  {\bibfnamefont {K.~A.}\ \bibnamefont {Jackson}},\ }\bibfield  {title}
  {\enquote {\bibinfo {title} {The effect of self-interaction error on
  electrostatic dipoles calculated using density functional theory},}\ }\href
  {\doibase 10.1063/1.5125205} {\bibfield  {journal} {\bibinfo  {journal} {J.
  Chem. Phys.}\ }\textbf {\bibinfo {volume} {151}},\ \bibinfo {pages} {174106}
  (\bibinfo {year} {2019})}\BibitemShut {NoStop}%
\bibitem [{\citenamefont {Jackson}\ \emph {et~al.}(2019)\citenamefont
  {Jackson}, \citenamefont {Peralta}, \citenamefont {Joshi}, \citenamefont
  {Withanage}, \citenamefont {Trepte}, \citenamefont {Sharkas},\ and\
  \citenamefont {Johnson}}]{Jackson2019_012002}%
  \BibitemOpen
  \bibfield  {author} {\bibinfo {author} {\bibfnamefont {K.~A.}\ \bibnamefont
  {Jackson}}, \bibinfo {author} {\bibfnamefont {J.~E.}\ \bibnamefont
  {Peralta}}, \bibinfo {author} {\bibfnamefont {R.~P.}\ \bibnamefont {Joshi}},
  \bibinfo {author} {\bibfnamefont {K.~P.}\ \bibnamefont {Withanage}}, \bibinfo
  {author} {\bibfnamefont {K.}~\bibnamefont {Trepte}}, \bibinfo {author}
  {\bibfnamefont {K.}~\bibnamefont {Sharkas}}, \ and\ \bibinfo {author}
  {\bibfnamefont {A.~I.}\ \bibnamefont {Johnson}},\ }\bibfield  {title}
  {\enquote {\bibinfo {title} {{Towards efficient density functional theory
  calculations without self-interaction: The Fermi--L\"owdin orbital
  self-interaction correction}},}\ }\href {\doibase
  10.1088/1742-6596/1290/1/012002} {\bibfield  {journal} {\bibinfo  {journal}
  {{J. Phys.: Conf. Ser.}}\ }\textbf {\bibinfo {volume} {1290}},\ \bibinfo
  {pages} {012002} (\bibinfo {year} {2019})}\BibitemShut {NoStop}%
\bibitem [{\citenamefont {Zope}\ \emph {et~al.}(2019)\citenamefont {Zope},
  \citenamefont {Yamamoto}, \citenamefont {Diaz}, \citenamefont {Baruah},
  \citenamefont {Peralta}, \citenamefont {Jackson}, \citenamefont {Santra},\
  and\ \citenamefont {Perdew}}]{Zope2019_214108}%
  \BibitemOpen
  \bibfield  {author} {\bibinfo {author} {\bibfnamefont {R.~R.}\ \bibnamefont
  {Zope}}, \bibinfo {author} {\bibfnamefont {Y.}~\bibnamefont {Yamamoto}},
  \bibinfo {author} {\bibfnamefont {C.~M.}\ \bibnamefont {Diaz}}, \bibinfo
  {author} {\bibfnamefont {T.}~\bibnamefont {Baruah}}, \bibinfo {author}
  {\bibfnamefont {J.~E.}\ \bibnamefont {Peralta}}, \bibinfo {author}
  {\bibfnamefont {K.~A.}\ \bibnamefont {Jackson}}, \bibinfo {author}
  {\bibfnamefont {B.}~\bibnamefont {Santra}}, \ and\ \bibinfo {author}
  {\bibfnamefont {J.~P.}\ \bibnamefont {Perdew}},\ }\bibfield  {title}
  {\enquote {\bibinfo {title} {{A step in the direction of resolving the
  paradox of Perdew--Zunger self-interaction correction}},}\ }\href {\doibase
  10.1063/1.5129533} {\bibfield  {journal} {\bibinfo  {journal} {J. Chem.
  Phys.}\ }\textbf {\bibinfo {volume} {151}},\ \bibinfo {pages} {214108}
  (\bibinfo {year} {2019})},\ \Eprint {http://arxiv.org/abs/1911.08659}
  {arXiv:1911.08659} \BibitemShut {NoStop}%
\bibitem [{\citenamefont {Withanage}\ \emph {et~al.}(2019)\citenamefont
  {Withanage}, \citenamefont {Akter}, \citenamefont {Shahi}, \citenamefont
  {Joshi}, \citenamefont {Diaz}, \citenamefont {Yamamoto}, \citenamefont
  {Zope}, \citenamefont {Baruah}, \citenamefont {Perdew}, \citenamefont
  {Peralta},\ and\ \citenamefont {Jackson}}]{Withanage2019_012505}%
  \BibitemOpen
  \bibfield  {author} {\bibinfo {author} {\bibfnamefont {K.~P.~K.}\
  \bibnamefont {Withanage}}, \bibinfo {author} {\bibfnamefont {S.}~\bibnamefont
  {Akter}}, \bibinfo {author} {\bibfnamefont {C.}~\bibnamefont {Shahi}},
  \bibinfo {author} {\bibfnamefont {R.~P.}\ \bibnamefont {Joshi}}, \bibinfo
  {author} {\bibfnamefont {C.}~\bibnamefont {Diaz}}, \bibinfo {author}
  {\bibfnamefont {Y.}~\bibnamefont {Yamamoto}}, \bibinfo {author}
  {\bibfnamefont {R.}~\bibnamefont {Zope}}, \bibinfo {author} {\bibfnamefont
  {T.}~\bibnamefont {Baruah}}, \bibinfo {author} {\bibfnamefont {J.~P.}\
  \bibnamefont {Perdew}}, \bibinfo {author} {\bibfnamefont {J.~E.}\
  \bibnamefont {Peralta}}, \ and\ \bibinfo {author} {\bibfnamefont {K.~A.}\
  \bibnamefont {Jackson}},\ }\bibfield  {title} {\enquote {\bibinfo {title}
  {{Self-interaction-free electric dipole polarizabilities for atoms and their
  ions using the Fermi--L\"owdin self-interaction correction}},}\ }\href
  {\doibase 10.1103/PhysRevA.100.012505} {\bibfield  {journal} {\bibinfo
  {journal} {Phys. Rev. A}\ }\textbf {\bibinfo {volume} {100}},\ \bibinfo
  {pages} {012505} (\bibinfo {year} {2019})}\BibitemShut {NoStop}%
\bibitem [{\citenamefont {Santra}\ and\ \citenamefont
  {Perdew}(2019)}]{Santra2019_174106}%
  \BibitemOpen
  \bibfield  {author} {\bibinfo {author} {\bibfnamefont {B.}~\bibnamefont
  {Santra}}\ and\ \bibinfo {author} {\bibfnamefont {J.~P.}\ \bibnamefont
  {Perdew}},\ }\bibfield  {title} {\enquote {\bibinfo {title} {{Perdew--Zunger
  self-interaction correction: How wrong for uniform densities and large-Z
  atoms?}}}\ }\href {\doibase 10.1063/1.5090534} {\bibfield  {journal}
  {\bibinfo  {journal} {J. Chem. Phys.}\ }\textbf {\bibinfo {volume} {150}},\
  \bibinfo {pages} {174106} (\bibinfo {year} {2019})},\ \Eprint
  {http://arxiv.org/abs/1902.00117} {arXiv:1902.00117} \BibitemShut {NoStop}%
\bibitem [{\citenamefont {Trepte}\ \emph {et~al.}(2019)\citenamefont {Trepte},
  \citenamefont {Schwalbe}, \citenamefont {Hahn}, \citenamefont {Kortus},
  \citenamefont {Kao}, \citenamefont {Yamamoto}, \citenamefont {Baruah},
  \citenamefont {Zope}, \citenamefont {Withanage}, \citenamefont {Peralta},\
  and\ \citenamefont {Jackson}}]{Trepte2019_820}%
  \BibitemOpen
  \bibfield  {author} {\bibinfo {author} {\bibfnamefont {K.}~\bibnamefont
  {Trepte}}, \bibinfo {author} {\bibfnamefont {S.}~\bibnamefont {Schwalbe}},
  \bibinfo {author} {\bibfnamefont {T.}~\bibnamefont {Hahn}}, \bibinfo {author}
  {\bibfnamefont {J.}~\bibnamefont {Kortus}}, \bibinfo {author} {\bibfnamefont
  {D.-y.}\ \bibnamefont {Kao}}, \bibinfo {author} {\bibfnamefont
  {Y.}~\bibnamefont {Yamamoto}}, \bibinfo {author} {\bibfnamefont
  {T.}~\bibnamefont {Baruah}}, \bibinfo {author} {\bibfnamefont {R.~R.}\
  \bibnamefont {Zope}}, \bibinfo {author} {\bibfnamefont {K.~P.~K.}\
  \bibnamefont {Withanage}}, \bibinfo {author} {\bibfnamefont {J.~E.}\
  \bibnamefont {Peralta}}, \ and\ \bibinfo {author} {\bibfnamefont {K.~A.}\
  \bibnamefont {Jackson}},\ }\bibfield  {title} {\enquote {\bibinfo {title}
  {{Analytic atomic gradients in the Fermi--L\"owdin orbital self-interaction
  correction}},}\ }\href {\doibase 10.1002/jcc.25767} {\bibfield  {journal}
  {\bibinfo  {journal} {J. Comput. Chem.}\ }\textbf {\bibinfo {volume} {40}},\
  \bibinfo {pages} {820} (\bibinfo {year} {2019})}\BibitemShut {NoStop}%
\bibitem [{\citenamefont {Schwalbe}\ \emph {et~al.}(2019)\citenamefont
  {Schwalbe}, \citenamefont {Trepte}, \citenamefont {Fiedler}, \citenamefont
  {Johnson}, \citenamefont {Kraus}, \citenamefont {Hahn}, \citenamefont
  {Peralta}, \citenamefont {Jackson},\ and\ \citenamefont
  {Kortus}}]{Schwalbe2019_2843}%
  \BibitemOpen
  \bibfield  {author} {\bibinfo {author} {\bibfnamefont {S.}~\bibnamefont
  {Schwalbe}}, \bibinfo {author} {\bibfnamefont {K.}~\bibnamefont {Trepte}},
  \bibinfo {author} {\bibfnamefont {L.}~\bibnamefont {Fiedler}}, \bibinfo
  {author} {\bibfnamefont {A.~I.}\ \bibnamefont {Johnson}}, \bibinfo {author}
  {\bibfnamefont {J.}~\bibnamefont {Kraus}}, \bibinfo {author} {\bibfnamefont
  {T.}~\bibnamefont {Hahn}}, \bibinfo {author} {\bibfnamefont {J.~E.}\
  \bibnamefont {Peralta}}, \bibinfo {author} {\bibfnamefont {K.~A.}\
  \bibnamefont {Jackson}}, \ and\ \bibinfo {author} {\bibfnamefont
  {J.}~\bibnamefont {Kortus}},\ }\bibfield  {title} {\enquote {\bibinfo {title}
  {{Interpretation and automatic generation of Fermi-orbital descriptors}},}\
  }\href {\doibase 10.1002/jcc.26062} {\bibfield  {journal} {\bibinfo
  {journal} {J. Comput. Chem.}\ }\textbf {\bibinfo {volume} {40}},\ \bibinfo
  {pages} {2843} (\bibinfo {year} {2019})}\BibitemShut {NoStop}%
\bibitem [{\citenamefont {Yamamoto}\ \emph {et~al.}(2019)\citenamefont
  {Yamamoto}, \citenamefont {Diaz}, \citenamefont {Basurto}, \citenamefont
  {Jackson}, \citenamefont {Baruah},\ and\ \citenamefont
  {Zope}}]{Yamamoto2019_154105}%
  \BibitemOpen
  \bibfield  {author} {\bibinfo {author} {\bibfnamefont {Y.}~\bibnamefont
  {Yamamoto}}, \bibinfo {author} {\bibfnamefont {C.~M.}\ \bibnamefont {Diaz}},
  \bibinfo {author} {\bibfnamefont {L.}~\bibnamefont {Basurto}}, \bibinfo
  {author} {\bibfnamefont {K.~A.}\ \bibnamefont {Jackson}}, \bibinfo {author}
  {\bibfnamefont {T.}~\bibnamefont {Baruah}}, \ and\ \bibinfo {author}
  {\bibfnamefont {R.~R.}\ \bibnamefont {Zope}},\ }\bibfield  {title} {\enquote
  {\bibinfo {title} {{Fermi--L\"owdin orbital self-interaction correction using
  the strongly constrained and appropriately normed meta-GGA functional}},}\
  }\href {\doibase 10.1063/1.5120532} {\bibfield  {journal} {\bibinfo
  {journal} {J. Chem. Phys.}\ }\textbf {\bibinfo {volume} {151}},\ \bibinfo
  {pages} {154105} (\bibinfo {year} {2019})}\BibitemShut {NoStop}%
\bibitem [{\citenamefont {Vargas}\ \emph {et~al.}(2020)\citenamefont {Vargas},
  \citenamefont {Ufondu}, \citenamefont {Baruah}, \citenamefont {Yamamoto},
  \citenamefont {Jackson},\ and\ \citenamefont {Zope}}]{Vargas2020_3789}%
  \BibitemOpen
  \bibfield  {author} {\bibinfo {author} {\bibfnamefont {J.}~\bibnamefont
  {Vargas}}, \bibinfo {author} {\bibfnamefont {P.}~\bibnamefont {Ufondu}},
  \bibinfo {author} {\bibfnamefont {T.}~\bibnamefont {Baruah}}, \bibinfo
  {author} {\bibfnamefont {Y.}~\bibnamefont {Yamamoto}}, \bibinfo {author}
  {\bibfnamefont {K.~A.}\ \bibnamefont {Jackson}}, \ and\ \bibinfo {author}
  {\bibfnamefont {R.~R.}\ \bibnamefont {Zope}},\ }\bibfield  {title} {\enquote
  {\bibinfo {title} {Importance of self-interaction-error removal in density
  functional calculations on water cluster anions},}\ }\href {\doibase
  10.1039/c9cp06106a} {\bibfield  {journal} {\bibinfo  {journal} {Phys. Chem.
  Chem. Phys.}\ }\textbf {\bibinfo {volume} {22}},\ \bibinfo {pages} {3789}
  (\bibinfo {year} {2020})}\BibitemShut {NoStop}%
\bibitem [{\citenamefont {Lehtola}, \citenamefont {J\'onsson},\ and\
  \citenamefont {J\'onsson}(2016)}]{Lehtola2016_4296}%
  \BibitemOpen
  \bibfield  {author} {\bibinfo {author} {\bibfnamefont {S.}~\bibnamefont
  {Lehtola}}, \bibinfo {author} {\bibfnamefont {E.~{\"O}.}\ \bibnamefont
  {J\'onsson}}, \ and\ \bibinfo {author} {\bibfnamefont {H.}~\bibnamefont
  {J\'onsson}},\ }\bibfield  {title} {\enquote {\bibinfo {title} {{Effect of
  complex-valued optimal orbitals on atomization energies with the
  Perdew–-Zunger self-interaction correction to density functional
  theory}},}\ }\href {\doibase 10.1021/acs.jctc.6b00622} {\bibfield  {journal}
  {\bibinfo  {journal} {J. Chem. Theory Comput.}\ }\textbf {\bibinfo {volume}
  {12}},\ \bibinfo {pages} {4296} (\bibinfo {year} {2016})}\BibitemShut
  {NoStop}%
\bibitem [{\citenamefont {Lundin}\ and\ \citenamefont
  {Eriksson}(2001)}]{Lundin2001_247}%
  \BibitemOpen
  \bibfield  {author} {\bibinfo {author} {\bibfnamefont {U.}~\bibnamefont
  {Lundin}}\ and\ \bibinfo {author} {\bibfnamefont {O.}~\bibnamefont
  {Eriksson}},\ }\bibfield  {title} {\enquote {\bibinfo {title} {Novel method
  of self-interaction corrections in density functional calculations},}\ }\href
  {\doibase 10.1002/1097-461X(2001)81:4<247::AID-QUA1>3.0.CO;2-9} {\bibfield
  {journal} {\bibinfo  {journal} {{Int. J. Quantum Chem.}}\ }\textbf {\bibinfo
  {volume} {81}},\ \bibinfo {pages} {247} (\bibinfo {year} {2001})}\BibitemShut
  {NoStop}%
\bibitem [{\citenamefont {Lehtola}, \citenamefont {Blockhuys},\ and\
  \citenamefont {Van~Alsenoy}(2020)}]{Lehtola2020_1218}%
  \BibitemOpen
  \bibfield  {author} {\bibinfo {author} {\bibfnamefont {S.}~\bibnamefont
  {Lehtola}}, \bibinfo {author} {\bibfnamefont {F.}~\bibnamefont {Blockhuys}},
  \ and\ \bibinfo {author} {\bibfnamefont {C.}~\bibnamefont {Van~Alsenoy}},\
  }\bibfield  {title} {\enquote {\bibinfo {title} {An overview of
  self-consistent field calculations within finite basis sets},}\ }\href
  {\doibase 10.3390/molecules25051218} {\bibfield  {journal} {\bibinfo
  {journal} {{Molecules}}\ }\textbf {\bibinfo {volume} {25}},\ \bibinfo {pages}
  {1218} (\bibinfo {year} {2020})},\ \Eprint {http://arxiv.org/abs/1912.12029}
  {arXiv:1912.12029} \BibitemShut {NoStop}%
\bibitem [{\citenamefont {Harrison}, \citenamefont {Heaton},\ and\
  \citenamefont {Lin}(1983)}]{Harrison1983_2079}%
  \BibitemOpen
  \bibfield  {author} {\bibinfo {author} {\bibfnamefont {J.~G.}\ \bibnamefont
  {Harrison}}, \bibinfo {author} {\bibfnamefont {R.~A.}\ \bibnamefont
  {Heaton}}, \ and\ \bibinfo {author} {\bibfnamefont {C.~C.}\ \bibnamefont
  {Lin}},\ }\bibfield  {title} {\enquote {\bibinfo {title} {{Self-interaction
  correction to the local density Hartree--Fock atomic calculations of excited
  and ground states}},}\ }\href {\doibase 10.1088/0022-3700/16/12/006}
  {\bibfield  {journal} {\bibinfo  {journal} {{J. Phys. B: At. Mol. Phys.}}\
  }\textbf {\bibinfo {volume} {16}},\ \bibinfo {pages} {2079} (\bibinfo {year}
  {1983})}\BibitemShut {NoStop}%
\bibitem [{\citenamefont {Pederson}, \citenamefont {Heaton},\ and\
  \citenamefont {Lin}(1984)}]{Pederson1984_1972}%
  \BibitemOpen
  \bibfield  {author} {\bibinfo {author} {\bibfnamefont {M.~R.}\ \bibnamefont
  {Pederson}}, \bibinfo {author} {\bibfnamefont {R.~A.}\ \bibnamefont
  {Heaton}}, \ and\ \bibinfo {author} {\bibfnamefont {C.~C.}\ \bibnamefont
  {Lin}},\ }\bibfield  {title} {\enquote {\bibinfo {title} {{Local-density
  Hartree--Fock theory of electronic states of molecules with self-interaction
  correction}},}\ }\href {\doibase 10.1063/1.446959} {\bibfield  {journal}
  {\bibinfo  {journal} {{J. Chem. Phys.}}\ }\textbf {\bibinfo {volume} {80}},\
  \bibinfo {pages} {1972} (\bibinfo {year} {1984})}\BibitemShut {NoStop}%
\bibitem [{\citenamefont {Lehtola}\ and\ \citenamefont
  {J{\'{o}}nsson}(2013)}]{Lehtola2013_5365}%
  \BibitemOpen
  \bibfield  {author} {\bibinfo {author} {\bibfnamefont {S.}~\bibnamefont
  {Lehtola}}\ and\ \bibinfo {author} {\bibfnamefont {H.}~\bibnamefont
  {J{\'{o}}nsson}},\ }\bibfield  {title} {\enquote {\bibinfo {title} {{Unitary
  optimization of localized molecular orbitals}},}\ }\href {\doibase
  10.1021/ct400793q} {\bibfield  {journal} {\bibinfo  {journal} {J. Chem.
  Theory Comput.}\ }\textbf {\bibinfo {volume} {9}},\ \bibinfo {pages} {5365}
  (\bibinfo {year} {2013})}\BibitemShut {NoStop}%
\bibitem [{\citenamefont {Boys}(1960)}]{Boys1960_296}%
  \BibitemOpen
  \bibfield  {author} {\bibinfo {author} {\bibfnamefont {S.~F.}\ \bibnamefont
  {Boys}},\ }\bibfield  {title} {\enquote {\bibinfo {title} {{Construction of
  some molecular orbitals to be approximately invariant for changes from one
  molecule to another}},}\ }\href {\doibase 10.1103/RevModPhys.32.296}
  {\bibfield  {journal} {\bibinfo  {journal} {Rev. Mod. Phys.}\ }\textbf
  {\bibinfo {volume} {32}},\ \bibinfo {pages} {296} (\bibinfo {year}
  {1960})}\BibitemShut {NoStop}%
\bibitem [{\citenamefont {Edmiston}\ and\ \citenamefont
  {Ruedenberg}(1963)}]{Edmiston1963_457}%
  \BibitemOpen
  \bibfield  {author} {\bibinfo {author} {\bibfnamefont {C.}~\bibnamefont
  {Edmiston}}\ and\ \bibinfo {author} {\bibfnamefont {K.}~\bibnamefont
  {Ruedenberg}},\ }\bibfield  {title} {\enquote {\bibinfo {title} {{Localized
  atomic and molecular orbitals}},}\ }\href {\doibase
  10.1103/RevModPhys.35.457} {\bibfield  {journal} {\bibinfo  {journal} {Rev.
  Mod. Phys.}\ }\textbf {\bibinfo {volume} {35}},\ \bibinfo {pages} {457}
  (\bibinfo {year} {1963})}\BibitemShut {NoStop}%
\bibitem [{\citenamefont {Pipek}\ and\ \citenamefont
  {Mezey}(1989)}]{Pipek1989_4916}%
  \BibitemOpen
  \bibfield  {author} {\bibinfo {author} {\bibfnamefont {J.}~\bibnamefont
  {Pipek}}\ and\ \bibinfo {author} {\bibfnamefont {P.~G.}\ \bibnamefont
  {Mezey}},\ }\bibfield  {title} {\enquote {\bibinfo {title} {{A fast intrinsic
  localization procedure applicable for ab initio and semiempirical linear
  combination of atomic orbital wave functions}},}\ }\href {\doibase
  10.1063/1.456588} {\bibfield  {journal} {\bibinfo  {journal} {J. Chem.
  Phys.}\ }\textbf {\bibinfo {volume} {90}},\ \bibinfo {pages} {4916} (\bibinfo
  {year} {1989})}\BibitemShut {NoStop}%
\bibitem [{\citenamefont {Lehtola}\ and\ \citenamefont
  {J{\'{o}}nsson}(2014)}]{Lehtola2014_642}%
  \BibitemOpen
  \bibfield  {author} {\bibinfo {author} {\bibfnamefont {S.}~\bibnamefont
  {Lehtola}}\ and\ \bibinfo {author} {\bibfnamefont {H.}~\bibnamefont
  {J{\'{o}}nsson}},\ }\bibfield  {title} {\enquote {\bibinfo {title}
  {{Pipek--Mezey orbital localization using various partial charge
  estimates}},}\ }\href {\doibase 10.1021/ct401016x} {\bibfield  {journal}
  {\bibinfo  {journal} {J. Chem. Theory Comput.}\ }\textbf {\bibinfo {volume}
  {10}},\ \bibinfo {pages} {642} (\bibinfo {year} {2014})}\BibitemShut
  {NoStop}%
\bibitem [{\citenamefont {Luken}\ and\ \citenamefont
  {Culberson}(1982)}]{Luken1982_265}%
  \BibitemOpen
  \bibfield  {author} {\bibinfo {author} {\bibfnamefont {W.~L.}\ \bibnamefont
  {Luken}}\ and\ \bibinfo {author} {\bibfnamefont {J.~C.}\ \bibnamefont
  {Culberson}},\ }\bibfield  {title} {\enquote {\bibinfo {title} {{Mobility of
  the Fermi hole in a single-determinant wavefunction}},}\ }\href {\doibase
  10.1002/qua.560220828} {\bibfield  {journal} {\bibinfo  {journal} {{Int. J.
  Quantum Chem.}}\ }\textbf {\bibinfo {volume} {22}},\ \bibinfo {pages} {265}
  (\bibinfo {year} {1982})}\BibitemShut {NoStop}%
\bibitem [{\citenamefont {Luken}\ and\ \citenamefont
  {Beratan}(1982)}]{Luken1982_265b}%
  \BibitemOpen
  \bibfield  {author} {\bibinfo {author} {\bibfnamefont {W.~L.}\ \bibnamefont
  {Luken}}\ and\ \bibinfo {author} {\bibfnamefont {D.~N.}\ \bibnamefont
  {Beratan}},\ }\bibfield  {title} {\enquote {\bibinfo {title} {{Localized
  orbitals and the Fermi hole}},}\ }\href {\doibase 10.1007/BF00550971}
  {\bibfield  {journal} {\bibinfo  {journal} {{Theor. Chim. Acta}}\ }\textbf
  {\bibinfo {volume} {61}},\ \bibinfo {pages} {265} (\bibinfo {year}
  {1982})}\BibitemShut {NoStop}%
\bibitem [{\citenamefont {Luken}(1984)}]{Luken1984_1283}%
  \BibitemOpen
  \bibfield  {author} {\bibinfo {author} {\bibfnamefont {W.~L.}\ \bibnamefont
  {Luken}},\ }\bibfield  {title} {\enquote {\bibinfo {title} {{Properties of
  the Fermi hole in molecules}},}\ }\href@noop {} {\bibfield  {journal}
  {\bibinfo  {journal} {{Croat. Chem. Acta}}\ }\textbf {\bibinfo {volume}
  {57}},\ \bibinfo {pages} {1283} (\bibinfo {year} {1984})}\BibitemShut
  {NoStop}%
\bibitem [{\citenamefont {Luken}\ and\ \citenamefont
  {Culberson}(1984)}]{Luken1984_279}%
  \BibitemOpen
  \bibfield  {author} {\bibinfo {author} {\bibfnamefont {W.~L.}\ \bibnamefont
  {Luken}}\ and\ \bibinfo {author} {\bibfnamefont {J.~C.}\ \bibnamefont
  {Culberson}},\ }\bibfield  {title} {\enquote {\bibinfo {title} {{Localized
  orbitals based on the Fermi hole}},}\ }\href {\doibase 10.1007/BF00554785}
  {\bibfield  {journal} {\bibinfo  {journal} {{Theor. Chim. Acta}}\ }\textbf
  {\bibinfo {volume} {66}},\ \bibinfo {pages} {279} (\bibinfo {year}
  {1984})}\BibitemShut {NoStop}%
\bibitem [{\citenamefont {L{\"o}wdin}(1950)}]{Lowdin1950_365}%
  \BibitemOpen
  \bibfield  {author} {\bibinfo {author} {\bibfnamefont {P.-O.}\ \bibnamefont
  {L{\"o}wdin}},\ }\bibfield  {title} {\enquote {\bibinfo {title} {On the
  non-orthogonality problem connected with the use of atomic wave functions in
  the theory of molecules and crystals},}\ }\href {\doibase 10.1063/1.1747632}
  {\bibfield  {journal} {\bibinfo  {journal} {J. Chem. Phys.}\ }\textbf
  {\bibinfo {volume} {18}},\ \bibinfo {pages} {365} (\bibinfo {year}
  {1950})}\BibitemShut {NoStop}%
\bibitem [{\citenamefont {Lewis}(1916)}]{Lewis1916_762}%
  \BibitemOpen
  \bibfield  {author} {\bibinfo {author} {\bibfnamefont {G.~N.}\ \bibnamefont
  {Lewis}},\ }\bibfield  {title} {\enquote {\bibinfo {title} {The atom and the
  molecule},}\ }\href {\doibase 10.1021/ja02261a002} {\bibfield  {journal}
  {\bibinfo  {journal} {{J. Am. Chem. Soc.}}\ }\textbf {\bibinfo {volume}
  {38}},\ \bibinfo {pages} {762} (\bibinfo {year} {1916})}\BibitemShut
  {NoStop}%
\bibitem [{\citenamefont {Linnett}(1960)}]{Linnett1960_859}%
  \BibitemOpen
  \bibfield  {author} {\bibinfo {author} {\bibfnamefont {J.~W.}\ \bibnamefont
  {Linnett}},\ }\bibfield  {title} {\enquote {\bibinfo {title} {{Valence-bond
  structures: A new proposal}},}\ }\href {\doibase 10.1038/187859a0} {\bibfield
   {journal} {\bibinfo  {journal} {Nature}\ }\textbf {\bibinfo {volume}
  {187}},\ \bibinfo {pages} {859} (\bibinfo {year} {1960})}\BibitemShut
  {NoStop}%
\bibitem [{\citenamefont {Linnett}(1961)}]{Linnett1961_2643}%
  \BibitemOpen
  \bibfield  {author} {\bibinfo {author} {\bibfnamefont {J.~W.}\ \bibnamefont
  {Linnett}},\ }\bibfield  {title} {\enquote {\bibinfo {title} {{A modification
  of the Lewis--Langmuir octet rule}},}\ }\href {\doibase 10.1021/ja01473a011}
  {\bibfield  {journal} {\bibinfo  {journal} {J. Am. Chem. Soc.}\ }\textbf
  {\bibinfo {volume} {83}},\ \bibinfo {pages} {2643} (\bibinfo {year}
  {1961})}\BibitemShut {NoStop}%
\bibitem [{\citenamefont {Linnett}(1964)}]{Linnett1964_1}%
  \BibitemOpen
  \bibfield  {author} {\bibinfo {author} {\bibfnamefont {J.~W.}\ \bibnamefont
  {Linnett}},\ }\href@noop {} {\emph {\bibinfo {title} {Electronic structure of
  molecules}}}\ (\bibinfo  {publisher} {Methuen \& Co. Ltd.},\ \bibinfo
  {address} {London},\ \bibinfo {year} {1964})\BibitemShut {NoStop}%
\bibitem [{\citenamefont {Luder}(1964)}]{Luder1964_55}%
  \BibitemOpen
  \bibfield  {author} {\bibinfo {author} {\bibfnamefont {W.~F.}\ \bibnamefont
  {Luder}},\ }\bibfield  {title} {\enquote {\bibinfo {title} {{Electronic
  structure of molecules (Linnett, JW)}},}\ }\href {\doibase
  10.1021/ed043p55.5} {\bibfield  {journal} {\bibinfo  {journal} {J. Chem.
  Educ.}\ }\textbf {\bibinfo {volume} {43}},\ \bibinfo {pages} {55} (\bibinfo
  {year} {1964})}\BibitemShut {NoStop}%
\bibitem [{\citenamefont {Kraus}(2017)}]{Kraus2017_1}%
  \BibitemOpen
  \bibfield  {author} {\bibinfo {author} {\bibfnamefont {J.}~\bibnamefont
  {Kraus}},\ }\href {\doibase 10.13140/RG.2.2.15045.91362} {\enquote {\bibinfo
  {title} {{\emph{FLOSIC-DFT analysis of chemical bonding: application to
  diatomic molecules}, Bachelor's thesis, TU Bergakademie Freiberg}},}\ }
  (\bibinfo {year} {2017})\BibitemShut {NoStop}%
\bibitem [{\citenamefont {Larsen}\ \emph {et~al.}(2017)\citenamefont {Larsen},
  \citenamefont {Mortensen}, \citenamefont {Blomqvist}, \citenamefont
  {Castelli}, \citenamefont {Christensen}, \citenamefont {Du{\l}ak},
  \citenamefont {Friis}, \citenamefont {Groves}, \citenamefont {Hammer},
  \citenamefont {Hargus} \emph {et~al.}}]{Larsen2017_273002}%
  \BibitemOpen
  \bibfield  {author} {\bibinfo {author} {\bibfnamefont {A.~H.}\ \bibnamefont
  {Larsen}}, \bibinfo {author} {\bibfnamefont {J.~J.}\ \bibnamefont
  {Mortensen}}, \bibinfo {author} {\bibfnamefont {J.}~\bibnamefont
  {Blomqvist}}, \bibinfo {author} {\bibfnamefont {I.~E.}\ \bibnamefont
  {Castelli}}, \bibinfo {author} {\bibfnamefont {R.}~\bibnamefont
  {Christensen}}, \bibinfo {author} {\bibfnamefont {M.}~\bibnamefont
  {Du{\l}ak}}, \bibinfo {author} {\bibfnamefont {J.}~\bibnamefont {Friis}},
  \bibinfo {author} {\bibfnamefont {M.~N.}\ \bibnamefont {Groves}}, \bibinfo
  {author} {\bibfnamefont {B.}~\bibnamefont {Hammer}}, \bibinfo {author}
  {\bibfnamefont {C.}~\bibnamefont {Hargus}},  \emph {et~al.},\ }\bibfield
  {title} {\enquote {\bibinfo {title} {{The atomic simulation environment - a
  Python library for working with atoms}},}\ }\href {\doibase
  10.1088/1361-648X/aa680e} {\bibfield  {journal} {\bibinfo  {journal} {J.
  Phys. Condens. Matter}\ }\textbf {\bibinfo {volume} {29}},\ \bibinfo {pages}
  {273002} (\bibinfo {year} {2017})}\BibitemShut {NoStop}%
\bibitem [{\citenamefont {Broyden}(1970)}]{Broyden1970_76}%
  \BibitemOpen
  \bibfield  {author} {\bibinfo {author} {\bibfnamefont {C.~G.}\ \bibnamefont
  {Broyden}},\ }\bibfield  {title} {\enquote {\bibinfo {title} {{The
  convergence of a class of double-rank minimization algorithms 1. General
  considerations}},}\ }\href {\doibase 10.1093/imamat/6.1.76} {\bibfield
  {journal} {\bibinfo  {journal} {IMA J. Appl. Math.}\ }\textbf {\bibinfo
  {volume} {6}},\ \bibinfo {pages} {76} (\bibinfo {year} {1970})}\BibitemShut
  {NoStop}%
\bibitem [{\citenamefont {Fletcher}(1970)}]{Fletcher1970_317}%
  \BibitemOpen
  \bibfield  {author} {\bibinfo {author} {\bibfnamefont {R.}~\bibnamefont
  {Fletcher}},\ }\bibfield  {title} {\enquote {\bibinfo {title} {{A new
  approach to variable metric algorithms}},}\ }\href {\doibase
  10.1093/comjnl/13.3.317} {\bibfield  {journal} {\bibinfo  {journal} {Comput.
  J.}\ }\textbf {\bibinfo {volume} {13}},\ \bibinfo {pages} {317} (\bibinfo
  {year} {1970})}\BibitemShut {NoStop}%
\bibitem [{\citenamefont {Goldfarb}(1970)}]{Goldfarb1970_23}%
  \BibitemOpen
  \bibfield  {author} {\bibinfo {author} {\bibfnamefont {D.}~\bibnamefont
  {Goldfarb}},\ }\bibfield  {title} {\enquote {\bibinfo {title} {{A family of
  variable-metric methods derived by variational means}},}\ }\href {\doibase
  10.1090/s0025-5718-1970-0258249-6} {\bibfield  {journal} {\bibinfo  {journal}
  {Math. Comput.}\ }\textbf {\bibinfo {volume} {24}},\ \bibinfo {pages} {23}
  (\bibinfo {year} {1970})}\BibitemShut {NoStop}%
\bibitem [{\citenamefont {Shanno}(1970)}]{Shanno1970_647}%
  \BibitemOpen
  \bibfield  {author} {\bibinfo {author} {\bibfnamefont {D.~F.}\ \bibnamefont
  {Shanno}},\ }\bibfield  {title} {\enquote {\bibinfo {title} {{Conditioning of
  quasi-Newton methods for function minimization}},}\ }\href {\doibase
  10.1090/s0025-5718-1970-0274029-x} {\bibfield  {journal} {\bibinfo  {journal}
  {Math. Comput.}\ }\textbf {\bibinfo {volume} {24}},\ \bibinfo {pages} {647}
  (\bibinfo {year} {1970})}\BibitemShut {NoStop}%
\bibitem [{\citenamefont {Nocedal}(1980)}]{Nocedal1980_773}%
  \BibitemOpen
  \bibfield  {author} {\bibinfo {author} {\bibfnamefont {J.}~\bibnamefont
  {Nocedal}},\ }\bibfield  {title} {\enquote {\bibinfo {title} {{Updating
  quasi-Newton matrices with limited storage}},}\ }\href {\doibase
  10.1090/s0025-5718-1980-0572855-7} {\bibfield  {journal} {\bibinfo  {journal}
  {Math. Comput.}\ }\textbf {\bibinfo {volume} {35}},\ \bibinfo {pages} {773}
  (\bibinfo {year} {1980})}\BibitemShut {NoStop}%
\bibitem [{\citenamefont {Liu}\ and\ \citenamefont
  {Nocedal}(1989)}]{Liu1989_503}%
  \BibitemOpen
  \bibfield  {author} {\bibinfo {author} {\bibfnamefont {D.~C.}\ \bibnamefont
  {Liu}}\ and\ \bibinfo {author} {\bibfnamefont {J.}~\bibnamefont {Nocedal}},\
  }\bibfield  {title} {\enquote {\bibinfo {title} {{On the limited memory BFGS
  method for large scale optimization}},}\ }\href {\doibase 10.1007/BF01589116}
  {\bibfield  {journal} {\bibinfo  {journal} {{Math. Program.}}\ }\textbf
  {\bibinfo {volume} {45}},\ \bibinfo {pages} {503} (\bibinfo {year}
  {1989})}\BibitemShut {NoStop}%
\bibitem [{\citenamefont {Byrd}\ \emph {et~al.}(1995)\citenamefont {Byrd},
  \citenamefont {Lu}, \citenamefont {Nocedal},\ and\ \citenamefont
  {Zhu}}]{Byrd1995_1190}%
  \BibitemOpen
  \bibfield  {author} {\bibinfo {author} {\bibfnamefont {R.~H.}\ \bibnamefont
  {Byrd}}, \bibinfo {author} {\bibfnamefont {P.}~\bibnamefont {Lu}}, \bibinfo
  {author} {\bibfnamefont {J.}~\bibnamefont {Nocedal}}, \ and\ \bibinfo
  {author} {\bibfnamefont {C.}~\bibnamefont {Zhu}},\ }\bibfield  {title}
  {\enquote {\bibinfo {title} {A limited memory algorithm for bound constrained
  optimization},}\ }\href {\doibase 10.1137/0916069} {\bibfield  {journal}
  {\bibinfo  {journal} {{SIAM J. Sci. Comput.}}\ }\textbf {\bibinfo {volume}
  {16}},\ \bibinfo {pages} {1190} (\bibinfo {year} {1995})}\BibitemShut
  {NoStop}%
\bibitem [{\citenamefont {Zhu}\ \emph {et~al.}(1997)\citenamefont {Zhu},
  \citenamefont {Byrd}, \citenamefont {Lu},\ and\ \citenamefont
  {Nocedal}}]{Zhu1997_550}%
  \BibitemOpen
  \bibfield  {author} {\bibinfo {author} {\bibfnamefont {C.}~\bibnamefont
  {Zhu}}, \bibinfo {author} {\bibfnamefont {R.~H.}\ \bibnamefont {Byrd}},
  \bibinfo {author} {\bibfnamefont {P.}~\bibnamefont {Lu}}, \ and\ \bibinfo
  {author} {\bibfnamefont {J.}~\bibnamefont {Nocedal}},\ }\bibfield  {title}
  {\enquote {\bibinfo {title} {{Algorithm 778: L-BFGS-B: Fortran subroutines
  for large-scale bound-constrained optimization}},}\ }\href {\doibase
  10.1145/279232.279236} {\bibfield  {journal} {\bibinfo  {journal} {{ACM
  Trans. Math. Softw.}}\ }\textbf {\bibinfo {volume} {23}},\ \bibinfo {pages}
  {550} (\bibinfo {year} {1997})}\BibitemShut {NoStop}%
\bibitem [{\citenamefont {Bitzek}\ \emph {et~al.}(2006)\citenamefont {Bitzek},
  \citenamefont {Koskinen}, \citenamefont {G\"ahler}, \citenamefont {Moseler},\
  and\ \citenamefont {Gumbsch}}]{Bitzek2006_170201}%
  \BibitemOpen
  \bibfield  {author} {\bibinfo {author} {\bibfnamefont {E.}~\bibnamefont
  {Bitzek}}, \bibinfo {author} {\bibfnamefont {P.}~\bibnamefont {Koskinen}},
  \bibinfo {author} {\bibfnamefont {F.}~\bibnamefont {G\"ahler}}, \bibinfo
  {author} {\bibfnamefont {M.}~\bibnamefont {Moseler}}, \ and\ \bibinfo
  {author} {\bibfnamefont {P.}~\bibnamefont {Gumbsch}},\ }\bibfield  {title}
  {\enquote {\bibinfo {title} {{Structural relaxation made simple}},}\ }\href
  {\doibase 10.1103/PhysRevLett.97.170201} {\bibfield  {journal} {\bibinfo
  {journal} {Phys. Rev. Lett.}\ }\textbf {\bibinfo {volume} {97}},\ \bibinfo
  {pages} {170201} (\bibinfo {year} {2006})}\BibitemShut {NoStop}%
\bibitem [{\citenamefont {Heaton}, \citenamefont {Harrison},\ and\
  \citenamefont {Lin}(1982)}]{Heaton1982_827}%
  \BibitemOpen
  \bibfield  {author} {\bibinfo {author} {\bibfnamefont {R.~A.}\ \bibnamefont
  {Heaton}}, \bibinfo {author} {\bibfnamefont {J.~G.}\ \bibnamefont
  {Harrison}}, \ and\ \bibinfo {author} {\bibfnamefont {C.~C.}\ \bibnamefont
  {Lin}},\ }\bibfield  {title} {\enquote {\bibinfo {title} {{Self-interaction
  correction for energy band calculations: Application to LiCl}},}\ }\href
  {\doibase 10.1016/0038-1098(82)91257-1} {\bibfield  {journal} {\bibinfo
  {journal} {{Solid State Commun.}}\ }\textbf {\bibinfo {volume} {41}},\
  \bibinfo {pages} {827} (\bibinfo {year} {1982})}\BibitemShut {NoStop}%
\bibitem [{\citenamefont {Kim}\ \emph {et~al.}(2019)\citenamefont {Kim},
  \citenamefont {Chen}, \citenamefont {Cheng}, \citenamefont {Gindulyte},
  \citenamefont {He}, \citenamefont {He}, \citenamefont {Li}, \citenamefont
  {Shoemaker}, \citenamefont {Thiessen}, \citenamefont {Yu} \emph
  {et~al.}}]{Kim2019_D1102}%
  \BibitemOpen
  \bibfield  {author} {\bibinfo {author} {\bibfnamefont {S.}~\bibnamefont
  {Kim}}, \bibinfo {author} {\bibfnamefont {J.}~\bibnamefont {Chen}}, \bibinfo
  {author} {\bibfnamefont {T.}~\bibnamefont {Cheng}}, \bibinfo {author}
  {\bibfnamefont {A.}~\bibnamefont {Gindulyte}}, \bibinfo {author}
  {\bibfnamefont {J.}~\bibnamefont {He}}, \bibinfo {author} {\bibfnamefont
  {S.}~\bibnamefont {He}}, \bibinfo {author} {\bibfnamefont {Q.}~\bibnamefont
  {Li}}, \bibinfo {author} {\bibfnamefont {B.~A.}\ \bibnamefont {Shoemaker}},
  \bibinfo {author} {\bibfnamefont {P.~A.}\ \bibnamefont {Thiessen}}, \bibinfo
  {author} {\bibfnamefont {B.}~\bibnamefont {Yu}},  \emph {et~al.},\ }\bibfield
   {title} {\enquote {\bibinfo {title} {{PubChem 2019 update: Improved access
  to chemical data}},}\ }\href {\doibase 10.1093/nar/gky1033} {\bibfield
  {journal} {\bibinfo  {journal} {{Nucleic Acids Res.}}\ }\textbf {\bibinfo
  {volume} {47}},\ \bibinfo {pages} {D1102} (\bibinfo {year}
  {2019})}\BibitemShut {NoStop}%
\bibitem [{\citenamefont {Perdew}\ \emph {et~al.}(2008)\citenamefont {Perdew},
  \citenamefont {Ruzsinszky}, \citenamefont {Csonka}, \citenamefont {Vydrov},
  \citenamefont {Scuseria}, \citenamefont {Constantin}, \citenamefont {Zhou},\
  and\ \citenamefont {Burke}}]{Perdew2008_136406}%
  \BibitemOpen
  \bibfield  {author} {\bibinfo {author} {\bibfnamefont {J.~P.}\ \bibnamefont
  {Perdew}}, \bibinfo {author} {\bibfnamefont {A.}~\bibnamefont {Ruzsinszky}},
  \bibinfo {author} {\bibfnamefont {G.~I.}\ \bibnamefont {Csonka}}, \bibinfo
  {author} {\bibfnamefont {O.~A.}\ \bibnamefont {Vydrov}}, \bibinfo {author}
  {\bibfnamefont {G.~E.}\ \bibnamefont {Scuseria}}, \bibinfo {author}
  {\bibfnamefont {L.~A.}\ \bibnamefont {Constantin}}, \bibinfo {author}
  {\bibfnamefont {X.}~\bibnamefont {Zhou}}, \ and\ \bibinfo {author}
  {\bibfnamefont {K.}~\bibnamefont {Burke}},\ }\bibfield  {title} {\enquote
  {\bibinfo {title} {{Restoring the density-gradient expansion for exchange in
  solids and surfaces}},}\ }\href {\doibase 10.1103/PhysRevLett.100.136406}
  {\bibfield  {journal} {\bibinfo  {journal} {Phys. Rev. Lett.}\ }\textbf
  {\bibinfo {volume} {100}},\ \bibinfo {pages} {136406} (\bibinfo {year}
  {2008})},\ \Eprint {http://arxiv.org/abs/0711.0156} {arXiv:0711.0156}
  \BibitemShut {NoStop}%
\bibitem [{\citenamefont {Jensen}(2001)}]{Jensen2001_9113}%
  \BibitemOpen
  \bibfield  {author} {\bibinfo {author} {\bibfnamefont {F.}~\bibnamefont
  {Jensen}},\ }\bibfield  {title} {\enquote {\bibinfo {title} {{Polarization
  consistent basis sets: Principles}},}\ }\href {\doibase 10.1063/1.1413524}
  {\bibfield  {journal} {\bibinfo  {journal} {J. Chem. Phys.}\ }\textbf
  {\bibinfo {volume} {115}},\ \bibinfo {pages} {9113} (\bibinfo {year}
  {2001})}\BibitemShut {NoStop}%
\bibitem [{\citenamefont {Jensen}(2002)}]{Jensen2002_7372}%
  \BibitemOpen
  \bibfield  {author} {\bibinfo {author} {\bibfnamefont {F.}~\bibnamefont
  {Jensen}},\ }\bibfield  {title} {\enquote {\bibinfo {title} {{Polarization
  consistent basis sets. II. Estimating the Kohn--Sham basis set limit}},}\
  }\href {\doibase 10.1063/1.1465405} {\bibfield  {journal} {\bibinfo
  {journal} {J. Chem. Phys.}\ }\textbf {\bibinfo {volume} {116}},\ \bibinfo
  {pages} {7372} (\bibinfo {year} {2002})}\BibitemShut {NoStop}%
\bibitem [{\citenamefont {Shahi}\ \emph {et~al.}(2019)\citenamefont {Shahi},
  \citenamefont {Bhattarai}, \citenamefont {Wagle}, \citenamefont {Santra},
  \citenamefont {Schwalbe}, \citenamefont {Hahn}, \citenamefont {Kortus},
  \citenamefont {Jackson}, \citenamefont {Peralta}, \citenamefont {Trepte},
  \citenamefont {Lehtola}, \citenamefont {Nepal}, \citenamefont {Myneni},
  \citenamefont {Neupane}, \citenamefont {Adhikari}, \citenamefont
  {Ruzsinszky}, \citenamefont {Yamamoto}, \citenamefont {Baruah}, \citenamefont
  {Zope},\ and\ \citenamefont {Perdew}}]{Shahi2019_174102}%
  \BibitemOpen
  \bibfield  {author} {\bibinfo {author} {\bibfnamefont {C.}~\bibnamefont
  {Shahi}}, \bibinfo {author} {\bibfnamefont {P.}~\bibnamefont {Bhattarai}},
  \bibinfo {author} {\bibfnamefont {K.}~\bibnamefont {Wagle}}, \bibinfo
  {author} {\bibfnamefont {B.}~\bibnamefont {Santra}}, \bibinfo {author}
  {\bibfnamefont {S.}~\bibnamefont {Schwalbe}}, \bibinfo {author}
  {\bibfnamefont {T.}~\bibnamefont {Hahn}}, \bibinfo {author} {\bibfnamefont
  {J.}~\bibnamefont {Kortus}}, \bibinfo {author} {\bibfnamefont {K.~A.}\
  \bibnamefont {Jackson}}, \bibinfo {author} {\bibfnamefont {J.~E.}\
  \bibnamefont {Peralta}}, \bibinfo {author} {\bibfnamefont {K.}~\bibnamefont
  {Trepte}}, \bibinfo {author} {\bibfnamefont {S.}~\bibnamefont {Lehtola}},
  \bibinfo {author} {\bibfnamefont {N.~K.}\ \bibnamefont {Nepal}}, \bibinfo
  {author} {\bibfnamefont {H.}~\bibnamefont {Myneni}}, \bibinfo {author}
  {\bibfnamefont {B.}~\bibnamefont {Neupane}}, \bibinfo {author} {\bibfnamefont
  {S.}~\bibnamefont {Adhikari}}, \bibinfo {author} {\bibfnamefont
  {A.}~\bibnamefont {Ruzsinszky}}, \bibinfo {author} {\bibfnamefont
  {Y.}~\bibnamefont {Yamamoto}}, \bibinfo {author} {\bibfnamefont
  {T.}~\bibnamefont {Baruah}}, \bibinfo {author} {\bibfnamefont {R.~R.}\
  \bibnamefont {Zope}}, \ and\ \bibinfo {author} {\bibfnamefont {J.~P.}\
  \bibnamefont {Perdew}},\ }\bibfield  {title} {\enquote {\bibinfo {title}
  {Stretched or noded orbital densities and self-interaction correction in
  density functional theory},}\ }\href {\doibase 10.1063/1.5087065} {\bibfield
  {journal} {\bibinfo  {journal} {J. Chem. Phys.}\ }\textbf {\bibinfo {volume}
  {150}},\ \bibinfo {pages} {174102} (\bibinfo {year} {2019})}\BibitemShut
  {NoStop}%
\bibitem [{\citenamefont {Perdew}, \citenamefont {Burke},\ and\ \citenamefont
  {Ernzerhof}(1996)}]{Perdew1996_3865}%
  \BibitemOpen
  \bibfield  {author} {\bibinfo {author} {\bibfnamefont {J.~P.}\ \bibnamefont
  {Perdew}}, \bibinfo {author} {\bibfnamefont {K.}~\bibnamefont {Burke}}, \
  and\ \bibinfo {author} {\bibfnamefont {M.}~\bibnamefont {Ernzerhof}},\
  }\bibfield  {title} {\enquote {\bibinfo {title} {Generalized gradient
  approximation made simple},}\ }\href {\doibase 10.1103/PhysRevLett.77.3865}
  {\bibfield  {journal} {\bibinfo  {journal} {Phys. Rev. Lett.}\ }\textbf
  {\bibinfo {volume} {77}},\ \bibinfo {pages} {3865} (\bibinfo {year}
  {1996})}\BibitemShut {NoStop}%
\bibitem [{\citenamefont {J{\'o}nsson}, \citenamefont {Lehtola},\ and\
  \citenamefont {J{\'o}nsson}(2015)}]{Jonsson2015_1858}%
  \BibitemOpen
  \bibfield  {author} {\bibinfo {author} {\bibfnamefont {E.~{\"O}.}\
  \bibnamefont {J{\'o}nsson}}, \bibinfo {author} {\bibfnamefont
  {S.}~\bibnamefont {Lehtola}}, \ and\ \bibinfo {author} {\bibfnamefont
  {H.}~\bibnamefont {J{\'o}nsson}},\ }\bibfield  {title} {\enquote {\bibinfo
  {title} {{Towards an optimal gradient-dependent energy functional of the
  PZ-SIC form}},}\ }\href {\doibase 10.1016/j.procs.2015.05.417} {\bibfield
  {journal} {\bibinfo  {journal} {Procedia Comput. Sci.}\ }\textbf {\bibinfo
  {volume} {51}},\ \bibinfo {pages} {1858} (\bibinfo {year}
  {2015})}\BibitemShut {NoStop}%
\bibitem [{\citenamefont {Dunning}(1989)}]{Dunning1989_1007}%
  \BibitemOpen
  \bibfield  {author} {\bibinfo {author} {\bibfnamefont {T.~H.}\ \bibnamefont
  {Dunning}},\ }\bibfield  {title} {\enquote {\bibinfo {title} {{Gaussian basis
  sets for use in correlated molecular calculations. I. The atoms boron through
  neon and hydrogen}},}\ }\href {\doibase 10.1063/1.456153} {\bibfield
  {journal} {\bibinfo  {journal} {J. Chem. Phys.}\ }\textbf {\bibinfo {volume}
  {90}},\ \bibinfo {pages} {1007} (\bibinfo {year} {1989})}\BibitemShut
  {NoStop}%
\bibitem [{\citenamefont {Karton}, \citenamefont {Sylvetsky},\ and\
  \citenamefont {Martin}(2017)}]{Karton2017_2063}%
  \BibitemOpen
  \bibfield  {author} {\bibinfo {author} {\bibfnamefont {A.}~\bibnamefont
  {Karton}}, \bibinfo {author} {\bibfnamefont {N.}~\bibnamefont {Sylvetsky}}, \
  and\ \bibinfo {author} {\bibfnamefont {J.~M.~L.}\ \bibnamefont {Martin}},\
  }\bibfield  {title} {\enquote {\bibinfo {title} {{W4-17: A diverse and
  high-confidence dataset of atomization energies for benchmarking high-level
  electronic structure methods}},}\ }\href {\doibase 10.1002/jcc.24854}
  {\bibfield  {journal} {\bibinfo  {journal} {J. Comput. Chem.}\ }\textbf
  {\bibinfo {volume} {38}},\ \bibinfo {pages} {2063} (\bibinfo {year}
  {2017})}\BibitemShut {NoStop}%
\bibitem [{\citenamefont {Canc{\`{e}}s}, \citenamefont {Maday},\ and\
  \citenamefont {Stamm}(2013)}]{Cances2013_054111}%
  \BibitemOpen
  \bibfield  {author} {\bibinfo {author} {\bibfnamefont {E.}~\bibnamefont
  {Canc{\`{e}}s}}, \bibinfo {author} {\bibfnamefont {Y.}~\bibnamefont {Maday}},
  \ and\ \bibinfo {author} {\bibfnamefont {B.}~\bibnamefont {Stamm}},\
  }\bibfield  {title} {\enquote {\bibinfo {title} {{Domain decomposition for
  implicit solvation models}},}\ }\href {\doibase 10.1063/1.4816767} {\bibfield
   {journal} {\bibinfo  {journal} {J. Chem. Phys.}\ }\textbf {\bibinfo {volume}
  {139}},\ \bibinfo {pages} {054111} (\bibinfo {year} {2013})}\BibitemShut
  {NoStop}%
\bibitem [{\citenamefont {Lipparini}\ \emph {et~al.}(2013)\citenamefont
  {Lipparini}, \citenamefont {Stamm}, \citenamefont {Canc{\`{e}}s},
  \citenamefont {Maday},\ and\ \citenamefont {Mennucci}}]{Lipparini2013_3637}%
  \BibitemOpen
  \bibfield  {author} {\bibinfo {author} {\bibfnamefont {F.}~\bibnamefont
  {Lipparini}}, \bibinfo {author} {\bibfnamefont {B.}~\bibnamefont {Stamm}},
  \bibinfo {author} {\bibfnamefont {E.}~\bibnamefont {Canc{\`{e}}s}}, \bibinfo
  {author} {\bibfnamefont {Y.}~\bibnamefont {Maday}}, \ and\ \bibinfo {author}
  {\bibfnamefont {B.}~\bibnamefont {Mennucci}},\ }\bibfield  {title} {\enquote
  {\bibinfo {title} {{Fast domain decomposition algorithm for continuum
  solvation models: Energy and first derivatives}},}\ }\href {\doibase
  10.1021/ct400280b} {\bibfield  {journal} {\bibinfo  {journal} {J. Chem.
  Theory Comput.}\ }\textbf {\bibinfo {volume} {9}},\ \bibinfo {pages} {3637}
  (\bibinfo {year} {2013})}\BibitemShut {NoStop}%
\bibitem [{\citenamefont {Lipparini}\ \emph {et~al.}(2014)\citenamefont
  {Lipparini}, \citenamefont {Scalmani}, \citenamefont {Lagard{\`{e}}re},
  \citenamefont {Stamm}, \citenamefont {Canc{\`{e}}s}, \citenamefont {Maday},
  \citenamefont {Piquemal}, \citenamefont {Frisch},\ and\ \citenamefont
  {Mennucci}}]{Lipparini2014_184108}%
  \BibitemOpen
  \bibfield  {author} {\bibinfo {author} {\bibfnamefont {F.}~\bibnamefont
  {Lipparini}}, \bibinfo {author} {\bibfnamefont {G.}~\bibnamefont {Scalmani}},
  \bibinfo {author} {\bibfnamefont {L.}~\bibnamefont {Lagard{\`{e}}re}},
  \bibinfo {author} {\bibfnamefont {B.}~\bibnamefont {Stamm}}, \bibinfo
  {author} {\bibfnamefont {E.}~\bibnamefont {Canc{\`{e}}s}}, \bibinfo {author}
  {\bibfnamefont {Y.}~\bibnamefont {Maday}}, \bibinfo {author} {\bibfnamefont
  {J.-P.}\ \bibnamefont {Piquemal}}, \bibinfo {author} {\bibfnamefont {M.~J.}\
  \bibnamefont {Frisch}}, \ and\ \bibinfo {author} {\bibfnamefont
  {B.}~\bibnamefont {Mennucci}},\ }\bibfield  {title} {\enquote {\bibinfo
  {title} {{Quantum, classical, and hybrid QM/MM calculations in solution:
  General implementation of the ddCOSMO linear scaling strategy}},}\ }\href
  {\doibase 10.1063/1.4901304} {\bibfield  {journal} {\bibinfo  {journal} {J.
  Chem. Phys.}\ }\textbf {\bibinfo {volume} {141}},\ \bibinfo {pages} {184108}
  (\bibinfo {year} {2014})}\BibitemShut {NoStop}%
\bibitem [{\citenamefont {Stamm}\ \emph {et~al.}(2016)\citenamefont {Stamm},
  \citenamefont {Canc{\`{e}}s}, \citenamefont {Lipparini},\ and\ \citenamefont
  {Maday}}]{Stamm2016_054101}%
  \BibitemOpen
  \bibfield  {author} {\bibinfo {author} {\bibfnamefont {B.}~\bibnamefont
  {Stamm}}, \bibinfo {author} {\bibfnamefont {E.}~\bibnamefont {Canc{\`{e}}s}},
  \bibinfo {author} {\bibfnamefont {F.}~\bibnamefont {Lipparini}}, \ and\
  \bibinfo {author} {\bibfnamefont {Y.}~\bibnamefont {Maday}},\ }\bibfield
  {title} {\enquote {\bibinfo {title} {{A new discretization for the
  polarizable continuum model within the domain decomposition paradigm}},}\
  }\href {\doibase 10.1063/1.4940136} {\bibfield  {journal} {\bibinfo
  {journal} {J. Chem. Phys.}\ }\textbf {\bibinfo {volume} {144}},\ \bibinfo
  {pages} {054101} (\bibinfo {year} {2016})}\BibitemShut {NoStop}%
\bibitem [{\citenamefont {Lipparini}\ and\ \citenamefont
  {Mennucci}(2016)}]{Lipparini2016_160901}%
  \BibitemOpen
  \bibfield  {author} {\bibinfo {author} {\bibfnamefont {F.}~\bibnamefont
  {Lipparini}}\ and\ \bibinfo {author} {\bibfnamefont {B.}~\bibnamefont
  {Mennucci}},\ }\bibfield  {title} {\enquote {\bibinfo {title} {{Perspective:
  Polarizable continuum models for quantum-mechanical descriptions}},}\ }\href
  {\doibase 10.1063/1.4947236} {\bibfield  {journal} {\bibinfo  {journal} {J.
  Chem. Phys.}\ }\textbf {\bibinfo {volume} {144}},\ \bibinfo {pages} {160901}
  (\bibinfo {year} {2016})}\BibitemShut {NoStop}%
\bibitem [{\citenamefont {Pisani}, \citenamefont {Dovesi},\ and\ \citenamefont
  {Roetti}(1988)}]{Pisani1988_1}%
  \BibitemOpen
  \bibfield  {author} {\bibinfo {author} {\bibfnamefont {C.}~\bibnamefont
  {Pisani}}, \bibinfo {author} {\bibfnamefont {R.}~\bibnamefont {Dovesi}}, \
  and\ \bibinfo {author} {\bibfnamefont {C.}~\bibnamefont {Roetti}},\ }\href
  {\doibase 10.1007/978-3-642-93385-1} {\emph {\bibinfo {title} {{Hartree--Fock
  ab initio treatment of crystalline systems}}}},\ \bibinfo {series} {Lecture
  Notes in Chemistry}, Vol.~\bibinfo {volume} {48}\ (\bibinfo  {publisher}
  {Springer Berlin Heidelberg},\ \bibinfo {address} {Berlin, Heidelberg},\
  \bibinfo {year} {1988})\BibitemShut {NoStop}%
\bibitem [{\citenamefont {Heaton}, \citenamefont {Harrison},\ and\
  \citenamefont {Lin}(1983)}]{Heaton1983_5992}%
  \BibitemOpen
  \bibfield  {author} {\bibinfo {author} {\bibfnamefont {R.~A.}\ \bibnamefont
  {Heaton}}, \bibinfo {author} {\bibfnamefont {J.~G.}\ \bibnamefont
  {Harrison}}, \ and\ \bibinfo {author} {\bibfnamefont {C.~C.}\ \bibnamefont
  {Lin}},\ }\bibfield  {title} {\enquote {\bibinfo {title} {{Self-interaction
  correction for density-functional theory of electronic energy bands of
  solids}},}\ }\href {\doibase 10.1103/PhysRevB.28.5992} {\bibfield  {journal}
  {\bibinfo  {journal} {Phys. Rev. B}\ }\textbf {\bibinfo {volume} {28}},\
  \bibinfo {pages} {5992} (\bibinfo {year} {1983})}\BibitemShut {NoStop}%
\bibitem [{\citenamefont {Heaton}\ and\ \citenamefont
  {Lin}(1984)}]{Heaton1984_1853}%
  \BibitemOpen
  \bibfield  {author} {\bibinfo {author} {\bibfnamefont {R.~A.}\ \bibnamefont
  {Heaton}}\ and\ \bibinfo {author} {\bibfnamefont {C.~C.}\ \bibnamefont
  {Lin}},\ }\bibfield  {title} {\enquote {\bibinfo {title}
  {{Self-interaction-correction theory for density functional calculations of
  electronic energy bands for the lithium chloride crystal}},}\ }\href
  {\doibase 10.1088/0022-3719/17/11/007} {\bibfield  {journal} {\bibinfo
  {journal} {J. Phys. C: Solid State Phys.}\ }\textbf {\bibinfo {volume}
  {17}},\ \bibinfo {pages} {1853} (\bibinfo {year} {1984})}\BibitemShut
  {NoStop}%
\bibitem [{\citenamefont {Svane}\ and\ \citenamefont
  {Gunnarsson}(1988{\natexlab{a}})}]{Svane1988_171}%
  \BibitemOpen
  \bibfield  {author} {\bibinfo {author} {\bibfnamefont {A.}~\bibnamefont
  {Svane}}\ and\ \bibinfo {author} {\bibfnamefont {O.}~\bibnamefont
  {Gunnarsson}},\ }\bibfield  {title} {\enquote {\bibinfo {title}
  {Anti-ferromagnetic moment formation in the self-interaction-corrected
  density functional formalism},}\ }\href {\doibase 10.1209/0295-5075/7/2/013}
  {\bibfield  {journal} {\bibinfo  {journal} {EPL}\ }\textbf {\bibinfo {volume}
  {7}},\ \bibinfo {pages} {171} (\bibinfo {year}
  {1988}{\natexlab{a}})}\BibitemShut {NoStop}%
\bibitem [{\citenamefont {Svane}\ and\ \citenamefont
  {Gunnarsson}(1988{\natexlab{b}})}]{Svane1988_9919}%
  \BibitemOpen
  \bibfield  {author} {\bibinfo {author} {\bibfnamefont {A.}~\bibnamefont
  {Svane}}\ and\ \bibinfo {author} {\bibfnamefont {O.}~\bibnamefont
  {Gunnarsson}},\ }\bibfield  {title} {\enquote {\bibinfo {title} {Localization
  in the self-interaction-corrected density-functional formalism},}\ }\href
  {\doibase 10.1103/PhysRevB.37.9919} {\bibfield  {journal} {\bibinfo
  {journal} {Phys. Rev. B}\ }\textbf {\bibinfo {volume} {37}},\ \bibinfo
  {pages} {9919} (\bibinfo {year} {1988}{\natexlab{b}})}\BibitemShut {NoStop}%
\bibitem [{\citenamefont {Erwin}\ and\ \citenamefont
  {Lin}(1988)}]{Erwin1988_4285}%
  \BibitemOpen
  \bibfield  {author} {\bibinfo {author} {\bibfnamefont {S.~C.}\ \bibnamefont
  {Erwin}}\ and\ \bibinfo {author} {\bibfnamefont {C.~C.}\ \bibnamefont
  {Lin}},\ }\bibfield  {title} {\enquote {\bibinfo {title} {The
  self-interaction-corrected electronic band structure of six alkali fluoride
  and chloride crystals},}\ }\href {\doibase 10.1088/0022-3719/21/23/013}
  {\bibfield  {journal} {\bibinfo  {journal} {J. Phys. C: Solid State Phys.}\
  }\textbf {\bibinfo {volume} {21}},\ \bibinfo {pages} {4285} (\bibinfo {year}
  {1988})}\BibitemShut {NoStop}%
\bibitem [{\citenamefont {Szotek}, \citenamefont {Temmerman},\ and\
  \citenamefont {Winter}(1990{\natexlab{a}})}]{Szotek1990_1031}%
  \BibitemOpen
  \bibfield  {author} {\bibinfo {author} {\bibfnamefont {Z.}~\bibnamefont
  {Szotek}}, \bibinfo {author} {\bibfnamefont {W.~M.}\ \bibnamefont
  {Temmerman}}, \ and\ \bibinfo {author} {\bibfnamefont {H.}~\bibnamefont
  {Winter}},\ }\bibfield  {title} {\enquote {\bibinfo {title} {{On the
  self-interaction correction of localized bands: Application to rare gas
  solids}},}\ }\href {\doibase 10.1016/0038-1098(90)90704-F} {\bibfield
  {journal} {\bibinfo  {journal} {Solid State Commun.}\ }\textbf {\bibinfo
  {volume} {74}},\ \bibinfo {pages} {1031} (\bibinfo {year}
  {1990}{\natexlab{a}})}\BibitemShut {NoStop}%
\bibitem [{\citenamefont {Szotek}, \citenamefont {Temmerman},\ and\
  \citenamefont {Winter}(1990{\natexlab{b}})}]{Szotek1990_275}%
  \BibitemOpen
  \bibfield  {author} {\bibinfo {author} {\bibfnamefont {Z.}~\bibnamefont
  {Szotek}}, \bibinfo {author} {\bibfnamefont {W.~M.}\ \bibnamefont
  {Temmerman}}, \ and\ \bibinfo {author} {\bibfnamefont {H.}~\bibnamefont
  {Winter}},\ }\bibfield  {title} {\enquote {\bibinfo {title} {{On the
  self-interaction correction of localized bands: Application to the 4p
  semi-core states in Y}},}\ }\href {\doibase 10.1016/S0921-4526(90)80987-T}
  {\bibfield  {journal} {\bibinfo  {journal} {Phys. B (Amsterdam, Neth.)}\
  }\textbf {\bibinfo {volume} {165}},\ \bibinfo {pages} {275} (\bibinfo {year}
  {1990}{\natexlab{b}})}\BibitemShut {NoStop}%
\bibitem [{\citenamefont {Svane}\ and\ \citenamefont
  {Gunnarsson}(1990{\natexlab{a}})}]{Svane1990_851}%
  \BibitemOpen
  \bibfield  {author} {\bibinfo {author} {\bibfnamefont {A.}~\bibnamefont
  {Svane}}\ and\ \bibinfo {author} {\bibfnamefont {O.}~\bibnamefont
  {Gunnarsson}},\ }\bibfield  {title} {\enquote {\bibinfo {title} {Hydrogen
  solid in self-interaction-corrected local-spin-density approximation},}\
  }\href {\doibase 10.1016/0038-1098(90)90641-N} {\bibfield  {journal}
  {\bibinfo  {journal} {Solid State Commun.}\ }\textbf {\bibinfo {volume}
  {76}},\ \bibinfo {pages} {851} (\bibinfo {year}
  {1990}{\natexlab{a}})}\BibitemShut {NoStop}%
\bibitem [{\citenamefont {Svane}\ and\ \citenamefont
  {Gunnarsson}(1990{\natexlab{b}})}]{Svane1990_1148}%
  \BibitemOpen
  \bibfield  {author} {\bibinfo {author} {\bibfnamefont {A.}~\bibnamefont
  {Svane}}\ and\ \bibinfo {author} {\bibfnamefont {O.}~\bibnamefont
  {Gunnarsson}},\ }\bibfield  {title} {\enquote {\bibinfo {title}
  {Transition-metal oxides in the self-interaction-corrected density-functional
  formalism},}\ }\href {\doibase 10.1103/PhysRevLett.65.1148} {\bibfield
  {journal} {\bibinfo  {journal} {Phys. Rev. Lett.}\ }\textbf {\bibinfo
  {volume} {65}},\ \bibinfo {pages} {1148} (\bibinfo {year}
  {1990}{\natexlab{b}})}\BibitemShut {NoStop}%
\bibitem [{\citenamefont {Svane}(1992)}]{Svane1992_1900}%
  \BibitemOpen
  \bibfield  {author} {\bibinfo {author} {\bibfnamefont {A.}~\bibnamefont
  {Svane}},\ }\bibfield  {title} {\enquote {\bibinfo {title} {Electronic
  structure of {La}$_{\textrm{2}}${CuO}$_{\textrm{4}}$ in the
  self-interaction-corrected density-functional formalism},}\ }\href {\doibase
  10.1103/PhysRevLett.68.1900} {\bibfield  {journal} {\bibinfo  {journal}
  {Phys. Rev. Lett.}\ }\textbf {\bibinfo {volume} {68}},\ \bibinfo {pages}
  {1900} (\bibinfo {year} {1992})}\BibitemShut {NoStop}%
\bibitem [{\citenamefont {Rieger}\ and\ \citenamefont
  {Vogl}(1995)}]{Rieger1995_16567}%
  \BibitemOpen
  \bibfield  {author} {\bibinfo {author} {\bibfnamefont {M.~M.}\ \bibnamefont
  {Rieger}}\ and\ \bibinfo {author} {\bibfnamefont {P.}~\bibnamefont {Vogl}},\
  }\bibfield  {title} {\enquote {\bibinfo {title} {Self-interaction corrections
  in semiconductors},}\ }\href {\doibase 10.1103/PhysRevB.52.16567} {\bibfield
  {journal} {\bibinfo  {journal} {Phys. Rev. B}\ }\textbf {\bibinfo {volume}
  {52}},\ \bibinfo {pages} {16567} (\bibinfo {year} {1995})}\BibitemShut
  {NoStop}%
\bibitem [{\citenamefont {Vogel}, \citenamefont {Kr{\"u}ger},\ and\
  \citenamefont {Pollmann}(1995)}]{Vogel1995_R14316}%
  \BibitemOpen
  \bibfield  {author} {\bibinfo {author} {\bibfnamefont {D.}~\bibnamefont
  {Vogel}}, \bibinfo {author} {\bibfnamefont {P.}~\bibnamefont {Kr{\"u}ger}}, \
  and\ \bibinfo {author} {\bibfnamefont {J.}~\bibnamefont {Pollmann}},\
  }\bibfield  {title} {\enquote {\bibinfo {title} {{Ab initio
  electronic-structure calculations for II-VI semiconductors using
  self-interaction-corrected pseudopotentials}},}\ }\href {\doibase
  10.1103/PhysRevB.52.R14316} {\bibfield  {journal} {\bibinfo  {journal} {Phys.
  Rev. B}\ }\textbf {\bibinfo {volume} {52}},\ \bibinfo {pages} {R14316}
  (\bibinfo {year} {1995})}\BibitemShut {NoStop}%
\bibitem [{\citenamefont {Arai}\ and\ \citenamefont
  {Fujiwara}(1995)}]{Arai1995_1477}%
  \BibitemOpen
  \bibfield  {author} {\bibinfo {author} {\bibfnamefont {M.}~\bibnamefont
  {Arai}}\ and\ \bibinfo {author} {\bibfnamefont {T.}~\bibnamefont
  {Fujiwara}},\ }\bibfield  {title} {\enquote {\bibinfo {title} {Electronic
  structures of transition-metal mono-oxides in the self-interaction-corrected
  local-spin-density approximation},}\ }\href {\doibase
  10.1103/PhysRevB.51.1477} {\bibfield  {journal} {\bibinfo  {journal} {Phys.
  Rev. B}\ }\textbf {\bibinfo {volume} {51}},\ \bibinfo {pages} {1477}
  (\bibinfo {year} {1995})}\BibitemShut {NoStop}%
\bibitem [{\citenamefont {Vogel}, \citenamefont {Kr{\"u}ger},\ and\
  \citenamefont {Pollmann}(1996)}]{Vogel1996_5495}%
  \BibitemOpen
  \bibfield  {author} {\bibinfo {author} {\bibfnamefont {D.}~\bibnamefont
  {Vogel}}, \bibinfo {author} {\bibfnamefont {P.}~\bibnamefont {Kr{\"u}ger}}, \
  and\ \bibinfo {author} {\bibfnamefont {J.}~\bibnamefont {Pollmann}},\
  }\bibfield  {title} {\enquote {\bibinfo {title} {{Self-interaction and
  relaxation-corrected pseudopotentials for II-VI semiconductors}},}\ }\href
  {\doibase 10.1103/PhysRevB.54.5495} {\bibfield  {journal} {\bibinfo
  {journal} {Phys. Rev. B}\ }\textbf {\bibinfo {volume} {54}},\ \bibinfo
  {pages} {5495} (\bibinfo {year} {1996})}\BibitemShut {NoStop}%
\bibitem [{\citenamefont {Svane}(1996)}]{Svane1996_4275}%
  \BibitemOpen
  \bibfield  {author} {\bibinfo {author} {\bibfnamefont {A.}~\bibnamefont
  {Svane}},\ }\bibfield  {title} {\enquote {\bibinfo {title} {Electronic
  structure of cerium in the self-interaction-corrected local-spin-density
  approximation},}\ }\href {\doibase 10.1103/physrevb.53.4275} {\bibfield
  {journal} {\bibinfo  {journal} {Phys. Rev. B}\ }\textbf {\bibinfo {volume}
  {53}},\ \bibinfo {pages} {4275} (\bibinfo {year} {1996})}\BibitemShut
  {NoStop}%
\bibitem [{\citenamefont {Svane}\ \emph {et~al.}(2000)\citenamefont {Svane},
  \citenamefont {Temmerman}, \citenamefont {Szotek}, \citenamefont
  {Laegsgaard},\ and\ \citenamefont {Winter}}]{Svane2000_799}%
  \BibitemOpen
  \bibfield  {author} {\bibinfo {author} {\bibfnamefont {A.}~\bibnamefont
  {Svane}}, \bibinfo {author} {\bibfnamefont {W.~M.}\ \bibnamefont
  {Temmerman}}, \bibinfo {author} {\bibfnamefont {Z.}~\bibnamefont {Szotek}},
  \bibinfo {author} {\bibfnamefont {J.}~\bibnamefont {Laegsgaard}}, \ and\
  \bibinfo {author} {\bibfnamefont {H.}~\bibnamefont {Winter}},\ }\bibfield
  {title} {\enquote {\bibinfo {title} {Self-interaction-corrected
  local-spin-density calculations for rare earth materials},}\ }\href {\doibase
  10.1002/(SICI)1097-461X(2000)77:5<799::AID-QUA2>3.0.CO;2-Z} {\bibfield
  {journal} {\bibinfo  {journal} {Int. J. Quantum Chem.}\ }\textbf {\bibinfo
  {volume} {77}},\ \bibinfo {pages} {799} (\bibinfo {year} {2000})}\BibitemShut
  {NoStop}%
\bibitem [{\citenamefont {Filippetti}\ and\ \citenamefont
  {Spaldin}(2003)}]{Filippetti2003_125109}%
  \BibitemOpen
  \bibfield  {author} {\bibinfo {author} {\bibfnamefont {A.}~\bibnamefont
  {Filippetti}}\ and\ \bibinfo {author} {\bibfnamefont {N.~A.}\ \bibnamefont
  {Spaldin}},\ }\bibfield  {title} {\enquote {\bibinfo {title}
  {Self-interaction-corrected pseudopotential scheme for magnetic and
  strongly-correlated systems},}\ }\href {\doibase 10.1103/PhysRevB.67.125109}
  {\bibfield  {journal} {\bibinfo  {journal} {Phys. Rev. B}\ }\textbf {\bibinfo
  {volume} {67}},\ \bibinfo {pages} {125109} (\bibinfo {year} {2003})},\
  \Eprint {http://arxiv.org/abs/cond-mat/0303042} {arXiv:cond-mat/0303042}
  \BibitemShut {NoStop}%
\bibitem [{\citenamefont {Bylaska}, \citenamefont {Tsemekhman},\ and\
  \citenamefont {J{\'{o}}nsson}(2004)}]{Bylaska2004_1}%
  \BibitemOpen
  \bibfield  {author} {\bibinfo {author} {\bibfnamefont {E.}~\bibnamefont
  {Bylaska}}, \bibinfo {author} {\bibfnamefont {K.}~\bibnamefont {Tsemekhman}},
  \ and\ \bibinfo {author} {\bibfnamefont {H.}~\bibnamefont {J{\'{o}}nsson}},\
  }\bibfield  {title} {\enquote {\bibinfo {title} {{Self-consistent
  self-interaction corrected DFT: The method and applications to extended and
  confined systems}},}\ }in\ \href
  {https://www.researchgate.net/publication/241352392_Self-Consistent_Self-Interaction_Corrected_DFT_The_Method_and_Applications_to_Extended_and_Confined_Systems}
  {\emph {\bibinfo {booktitle} {APS Meeting Abstracts}}}\ (\bibinfo {year}
  {2004})\BibitemShut {NoStop}%
\bibitem [{\citenamefont {L{\"u}ders}\ \emph {et~al.}(2005)\citenamefont
  {L{\"u}ders}, \citenamefont {Ernst}, \citenamefont {D{\"a}ne}, \citenamefont
  {Szotek}, \citenamefont {Svane}, \citenamefont {K{\"o}dderitzsch},
  \citenamefont {Hergert}, \citenamefont {Gy{\"o}rffy},\ and\ \citenamefont
  {Temmerman}}]{Luders2005_205109}%
  \BibitemOpen
  \bibfield  {author} {\bibinfo {author} {\bibfnamefont {M.}~\bibnamefont
  {L{\"u}ders}}, \bibinfo {author} {\bibfnamefont {A.}~\bibnamefont {Ernst}},
  \bibinfo {author} {\bibfnamefont {M.}~\bibnamefont {D{\"a}ne}}, \bibinfo
  {author} {\bibfnamefont {Z.}~\bibnamefont {Szotek}}, \bibinfo {author}
  {\bibfnamefont {A.}~\bibnamefont {Svane}}, \bibinfo {author} {\bibfnamefont
  {D.}~\bibnamefont {K{\"o}dderitzsch}}, \bibinfo {author} {\bibfnamefont
  {W.}~\bibnamefont {Hergert}}, \bibinfo {author} {\bibfnamefont {B.~L.}\
  \bibnamefont {Gy{\"o}rffy}}, \ and\ \bibinfo {author} {\bibfnamefont {W.~M.}\
  \bibnamefont {Temmerman}},\ }\bibfield  {title} {\enquote {\bibinfo {title}
  {Self-interaction correction in multiple scattering theory},}\ }\href
  {\doibase 10.1103/PhysRevB.71.205109} {\bibfield  {journal} {\bibinfo
  {journal} {Phys. Rev. B}\ }\textbf {\bibinfo {volume} {71}},\ \bibinfo
  {pages} {205109} (\bibinfo {year} {2005})}\BibitemShut {NoStop}%
\bibitem [{\citenamefont {Bylaska}, \citenamefont {Tsemekhman},\ and\
  \citenamefont {Gao}(2006)}]{Bylaska2006_86}%
  \BibitemOpen
  \bibfield  {author} {\bibinfo {author} {\bibfnamefont {E.~J.}\ \bibnamefont
  {Bylaska}}, \bibinfo {author} {\bibfnamefont {K.}~\bibnamefont {Tsemekhman}},
  \ and\ \bibinfo {author} {\bibfnamefont {F.}~\bibnamefont {Gao}},\ }\bibfield
   {title} {\enquote {\bibinfo {title} {{New development of self-interaction
  corrected DFT for extended systems applied to the calculation of native
  defects in 3C--SiC}},}\ }\href {\doibase 10.1088/0031-8949/2006/T124/017}
  {\bibfield  {journal} {\bibinfo  {journal} {Phys. Scr.}\ }\textbf {\bibinfo
  {volume} {T124}},\ \bibinfo {pages} {86} (\bibinfo {year}
  {2006})}\BibitemShut {NoStop}%
\bibitem [{\citenamefont {Hourahine}\ \emph {et~al.}(2007)\citenamefont
  {Hourahine}, \citenamefont {Sanna}, \citenamefont {Aradi}, \citenamefont
  {K{\"o}hler}, \citenamefont {Niehaus},\ and\ \citenamefont
  {Frauenheim}}]{Hourahine2007_5671}%
  \BibitemOpen
  \bibfield  {author} {\bibinfo {author} {\bibfnamefont {B.}~\bibnamefont
  {Hourahine}}, \bibinfo {author} {\bibfnamefont {S.}~\bibnamefont {Sanna}},
  \bibinfo {author} {\bibfnamefont {B.}~\bibnamefont {Aradi}}, \bibinfo
  {author} {\bibfnamefont {C.}~\bibnamefont {K{\"o}hler}}, \bibinfo {author}
  {\bibfnamefont {T.}~\bibnamefont {Niehaus}}, \ and\ \bibinfo {author}
  {\bibfnamefont {T.}~\bibnamefont {Frauenheim}},\ }\bibfield  {title}
  {\enquote {\bibinfo {title} {{Self-interaction and strong correlation in
  DFTB}},}\ }\href {\doibase 10.1021/jp070173b} {\bibfield  {journal} {\bibinfo
   {journal} {J. Phys. Chem. A}\ }\textbf {\bibinfo {volume} {111}},\ \bibinfo
  {pages} {5671} (\bibinfo {year} {2007})}\BibitemShut {NoStop}%
\bibitem [{\citenamefont {Stengel}\ and\ \citenamefont
  {Spaldin}(2008)}]{Stengel2008_155106}%
  \BibitemOpen
  \bibfield  {author} {\bibinfo {author} {\bibfnamefont {M.}~\bibnamefont
  {Stengel}}\ and\ \bibinfo {author} {\bibfnamefont {N.~A.}\ \bibnamefont
  {Spaldin}},\ }\bibfield  {title} {\enquote {\bibinfo {title}
  {{Self-interaction correction with Wannier functions}},}\ }\href {\doibase
  10.1103/physrevb.77.155106} {\bibfield  {journal} {\bibinfo  {journal} {Phys.
  Rev. B}\ }\textbf {\bibinfo {volume} {77}},\ \bibinfo {pages} {155106}
  (\bibinfo {year} {2008})}\BibitemShut {NoStop}%
\bibitem [{\citenamefont {D\"ane}\ \emph {et~al.}(2009)\citenamefont {D\"ane},
  \citenamefont {Lueders}, \citenamefont {Ernst}, \citenamefont
  {K{\"o}dderitzsch}, \citenamefont {Temmerman}, \citenamefont {Szotek},\ and\
  \citenamefont {Hergert}}]{Dane2009_045604}%
  \BibitemOpen
  \bibfield  {author} {\bibinfo {author} {\bibfnamefont {M.}~\bibnamefont
  {D\"ane}}, \bibinfo {author} {\bibfnamefont {M.}~\bibnamefont {Lueders}},
  \bibinfo {author} {\bibfnamefont {A.}~\bibnamefont {Ernst}}, \bibinfo
  {author} {\bibfnamefont {D.}~\bibnamefont {K{\"o}dderitzsch}}, \bibinfo
  {author} {\bibfnamefont {W.~M.}\ \bibnamefont {Temmerman}}, \bibinfo {author}
  {\bibfnamefont {Z.}~\bibnamefont {Szotek}}, \ and\ \bibinfo {author}
  {\bibfnamefont {W.}~\bibnamefont {Hergert}},\ }\bibfield  {title} {\enquote
  {\bibinfo {title} {{Self-interaction correction in multiple scattering
  theory: Application to transition metal oxides}},}\ }\href {\doibase
  10.1088/0953-8984/21/4/045604} {\bibfield  {journal} {\bibinfo  {journal} {J.
  Phys.: Condens. Matter}\ }\textbf {\bibinfo {volume} {21}},\ \bibinfo {pages}
  {045604} (\bibinfo {year} {2009})}\BibitemShut {NoStop}%
\bibitem [{\citenamefont {Nguyen}\ \emph {et~al.}(2018)\citenamefont {Nguyen},
  \citenamefont {Colonna}, \citenamefont {Ferretti},\ and\ \citenamefont
  {Marzari}}]{Nguyen2018_021051}%
  \BibitemOpen
  \bibfield  {author} {\bibinfo {author} {\bibfnamefont {N.~L.}\ \bibnamefont
  {Nguyen}}, \bibinfo {author} {\bibfnamefont {N.}~\bibnamefont {Colonna}},
  \bibinfo {author} {\bibfnamefont {A.}~\bibnamefont {Ferretti}}, \ and\
  \bibinfo {author} {\bibfnamefont {N.}~\bibnamefont {Marzari}},\ }\bibfield
  {title} {\enquote {\bibinfo {title} {Koopmans-compliant spectral functionals
  for extended systems},}\ }\href {\doibase 10.1103/PhysRevX.8.021051}
  {\bibfield  {journal} {\bibinfo  {journal} {Phys. Rev. X}\ }\textbf {\bibinfo
  {volume} {8}},\ \bibinfo {pages} {021051} (\bibinfo {year} {2018})},\ \Eprint
  {http://arxiv.org/abs/1708.08518} {arXiv:1708.08518} \BibitemShut {NoStop}%
\bibitem [{\citenamefont {Marzari}\ and\ \citenamefont
  {Vanderbilt}(1997)}]{Marzari1997_12847}%
  \BibitemOpen
  \bibfield  {author} {\bibinfo {author} {\bibfnamefont {N.}~\bibnamefont
  {Marzari}}\ and\ \bibinfo {author} {\bibfnamefont {D.}~\bibnamefont
  {Vanderbilt}},\ }\bibfield  {title} {\enquote {\bibinfo {title} {{Maximally
  localized generalized Wannier functions for composite energy bands}},}\
  }\href {\doibase 10.1103/PhysRevB.56.12847} {\bibfield  {journal} {\bibinfo
  {journal} {Phys. Rev. B}\ }\textbf {\bibinfo {volume} {56}},\ \bibinfo
  {pages} {12847} (\bibinfo {year} {1997})},\ \Eprint
  {http://arxiv.org/abs/cond-mat/9707145} {arXiv:cond-mat/9707145} \BibitemShut
  {NoStop}%
\bibitem [{\citenamefont {J{\'{o}}nsson}\ \emph {et~al.}(2017)\citenamefont
  {J{\'{o}}nsson}, \citenamefont {Lehtola}, \citenamefont {Puska},\ and\
  \citenamefont {J{\'{o}}nsson}}]{Jonsson2017_460}%
  \BibitemOpen
  \bibfield  {author} {\bibinfo {author} {\bibfnamefont {E.~{\"{O}}.}\
  \bibnamefont {J{\'{o}}nsson}}, \bibinfo {author} {\bibfnamefont
  {S.}~\bibnamefont {Lehtola}}, \bibinfo {author} {\bibfnamefont
  {M.}~\bibnamefont {Puska}}, \ and\ \bibinfo {author} {\bibfnamefont
  {H.}~\bibnamefont {J{\'{o}}nsson}},\ }\bibfield  {title} {\enquote {\bibinfo
  {title} {{Theory and applications of generalized Pipek--Mezey Wannier
  functions}},}\ }\href {\doibase 10.1021/acs.jctc.6b00809} {\bibfield
  {journal} {\bibinfo  {journal} {J. Chem. Theory Comput.}\ }\textbf {\bibinfo
  {volume} {13}},\ \bibinfo {pages} {460} (\bibinfo {year} {2017})},\ \Eprint
  {http://arxiv.org/abs/1608.06396} {arXiv:1608.06396} \BibitemShut {NoStop}%
\bibitem [{\citenamefont {Su}\ and\ \citenamefont
  {Goddard}(2007)}]{Su2007_185003}%
  \BibitemOpen
  \bibfield  {author} {\bibinfo {author} {\bibfnamefont {J.~T.}\ \bibnamefont
  {Su}}\ and\ \bibinfo {author} {\bibfnamefont {W.~A.}\ \bibnamefont
  {Goddard}},\ }\bibfield  {title} {\enquote {\bibinfo {title} {Excited
  electron dynamics modeling of warm dense matter},}\ }\href {\doibase
  10.1103/PhysRevLett.99.185003} {\bibfield  {journal} {\bibinfo  {journal}
  {Phys. Rev. Lett.}\ }\textbf {\bibinfo {volume} {99}},\ \bibinfo {pages}
  {185003} (\bibinfo {year} {2007})}\BibitemShut {NoStop}%
\bibitem [{\citenamefont {Trepte}, \citenamefont {Schwalbe},\ and\
  \citenamefont {Seifert}(2015)}]{Trepte2015_17122}%
  \BibitemOpen
  \bibfield  {author} {\bibinfo {author} {\bibfnamefont {K.}~\bibnamefont
  {Trepte}}, \bibinfo {author} {\bibfnamefont {S.}~\bibnamefont {Schwalbe}}, \
  and\ \bibinfo {author} {\bibfnamefont {G.}~\bibnamefont {Seifert}},\
  }\bibfield  {title} {\enquote {\bibinfo {title} {{Electronic and magnetic
  properties of DUT-8 (Ni)}},}\ }\href {\doibase 10.1039/c5cp01881a} {\bibfield
   {journal} {\bibinfo  {journal} {Phys. Chem. Chem. Phys.}\ }\textbf {\bibinfo
  {volume} {17}},\ \bibinfo {pages} {17122} (\bibinfo {year}
  {2015})}\BibitemShut {NoStop}%
\bibitem [{\citenamefont {Schwalbe}\ \emph {et~al.}(2016)\citenamefont
  {Schwalbe}, \citenamefont {Trepte}, \citenamefont {Seifert},\ and\
  \citenamefont {Kortus}}]{Schwalbe2016_8075}%
  \BibitemOpen
  \bibfield  {author} {\bibinfo {author} {\bibfnamefont {S.}~\bibnamefont
  {Schwalbe}}, \bibinfo {author} {\bibfnamefont {K.}~\bibnamefont {Trepte}},
  \bibinfo {author} {\bibfnamefont {G.}~\bibnamefont {Seifert}}, \ and\
  \bibinfo {author} {\bibfnamefont {J.}~\bibnamefont {Kortus}},\ }\bibfield
  {title} {\enquote {\bibinfo {title} {{Screening for high-spin metal organic
  frameworks (MOFs): density functional theory study on DUT-8(M1{,}M2) (with Mi
  = V{,}...{,}Cu)}},}\ }\href {\doibase 10.1039/c5cp07662e} {\bibfield
  {journal} {\bibinfo  {journal} {Phys. Chem. Chem. Phys.}\ }\textbf {\bibinfo
  {volume} {18}},\ \bibinfo {pages} {8075} (\bibinfo {year}
  {2016})}\BibitemShut {NoStop}%
\bibitem [{\citenamefont {Trepte}\ \emph {et~al.}(2017)\citenamefont {Trepte},
  \citenamefont {Schaber}, \citenamefont {Schwalbe}, \citenamefont {Drache},
  \citenamefont {Senkovska}, \citenamefont {Kaskel}, \citenamefont {Kortus},
  \citenamefont {Brunner},\ and\ \citenamefont {Seifert}}]{Trepte2017_10020}%
  \BibitemOpen
  \bibfield  {author} {\bibinfo {author} {\bibfnamefont {K.}~\bibnamefont
  {Trepte}}, \bibinfo {author} {\bibfnamefont {J.}~\bibnamefont {Schaber}},
  \bibinfo {author} {\bibfnamefont {S.}~\bibnamefont {Schwalbe}}, \bibinfo
  {author} {\bibfnamefont {F.}~\bibnamefont {Drache}}, \bibinfo {author}
  {\bibfnamefont {I.}~\bibnamefont {Senkovska}}, \bibinfo {author}
  {\bibfnamefont {S.}~\bibnamefont {Kaskel}}, \bibinfo {author} {\bibfnamefont
  {J.}~\bibnamefont {Kortus}}, \bibinfo {author} {\bibfnamefont
  {E.}~\bibnamefont {Brunner}}, \ and\ \bibinfo {author} {\bibfnamefont
  {G.}~\bibnamefont {Seifert}},\ }\bibfield  {title} {\enquote {\bibinfo
  {title} {{The origin of the measured chemical shift of $^{129}$Xe in UiO-66
  and UiO-67 revealed by DFT investigations}},}\ }\href {\doibase
  10.1039/c7cp00852j} {\bibfield  {journal} {\bibinfo  {journal} {Phys. Chem.
  Chem. Phys.}\ }\textbf {\bibinfo {volume} {19}},\ \bibinfo {pages} {10020}
  (\bibinfo {year} {2017})}\BibitemShut {NoStop}%
\bibitem [{\citenamefont {Trepte}\ \emph {et~al.}(2018)\citenamefont {Trepte},
  \citenamefont {Schwalbe}, \citenamefont {Schaber}, \citenamefont {Krause},
  \citenamefont {Senkovska}, \citenamefont {Kaskel}, \citenamefont {Brunner},
  \citenamefont {Kortus},\ and\ \citenamefont {Seifert}}]{Trepte2018_25039}%
  \BibitemOpen
  \bibfield  {author} {\bibinfo {author} {\bibfnamefont {K.}~\bibnamefont
  {Trepte}}, \bibinfo {author} {\bibfnamefont {S.}~\bibnamefont {Schwalbe}},
  \bibinfo {author} {\bibfnamefont {J.}~\bibnamefont {Schaber}}, \bibinfo
  {author} {\bibfnamefont {S.}~\bibnamefont {Krause}}, \bibinfo {author}
  {\bibfnamefont {I.}~\bibnamefont {Senkovska}}, \bibinfo {author}
  {\bibfnamefont {S.}~\bibnamefont {Kaskel}}, \bibinfo {author} {\bibfnamefont
  {E.}~\bibnamefont {Brunner}}, \bibinfo {author} {\bibfnamefont
  {J.}~\bibnamefont {Kortus}}, \ and\ \bibinfo {author} {\bibfnamefont
  {G.}~\bibnamefont {Seifert}},\ }\bibfield  {title} {\enquote {\bibinfo
  {title} {{Theoretical and experimental investigations of $^{129}$Xe NMR
  chemical shift isotherms in metal-organic frameworks}},}\ }\href {\doibase
  10.1039/c8cp04025g} {\bibfield  {journal} {\bibinfo  {journal} {{Phys. Chem.
  Chem. Phys.}}\ }\textbf {\bibinfo {volume} {20}},\ \bibinfo {pages} {25039}
  (\bibinfo {year} {2018})}\BibitemShut {NoStop}%
\bibitem [{\citenamefont {Trepte}\ and\ \citenamefont
  {Schwalbe}(2019)}]{Trepte2019_1}%
  \BibitemOpen
  \bibfield  {author} {\bibinfo {author} {\bibfnamefont {K.}~\bibnamefont
  {Trepte}}\ and\ \bibinfo {author} {\bibfnamefont {S.}~\bibnamefont
  {Schwalbe}},\ }\bibfield  {title} {\enquote {\bibinfo {title} {{Systematic
  analysis of porosities in metal-organic frameworks}},}\ }\href {\doibase
  10.26434/chemrxiv.10060331.v1} {\bibfield  {journal} {\bibinfo  {journal}
  {{ChemRxiv}}\ } (\bibinfo {year} {2019}),\
  10.26434/chemrxiv.10060331.v1}\BibitemShut {NoStop}%
\end{thebibliography}%
\end{document}